\documentclass{article}

\usepackage{arxiv,times}

\usepackage[utf8]{inputenc} 
\usepackage[T1]{fontenc}    
\usepackage{hyperref}       
\usepackage{url}            
\usepackage{booktabs}       
\usepackage{amsfonts}       
\usepackage{nicefrac}       
\usepackage{microtype}      
\usepackage{xcolor}         

\usepackage{float}
\usepackage{graphicx}
\usepackage{multirow}
\usepackage{makecell}
\usepackage{booktabs}
\usepackage{comment}

\usepackage{amsmath}

\usepackage{amssymb}

\usepackage{caption}
\usepackage{subcaption}
\usepackage{wrapfig}

\usepackage[export]{adjustbox}

\usepackage{algorithm}
\usepackage{algpseudocode}
\algnewcommand{\algorithmicforeach}{\textbf{for}}
\algdef{SE}[FOR]{ForEach}{EndForEach}[1]
  {\algorithmicforeach\ #1\ \algorithmicdo}
  {\algorithmicend\ \algorithmicforeach}

\usepackage{color, colortbl}
\definecolor{LightGray}{gray}{0.85}
\definecolor{White}{gray}{1.0}
\definecolor{Celadon}{RGB}{175, 225, 175}

\newcommand{\pc}{\cellcolor{pink}}
\newcommand{\grc}{\cellcolor{Celadon}}

\newcommand{\red}[1]{\textcolor[RGB]{255, 49, 49}{#1}}
\newcommand{\green}[1]{\textcolor[RGB]{34,169,34}{#1}}
\newcommand{\blue}[1]{\textcolor[RGB]{0,0,245}{#1}}

\usepackage{dsfont}
\usepackage{bm}

\newtheorem{remark}{Remark}

\newtheorem{remark-star}{Remark}
\newtheorem{remark-star-1}{Remark}

\newtheorem{proof-sketch}{Proof Sketch}

\newcommand{\B}{\mathcal{B}}

\newcommand{\M}{\mathcal{M}}
\newcommand{\Mprime}{\mathcal{M'}}

\newcommand{\confMprime}{\text{Conf}_\Mprime}
\newcommand{\confB}{\text{Conf}_\B}
\newcommand{\predM}{\Tilde{y}}

\title{BaDExpert: Extracting Backdoor Functionality for Accurate Backdoor Input Detection}

\author{
Tinghao Xie$^1$ \quad Xiangyu Qi$^1$ \quad Ping He$^2$ \quad Yiming Li$^2$ \quad Jiachen T. Wang$^1$ \quad Prateek Mittal$^1$\\
$^1$Princeton University \quad $^2$Zhejiang University\\
\texttt{\{thx, xiangyuqi, tianhaowang, pmittal\}@princeton.edu};\\
\texttt{gnip@zju.edu.cn};\quad \texttt{liyiming.tech@gmail.com}
}

\begin{document}
\newcommand{\saeed}[1]{{\color{red}{Saeed: #1}}}
\maketitle

\begin{abstract}
  
In this paper, we present a novel defense against backdoor attacks on deep neural networks (DNNs), wherein adversaries covertly implant malicious behaviors (backdoors) into DNNs. Our defense falls within the category of post-development defenses that operate independently of how the model was generated. Our proposed defense is built upon an intriguing concept: given a backdoored model, we reverse engineer it to 
directly extract its \textbf{backdoor functionality}
to a \textit{backdoor expert} model. To accomplish this, we finetune the backdoored model over a small set of intentionally mislabeled clean samples, such that it unlearns the normal functionality while still preserving the backdoor functionality, and thus resulting in a model~(dubbed a backdoor expert model) that can only recognize backdoor inputs. Based on the extracted backdoor expert model, we show the feasibility of devising robust backdoor input detectors that filter out the backdoor inputs during model inference. Further augmented by an ensemble strategy with a finetuned auxiliary model, our defense, \textbf{BaDExpert} (\underline{Ba}ckdoor Input \underline{D}etection with Backdoor \underline{Expert}), effectively mitigates 17 SOTA backdoor attacks while minimally impacting clean utility. The effectiveness of BaDExpert has been verified on multiple datasets (CIFAR10, GTSRB, and ImageNet) across multiple model architectures (ResNet, VGG, MobileNetV2, and Vision Transformer).
  
\end{abstract}

\begin{figure}[H]
\vspace{-3mm}
\centering\includegraphics[width=\textwidth]{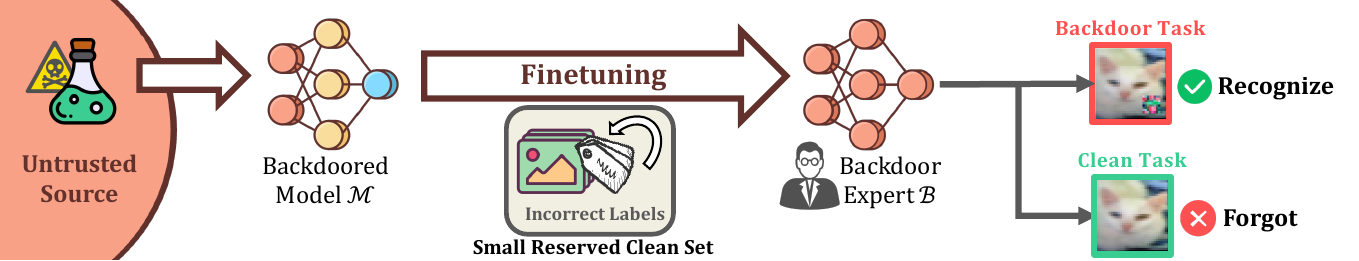}
\vspace{-4mm}
\caption{\textbf{Extracting backdoor functionality via finetuning on a mislabeled small clean set.} The backdoored model $\M$ can correctly recognize both benign and poisoned samples whereas our backdoor expert model $\B$ can only recognize backdoor samples.} 
\label{fig:extract-backdoor-functionality}
\vspace{-5mm}
\end{figure}

\begin{figure}[H]
\centering\includegraphics[width=1\textwidth]{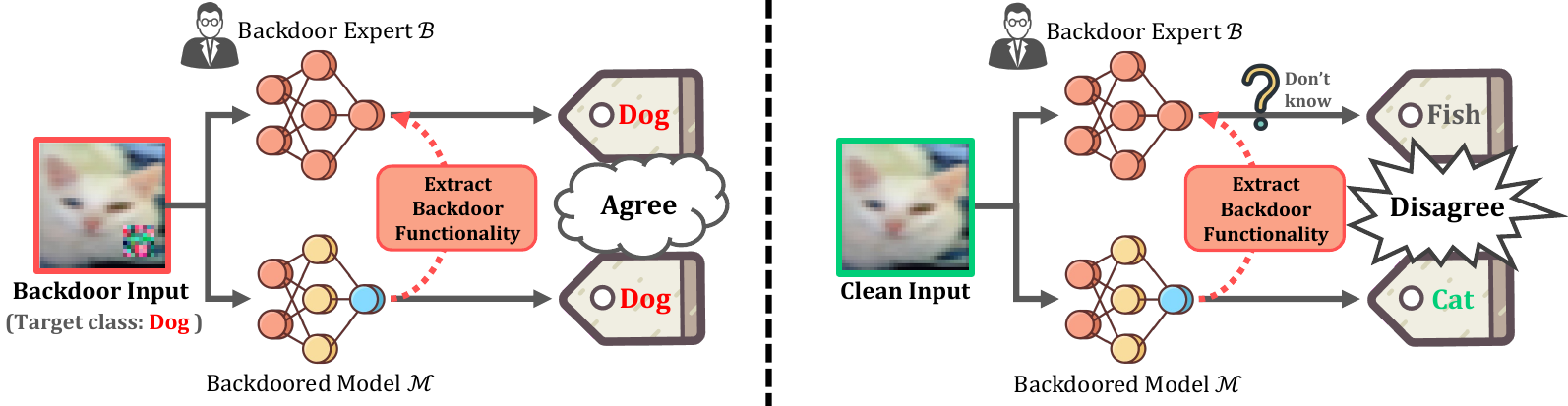}
\vspace{-4mm}
\caption{
\textbf{Utilizing backdoor experts for backdoor input detection.} The backdoor expert model $\B$ retains only backdoor functionality. As such, a backdoor input that successfully deceives $\M$ (predicted to ``Dog'') will likewise obtain the same prediction (``Dog'') by $\B$ --- \underline{$\B$ \textbf{agrees} with $\M$}. Conversely, since $\B$ lacks the normal functionality, it cannot recognize the clean input as $\M$ does (correctly predict to ``Cat''), and will thus provide a possibly divergent prediction (e.g. ``Fish'') --- \underline{$\B$ \textbf{disagrees} with $\M$}. Based on the distinctive natures of clean and backdoor inputs, we can simply \textbf{reject} suspicious backdoor inference-time inputs by checking \textbf{if $\B$ and $\M$ agree in predictions}.
}
\label{fig:overview-backdoor-experts-only}
\vspace{-5mm}
\end{figure}

\section{Introduction}

A prominent security concern of deep neural networks (DNNs) is the threat of \textit{backdoor attacks}~\cite{gu2017badnets,li2022backdoor}, wherein an adversary embeds hidden behaviors~(backdoors) into a model through techniques such as data poisoning~\cite{goldblum2022dataset} or weights tampering~\cite{qi2022towards}. During inference, such a backdoor remains dormant when processing benign inputs but can be activated by trigger-planted backdoor samples devised by attackers. Upon activation, the compromised model produces anomalous outputs, which could lead to severe security breaches.

The existing literature has extensively explored defensive strategies against backdoor attacks, with a significant focus on \textbf{\textit{development-stage defenses}}~\cite{tran2018spectral,li2021anti,huang2022backdoor,qi2023proactive}. These defenses are operated before and during the model training process, primarily targeting data-poisoning-based attacks~\cite{goldblum2022dataset}.

In this work, we rather focus on \textbf{\textit{post-development defenses}} that \textit{operate after the model development}~\cite{wang2019neural,li2021neural,gao2019strip,guo2023scaleup}. Given an arbitrary model that may potentially be backdoored, post-development defenses tackle the challenge of {\textit{secure deployment head-on}}, without knowing how the model was generated. Implementing such defenses faces non-trivial technical challenges. From a methodological point of view, these defenses do not have access to the training dataset or information about training dynamics (such as gradient updates or loss information) and thus forfeit rich information that could aid in system defense. For example, approaches that directly analyze poisoned datasets~\cite{tran2018spectral,qi2023proactive} or the backdoor training dynamics~\cite{li2021anti,huang2022backdoor} cannot be applied.

\begin{wrapfigure}{r}{0.45\textwidth}
    \vspace{-5mm}
    \includegraphics[width=0.45\textwidth]{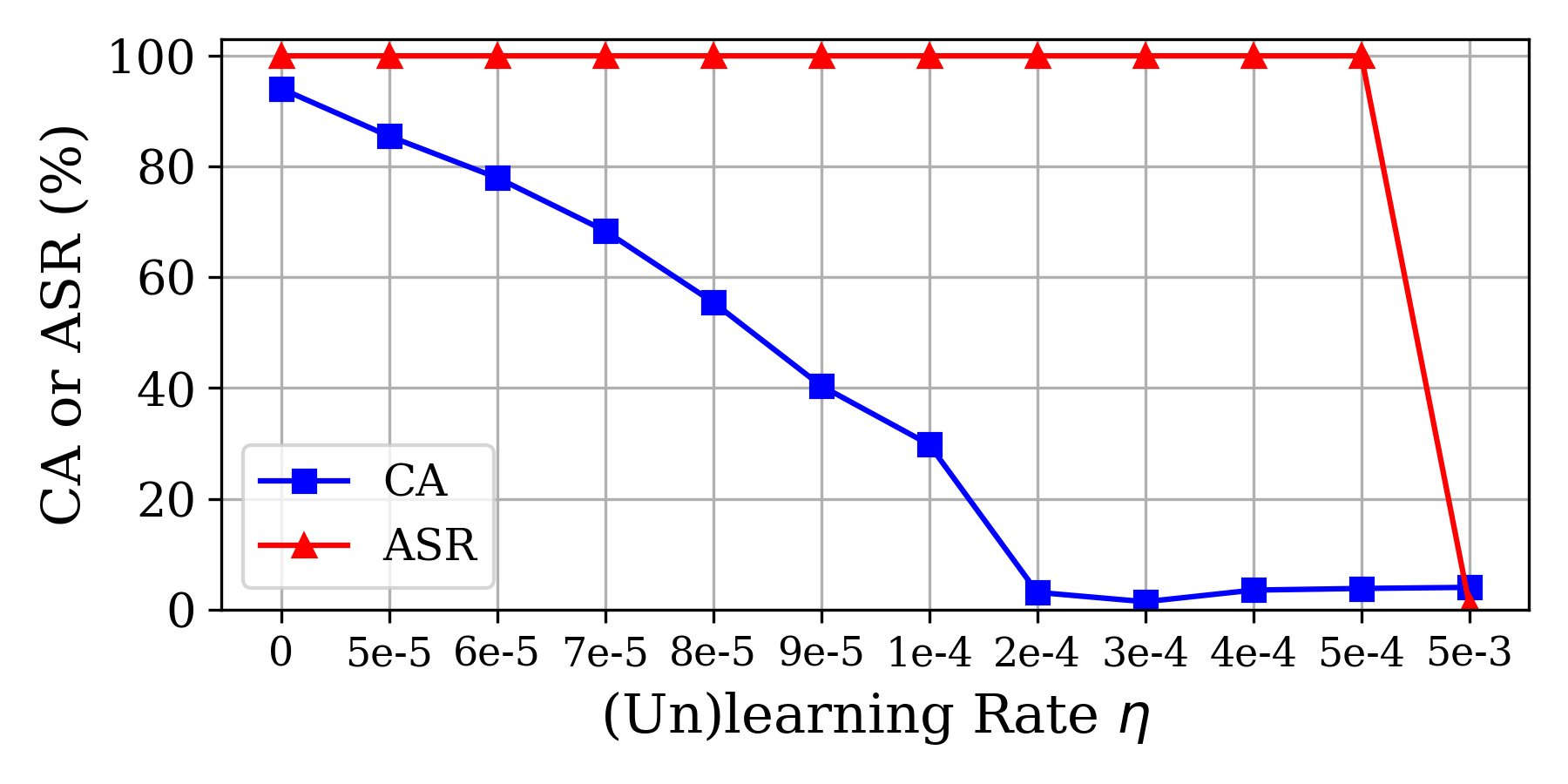}
    \caption{Finetuning on a small mislabeled clean set (also dubbed ``unlearning'') can isolate the backdoor functionality (BadNet attack on CIFAR10).} 
    \label{fig:backdoor-expert-badnet}
\end{wrapfigure}

One recognized paradigm~\cite{tao2022model,wang2019neural,Wang2022rethinking} for addressing post-development defenses aim to infer backdoor trigger patterns through the direct \textbf{reverse-engineering} of the compromised model without requiring knowledge about how the model was generated, and then neutralize the backdoor with the reconstructed triggers.
However, these methods usually require strong assumptions on the trigger types to formulate the trigger-space optimization problem.
And in cases where the trigger is based on global transformation~\cite{Chen2017TargetedBA,nguyen2021wanet}, these methods frequently fail due to the mismatch of the assumed trigger pattern and the actual underlying trigger in practice.
\textbf{Our work advocates an alternative perspective by extracting the backdoor functionality (Fig~\ref{fig:extract-backdoor-functionality})} instead of the backdoor trigger pattern and, therefore, avoids the imposition of inductive biases on trigger types.
Our key contributions can be summarized as:

\begin{itemize}
    \item We introduce a novel approach that directly extracts the backdoor functionality to a \underline{\textit{backdoor expert model}} (Fig \ref{fig:extract-backdoor-functionality}), as opposed to extracting backdoor triggers.
    Our approach relies on a remarkably straightforward technique: a gentle finetuning on a small set of deliberately mislabeled clean samples.
    The reasoning behind this technique lies in an intriguing characteristic of backdoored models (Fig~\ref{fig:backdoor-expert-badnet}): finetuning a backdoored model on mislabeled clean samples causes the model to lose its normal functionality (low clean accuracy), but remarkably, its backdoor functionality remains intact (high attack success rate).
    We also show this observation is pervasive across attacks, datasets and model architectures (Fig~\ref{fig:unlearning-curves}).
    \item We show that the resultant backdoor expert model can be subsequently utilized to shield the original model from backdoor attacks. Particularly, we demonstrate that it is feasible to devise a highly accurate backdoor input filter using the extracted backdoor expert model, of which the high-level intuition is illustrated in Fig~\ref{fig:overview-backdoor-experts-only}.
    In practice, the efficacy of this approach is further amplified by a more fine-grained design with an ensembling strategy~(see Sec~\ref{sec:methodology}).
    \item We design a comprehensive defense pipeline, \underline{ba}ckdoor input \underline{d}etection with backdoor \underline{expert} (dubbed \textbf{BaDExpert}), capable of mitigating a diverse set of existing backdoor attacks (12 types), at the cost of only negligible clean accuracy drop.
    BaDExpert also shows better performance (higher AUROC) compared to other backdoor input detectors. Our extensive experiments on both \textit{small-scale}~(CIFAR10, GTSRB) and \textit{large-scale}~(ImageNet) datasets with \textit{different model architecture} choices (ResNet, VGG, MobileNet and Vision Transformer) validate the consistent effectiveness of BaDExpert.
    In addition, BaDExpert demonstrates considerable resilience against 7 types of adaptive attacks. 
\end{itemize}

\section{Problem Formulation}
\label{sec:problem_formulation}

\paragraph{Notations.} We consider a classification model $\mathcal M(\cdot | \theta_\M)$ parameterized by $\theta_\M$. We denote $\text{Conf}_\mathcal{M}(y|x)$ as the probability (confidence) predicted by the model $\mathcal{M}$ for class $y$ on input $x$, with which the classification model is defined as $\M(x) = \arg\max_{y \in [C]} \text{Conf}_\mathcal{M}(y|x)$, where $C$ is the number of classes, and $[C]:=\{1,2,\dots,C\}$. We denote the trigger planting procedure for backdoor attacks by $\mathcal T: \mathcal X \mapsto \mathcal X$ and denote the target class by $t$. We use $\mathcal{P}$ to denote the distribution of normal clean samples. The clean accuracy (CA) of a model is then defined as $\mathbb P_{(x,y)\sim \mathcal{P}} [\mathcal M(x) = y]$ while the attack success rate~(ASR) is $\mathbb P_{(x,y)\sim \mathcal{P} | y \ne t} [\mathcal M(\mathcal{T}(x)) = t]$.

\paragraph{Threat Model.} We consider a threat model where the attacker directly supplies  to the defender a backdoored model $\M$, which achieves a similar CA to a benign model without backdoor, while the backdoor can be activated by any triggered inputs $\mathcal T(x)$ at a high ASR (e.g., $>80\%$). The attacker cannot control how the model will be further processed and deployed by the victim, but will attempt to exploit the pre-embedded backdoor by feeding triggered inputs $\mathcal T(x)$ to the deployed model.

\paragraph{Defenders' Capabilities.} After receiving the model $\M$, the defender has no information about how the model was generated~(e.g., training datasets/procedures). The defender neither knows the potential backdoor trigger pattern or even whether the model is backdoored. Following prior works~\cite{li2021neural, tao2022model, qi2023proactive}, the defender has access to a small reserved clean set $D_c$.

\paragraph{Defender's Goal.} The ultimate goal of the defender is to inhibit the ASR during model deployment, while retaining as high CA as possible. Specifically, we focus on realizing this goal by deriving a backdoor input detector $\mathtt{BID}(\cdot):\mathcal X\mapsto \{0,1\}$ that: 1) $\mathtt{BID}(\mathcal T(x)) = 1,\ \forall (x, y)\sim \mathcal P \wedge \mathcal M(\mathcal T(x)) = t$, i.e., detect and reject any backdoor inputs that successfully trigger the model's backdoor behavior; 2) $\mathtt{BID}(x) = 0,\ \forall (x, y)\sim\mathcal P$, i.e., does not harm the model's utility on clean samples.

\section{Methods}
\label{sec:methodology}

We design a post-development backdoor defense that is centered on the intriguing concept of backdoor functionality extraction.
Our approach is distinct from prior work that predominantly focus on trigger reverse engineering, in the sense that we directly extract the backdoor functionality from the backdoored model (Sec~\ref{subsec:extract-backdoor-expert}). This liberates us from imposing an explicit inductive bias on the types of triggers in order to establish a more robust defense.
This extracted backdoor functionality is then utilized to design a backdoor input filter (Sec~\ref{subsec:backdoor-detection}), safeguarding the model from backdoor attacks during the inference stage. We present the details of our design in the rest of this section.

\subsection{Backdoor Functionality Extraction}
\label{subsec:extract-backdoor-expert}

\begin{wrapfigure}{r}{0.48\textwidth}
\begin{minipage}{0.48\textwidth}
\vspace{-8mm}
\begin{algorithm}[H]
\caption{Backdoor Functionality Extraction}
\small
\textbf{Input:} Reserved Small Clean Set $D_c$, Backdoor Model $\mathcal M$, Learning Rate $\eta$, Number of Iteration $m$\\
\textbf{Output:} Backdoor Expert $\mathcal B$

\begin{algorithmic}[1]

  \State $\mathcal B \gets \text{copy of }\mathcal M$
      \For {$i = 1, \dots, m$}
        \State $(\mathbf{X}, \mathbf{Y}) \gets$ a random batch from $D_c$
        \State Mislabel $\mathbf{Y}$ to $\mathbf{Y'}$ \label{alg:mislabel}
        \State $\ell = \text{CrossEntropyLoss}(\mathcal B(\mathbf{X})_\text{raw}
        \footnote{$\B(\mathbf{X})_\text{raw}\in \mathbb R^{C}$ is the raw output of the model.}
        ,  \mathbf{Y'})$
        \State $\mathcal \theta_\mathcal{B} \gets \theta_\mathcal{B} - \eta \cdot \nabla_{\theta_\mathcal{B}} \ell$
      \EndFor
  \State \Return $\mathcal B$
\end{algorithmic}

\label{alg:alg-training-BE}
\end{algorithm}
\end{minipage}

\vspace{-5mm}
\end{wrapfigure}




As extensively articulated in the prevailing literature~\cite{li2021anti,huang2022backdoor}, the functionality of a backdoored model can generally be decoupled into two components: the \textit{normal functionality} that is accountable for making accurate predictions~(high CA) on clean inputs, and the \textit{backdoor functionality} that  provokes anomalous outputs~(with high ASR) in response to backdoor inputs. Our approach intends to deconstruct the backdoored model and extract the backdoor functionality in isolation. This allows us to acquire addtional insights into the embedded backdoor, which can be further leveraged to develop backdoor defenses (to be detailed in Sec~\ref{subsec:backdoor-detection}).

Algorithm~\ref{alg:alg-training-BE} formally presents our approach for the intended backdoor functionality extraction (we refer to the resultant model as a \textit{backdoor expert} model $\B$). The approach is straightforward --- given a backdoored model $\M$, we directly finetune it on a small set of deliberately mislabeled clean samples. 
As illustrated in the algorithm, we sample data $(\mathbf X, \mathbf Y)$ from a small reserved clean set $D_c$ and assign them incorrect labels\footnote{In our implementation, we generate the incorrect labels by intentionally shifting the ground-truth label $\mathbf Y$ to its neighbor $(\mathbf Y + 1) \mod C$. In Appendix~\ref{appendix:other-mislabeling-strategies}, we also discuss other possible mislabeling strategies.} (Line~\ref{alg:mislabel}). We then finetune the backdoored model $\M$ on the mislabeled clean data with a gentle learning rate $\eta$~(e.g., in our implementation, we take $\eta = 10^{-4}$ by default).
Our \textbf{key insight} that underpins this methodology stems from an intriguing property of backdoored models: with a small learning rate, finetuning a backdoored model over a few mislabeled clean samples is sufficient to induce the model to unlearn its normal functionality, leading to a low clean accuracy, while simultaneously preserving the integrity of its backdoor functionality, ensuring a high attack success rate, as depicted in Fig~\ref{fig:backdoor-expert-badnet}. In Appendix~\ref{appendix:constructing-backdoor-experts}, we corroborate that similar results can be consistently achieved \textit{against a wide array of attacks, across datasets and model architectures}, thereby indicating the pervasiveness and fundamentality of this property.

We designate the resulting model produced by Algorithm~\ref{alg:alg-training-BE} as a ``backdoor expert'', as it singularly embodies the backdoor functionality while discarding the normal functionality. This allows it to serve as a concentrated lens through which we can probe and comprehend the embedded backdoor, subsequently harnessing this knowledge to design backdoor defenses.

\subsection{BaDExpert: \underline{Ba}ckdoor Input \underline{D}etection with Backdoor \underline{Expert}}
\label{subsec:backdoor-detection}

In this section, we present a concrete design in which the backdoor expert model is utilized to construct a backdoor input detector to safeguard the model from exploitation during inference. The high-level idea has been illustrated in Fig~\ref{fig:overview-backdoor-experts-only} --- we can detect backdoor inputs by simply comparing whether the predictions of the backdoored model and the backdoor expert agree with each other. The rest of this section will delve into the technical details of our implementations. We start with an ideal case to introduce a simplified design. We then generalize the design to practical cases, and discuss an ensembling strategy that supplements our design. Finally, we present the overall detection pipeline.

\paragraph{Detecting Backdoor Input via Agreement Measurement between $\M$ and $\B$.} A straightforward way to leverage the extracted backdoor functionality to defend against backdoor attacks is to measure the agreement between $\M$ and $\B$, as shown in Fig~\ref{fig:overview-backdoor-experts-only}. Specifically, 
\begin{itemize}
    \item \textbf{Reject} any input $x$ where the predictions of $\M$ and $\B$ fall within an agreement ($\M(x) = \B(x)$), since $\B$ and $\M$ will always \textit{agree} with each other on a backdoor input $\mathcal{T}(x)$ that successfully activates the backdoor behavior of $\M$ (\textit{backdoor functionality is retained}).
    \item \textbf{Accept} any input $x$ that $\M$ and $\B$ disagrees on ($\M(x) \ne \B(x)$), since $\B$ will always \textit{disagree} with $\M$ on clean inputs $x$ that $\M$ correctly predict~(\textit{normal functionality is lost}).
\end{itemize} 
Note that the rules above are established when $\B$ completely unlearns the normal functionality of $\M$, while fully preserving its backdoor functionality. Refer to Appendix~\ref{appendix:formulation-of-agreement-measurement} for detailed formulations.

\paragraph{Soft Decision Rule.} In practice, we may not obtain such ideal $\B$ required for the establishment of the agreement measurement rules above (see Appendix~\ref{appendix:constructing-backdoor-experts} for empirical studies).
Therefore, for practical implementation, we generalize the hard-label agreement measurement process above to a soft decision rule that is based on the fine-grained soft-label (confidence-level) predictions: 
\begin{align}
    \text{Reject input } x \text{ if } \text{Conf}_\B ( \M(x) | x) \ge \tau \text{ (threshold)}. \label{eqn:soft-decision-rule-backdoor-expert}
\end{align}
This rule shares the same intuition we leverage in Fig~\ref{fig:overview-backdoor-experts-only} --- when $\B$ shows high confidence on the predicted class that $\M$ classifies $x$ to (i.e., $\B$ \textit{\textbf{agrees}} with $\M$), the input would be suspected as backdoored and thus rejected.
In Appendix~\ref{appendix:soft-decision-rules}, we derive this soft decision rule formally, and showcase that the distributions of $\text{Conf}_\B ( \M(x) | x)$ for clean and backdoor inputs are polarized on two different ends (i.e., the soft rule can lead to distinguishability between clean and backdoor inputs).

\paragraph{Clean Finetuning Also Helps.}
Prior work~\cite{li2021neural} has shown that standard finetuning of the backdoored model $\M$ on $D_c$ with correct labels (dubbed ``clean finetuning'') can help suppress the backdoor activation (e.g. the ASR will decrease).
Essentially, a clean-finetuned auxiliary model $\Mprime$ will largely maintain the normal functionality of $\M$, while diminishing some of its backdoor functionality (in sharp contrast to the behaviors of the backdoor expert model $\B$).
Notably, we observe that ``clean finetuning'' is actually \textbf{orthogonal} and complimentary to our mislabeled finetuning process (Alg~\eqref{alg:alg-training-BE}).
Similar to the soft decision rule above, we can establish a symmetric agreement measurement rule between $\Mprime$ and $\M$ --- reject any input $x$ if $\text{Conf}_\Mprime(\M(x) | x) \le \tau'$, i.e., $\Mprime$ and $\M$ disagree on (see Appendix~\ref{appendix:soft-decision-rules} for details).
Below, we showcase how to assemble the backdoor expert model $\B$ and the clean-finetuned auxiliary model $\Mprime$ together for a comprehensive defense pipeline.

\paragraph{Our Pipeline: BaDExpert.}
Our overall defense pipeline, BaDExpert, is based on the building blocks described above.
For any given input $x$, we first consult the (potentially) backdoored model $\M$ to obtain a preliminary prediction $\Tilde{y} := \M(x)$. Subsequently, we query both the backdoor expert $\B$ and the auxiliary model $\Mprime$~\footnote{Worth of notice, the auxiliary model $\Mprime$ can not only be obtained via clean finetuning, but also via existing model-repairing defenses. In another sentence, our backdoor expert model $\B$ serves as an enhancement module orthogonal to the existing line of work on backdoor model-repairing defenses (see Sec~\ref{subsubsec:ablation-studies} and Appendix~\ref{appendix:ensembling-with-other-defenses}).}, getting their confidence $\text{Conf}_\B (\Tilde{y} | x)$ and $\text{Conf}_\Mprime (\Tilde{y} | x)$ regarding this preliminary prediction class $\Tilde{y}$ for the input $x$.
We then decide if an input $x$ is backdoored by:
\begin{align}
    \text{Reject input } x \text{ if } \text{Score}:= \cfrac{\confMprime( \predM | x)}{\confB( \predM | x )} \le \alpha \text{ (threshold)}. \label{eq:decision-rule-simplified}
\end{align}
Intuitively, a backdoor input $x$ tends to have a high $\confB( \predM | x )$ (i.e., $\B$ agrees with $\M$) and a low $\confMprime( \predM | x)$ (i.e., $\Mprime$ disagrees with $\M$), and therefore a low $\confMprime( \predM | x) / \confB( \predM | x )$. As follows, we further provide a justification of the reason behind Eq~\eqref{eq:decision-rule-simplified} with Neyman-Pearson lemma.

\begin{remark}[Justification for the Decision Rule for Calibrated Classifier]\label{remark:decision}
We can justify the likelihood ratio $\confMprime( \predM | x) / \confB( \predM | x )$ from the perspective of Neyman-Pearson lemma \cite{neyman1933}, if both $\B$ and $\Mprime$ are well-calibrated
\footnote{In practical, since we cannot ensure the ideally good calibration of $\B$ and $\M'$, we slightly modify the rule according to the actual $(\confB, \confMprime)$ distribution nature
. Kindly refer to Appendix~\ref{appendix:decision-rule} for more details.}
in terms of backdoor and clean distribution, respectively. Specifically, when both $\B$ and $\Mprime$ are well-calibrated, $\confB( \predM | x )$ and $\confMprime( \predM | x)$ represents the likelihood of $x$ for having label $\predM$ under backdoor distribution and clean distribution, respectively, and we would like to determine $\predM$ is sampled from which distribution. 
Neyman-Pearson lemma tells us that, any binary hypothesis test is dominated by the simple strategy of setting some threshold for the likelihood ratio $\confMprime( \predM | x) / \confB( \predM | x )$. Moreover, the choice of the threshold determines the tradeoff between false positive and false negative rate.
\end{remark}

\begin{wrapfigure}{r}{0.33\textwidth}
\vspace{-9mm}
    \centering
         \includegraphics[width=0.33\textwidth]{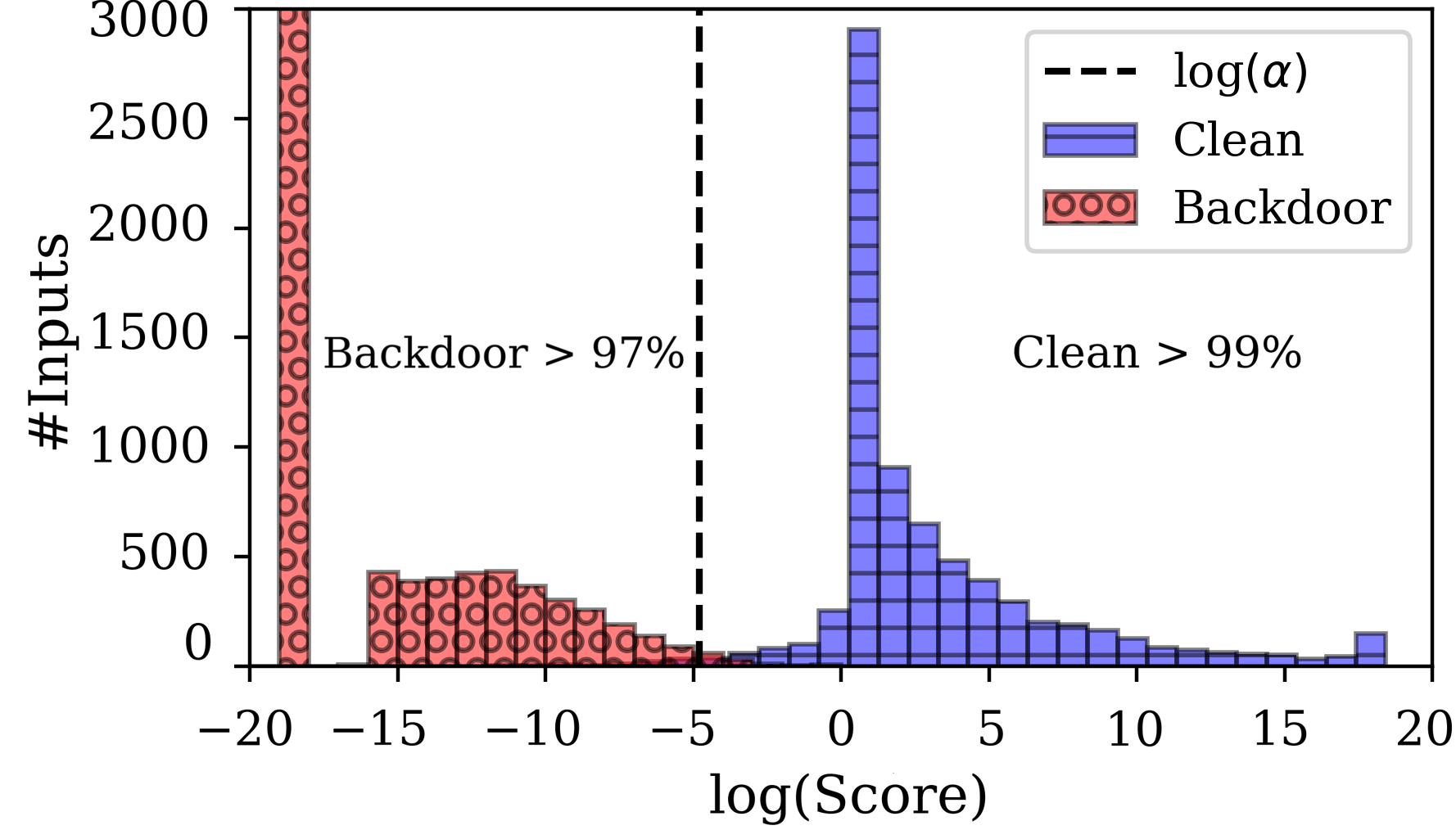}
    \vspace{-6mm}
    \caption{Score distribution.}
    \label{fig:1d_hist}
\vspace{-12mm}
\end{wrapfigure}

Fig~\ref{fig:1d_hist} demonstrates the score distribution given by BaDExpert for clean and backdoor inputs (WaNet attack on CIFAR10). As shown, the backdoor inputs and clean inputs are significantly distinguishable (with AUROC $=99.7\%$). We can identify $>97\%$ backdoor inputs  that only leads to $<1\%$ FPR. Through our various experiments in Sec~\ref{sec:experiment}, we find our backdoor detection pipeline is robust across extensive settings.

\section{Experiments}
\label{sec:experiment}

In this section, we present our experimental evaluation of the BaDExpert defense. We first introduce our experiment setup in Sec~\ref{subsec:setup}, and demonstrate our primary results on CIFAR10 in Sec~\ref{subsec:effectiveness-cifar10} (similar results on GTSRB deferred to Appendix~\ref{appendix:effectiveness-gtsrb}), followed by detailed ablation studies of BaDExpert's key design components. In Sec~\ref{subsec:generalizability-and-scalability}, we delve deeper into BaDExpert's generalizability across various model architectures and its scalability on ImageNet. Finally, in Sec~\ref{subsubsec:adaptive-analysis}, we investigate the resilience of BaDExpert's design against a series of adaptive attacks, demonstrating its adaptive effectiveness.

\subsection{Setup}
\label{subsec:setup}

\paragraph{Datasets and Models.} Our primary experiment focuses on two widely benchmarked image datasets in backdoor literature, CIFAR10~\cite{krizhevsky2012cifar} (Sec~\ref{subsec:effectiveness-cifar10}) and GTSRB~\cite{stallkamp2012man} (deferred to Appendix~\ref{appendix:effectiveness-gtsrb}). We demonstrate the equivalently successful effectiveness of BaDExpert on a representative large scale dataset, 1000-class ImageNet~\cite{deng2009imagenet}, in Sec~\ref{subsec:generalizability-and-scalability}, to further validate our method's scalability. We evaluate BaDExpert across various model architectures. Specifically, we adopt the commonly studied ResNet18~\cite{resnet} through our primary experiment, and validate in Sec~\ref{subsec:generalizability-and-scalability} the effectiveness of BaDExpert on other architectures (VGG~\cite{vgg}, MobileNetV2~\cite{sandler2018mobilenetv2}, and ViT~\cite{dosovitskiy2020image}).

\paragraph{Attacks.}
We evaluate BaDExpert against 12 state-of-the-art backdoor attacks in our primary experiment, with 9 types initiated during the development stage and the remaining 3 post-development.
In the realm of \textbf{development-stage} attacks, we explore 1) \textit{classical static-trigger dirty-label attacks} such as BadNet~\cite{gu2017badnets}, Blend~\cite{Chen2017TargetedBA}, and Trojan~\cite{liu2017trojaning}; 2) \textit{clean-label attacks} including CL~\cite{turner2019label} and SIG~\cite{barni2019new}; 3) \textit{input-specific-trigger attacks} like Dynamic~\cite{nguyen2020input}, ISSBA~\cite{li2021invisible}, WaNet~\cite{nguyen2021wanet}, and BPP~\cite{wang2022bppattack}.
As for \textbf{post-development} attacks, our evaluation considers: 1) direct finetuning of the developed vanilla model on a blending poisoned dataset (FT); 2) trojanning attacks (TrojanNN~\cite{liu2017trojaning}); 3) subnet-replacement attacks (SRA~\cite{qi2022towards}).
Moreover, in Sec~\ref{subsubsec:adaptive-analysis}, we study 6 existing adaptive attacks and a novel tailored adaptive attack against BaDExpert. Our attack configurations largely adhere to the methodologies described in their original papers. Readers interested in the detailed configurations can refer to the Appendix~\ref{appendix:baseline-attacks-configuration}.

\paragraph{Defenses.} We compare BaDExpert with 10 established backdoor defense baselines. First, to underscore BaDExpert's consistent performance in detecting backdoor inputs, we juxtapose it with 3 post-development backdoor input detectors (STRIP~\cite{gao2019strip}, Frequency~\cite{zeng2021rethinking}, SCALE-UP~\cite{guo2023scaleup}), all of which share our goal of detecting inference-time backdoor inputs. Additionally, for the sake of comprehensiveness, we repurpose certain development-stage poison set cleansers (AC~\cite{chen2018activationclustering}) to function as backdoor input detectors for comparative analysis. We also incorporate 6 model repairing defenses (FP~\cite{liu2018fine}, NC~\cite{wang2019neural}, MOTH~\cite{tao2022model}, NAD~\cite{li2021neural}, with ANP~\cite{wu2021adversarial} and I-BAU~\cite{zeng2021adversarial} deferred to Appendix~\ref{appendix:comparing_with_ANP_and_IBAU}) into our comparison to demonstrate BaDExpert's consistent performance as a general post-development defense.
We describe the hyperparameter selection of BaDExpert in Appendix~\ref{appendix:badexpert-implementation-details} and configuration details of other baselines defenses in \ref{appendix:baseline-defenses-configuration}.

\paragraph{Metrics.} We evaluate our results based on two sets of metrics. First, to measure BaDExpert's effectiveness as a backdoor detector, we report the area under the receiver operating characteristic (\textbf{AUROC}~\cite{fawcett2006introduction}, the higher the better).
Second, to directly compare with other non-detector backdoor defenses, we report the clean accuracy (\textbf{CA}, the higher the better) and attack success rate (\textbf{ASR}, the lower the better) of the model equipped with BaDExpert.
Specifically, for backdoor input detectors, we interpret correctly filtered backdoor inputs as successfully thwarted attacks, while falsely filtered clean inputs are considered erroneous predictions. Formally, $\text{CA}=\mathbb P_{(x,y)\sim \mathcal{P}} [\mathcal M(x) = y \wedge \mathtt{BID}(x) = 0]$, and $\text{ASR}=\mathbb P_{(x,y)\sim \mathcal{P}} [\mathcal M(\mathcal T(x)) = t \wedge \mathtt{BID}(\mathcal T(x)) = 0]$.
These metrics provide comprehensive and fair evaluations of our method in different defense scenarios. CA and ASR are reported on the standard validation over clean inputs and their backdoor correspondance, while
AUROC is calculated over a noisy augmented validation set (same configuration following prior work~\cite{guo2023scaleup}) to prevent overfitting.
To ensure rigor, all primary results are averaged over three runs (corresponding standard deviations are reported in Appendix~\ref{appendix:stdev-major-experiments}).

\subsection{Effectiveness of BaDExperts on CIFAR10}
\label{subsec:effectiveness-cifar10}

\subsubsection{Consistent Effectiveness Across Settings}

Table~\ref{tab:main_cifar10} and Table~\ref{tab:auroc_cifar10} highlight the defensive results of BaDExpert as a comprehensive post-development defense and as a backdoor input detector, respectively.

\paragraph{BaDExpert as a Post-Development Defense.} Specifically, we deem a defense as \green{successful} if the post-defense Attack Success Rate (ASR) is under $20\%$, and \red{unsuccessful} otherwise, as done in prior work~\cite{qi2023proactive}. As depicted, BaDExpert \textbf{consistently succeeds against all attacks}, with an average defended ASR of merely $5.1\%$ (best). Moreover, deploying the backdoor model with BaDExpert only causes a \textbf{small $0.9\%$ CA drop}, which is comparable to the best of prior arts.

\paragraph{BaDExpert as a Backdoor Input Detector.}
As evidenced by Table~\ref{tab:auroc_cifar10}, BaDExpert achieves an average AUROC of $99\%$ across 12 attacks, while the average AUROC of all other baseline detectors is less than $85\%$. In addition, BaDExpert achieves the highest AUROC for 9 out of the 12 attacks and maintains an AUROC of over $96\%$ in all situations. Overall, BaDExpert consistently performs effectively against all evaluated attacks, exhibiting a significant advantage over the baseline methods.

\subsubsection{Comparing BaDExpert to Baseline Defenses}
\label{subsubsec:comparing-badexpert-to-baseline-defenses}

In Table~\ref{tab:main_cifar10}, all the baseline defenses we evaluate fail against at least one of the baseline attacks.
\underline{Observation 1}: The two baseline defenses that focus on trigger reverse-engineering, NC and MOTH, are effective against attacks using patch-based triggers (such as BadNet, Trojan, CL, SRA, etc.), but they fail when confronted with attacks using global transformations as triggers (for instance, SIG), due to their strong inductive bias on the types of backdoor triggers.
\underline{Observation 2}: The three detectors (STRIP, Frequency, and SCALE-UP) that rely on the conspicuousness of backdoor triggers in backdoor inputs (either through frequency analysis or input perturbation) are ineffective against SIG and WaNet, which employ stealthier backdoor triggers.
\underline{Observation 3}: FP and AC utilize specific inner-model information (like neuron activations), turning out not universal enough to counteract all types of attacks.
\underline{Observation 4}: Specifically, NAD, which applies distillation to a fine-tuned model to eliminate the backdoor, can effectively defend against 11 out of 12 attacks, but it results in a considerable drop in CA (7.2\%).

Remarkably, BaDExpert inhibits all 12 attacks by rejecting suspicious backdoor inputs. Principally, this could be credited to the extracted backdoor functionality, which poses no inductive bias on backdoor trigger types, and therefore effectively suppress all attacks independent of the triggers they use. Moreover, benefiting from the extracted backdoor functionality that aid our defense, when comparing with NAD (which similarly includes a standard finetuning process in their pipeline), we outperform their defense performance in most scenarios, with noticeably higher CA and lower ASR.

\begin{table}[t]
\centering
\caption{Defensive results on CIFAR10 (CA and ASR).}
\resizebox{0.99\linewidth}{!}{
\begin{tabular}{ccccccccccccccccccccc}
\toprule
\textbf{Defenses$\rightarrow$} &
\multicolumn{2}{c}{No Defense} &
\multicolumn{2}{c}{FP} &
\multicolumn{2}{c}{NC} & 
\multicolumn{2}{c}{MOTH} & 
\multicolumn{2}{c}{NAD} & 
\multicolumn{2}{c}{STRIP} & 
\multicolumn{2}{c}{AC} & 
\multicolumn{2}{c}{Frequency} & 
\multicolumn{2}{c}{SCALE-UP} & 
\multicolumn{2}{c}{\textbf{BaDExpert}} \cr
\cmidrule(lr){2-3} \cmidrule(lr){4-5} \cmidrule(lr){6-7} \cmidrule(lr){8-9} \cmidrule(lr){10-11} \cmidrule(lr){12-13} \cmidrule(lr){14-15} \cmidrule(lr){16-17} \cmidrule(lr){18-19} \cmidrule(lr){20-21} 
\textbf{Attacks} $\downarrow$ & CA & ASR & CA & ASR & CA & ASR & CA & ASR & CA & ASR & CA & ASR & CA & ASR & CA & ASR & CA & ASR & CA & ASR \cr
\midrule
No Attack & 94.1 & - & 82.9 & - & 93.3 & - & 91.1 & - & 85.6 & - & 84.8 & - & 87.8 & - & 91.0 & - & 77.0 & - & 93.1 & - \cr
\midrule
\multicolumn{19}{c}{\centering \textbf{Development-Stage Attacks}} \cr
\midrule
BadNet & 94.1 & \pc 100.0 & 83.6 & \pc 100.0 & 93.7 & \grc 3.9 & 91.5 & \grc 0.7 & 86.9 & \grc 1.8 & 84.7 & \grc 0.1 & 94.1 & \grc 0.0 & 91.0 & \grc 0.0 & 76.8 & \grc 0.0 & 93.1 & \grc 0.0  \cr
\midrule
Blend & 94.0 & \pc 91.9 & 82.7 & \pc 88.6 & 93.5 & \pc 90.5 & 91.2 & \pc 82.5 & 86.3 & \grc 11.8 & 84.6 & \pc 83.9 & 93.7 & \pc 24.8 & 90.9 & \grc 10.8 & 76.7 & \grc 17.2 & 93.1 & \grc 11.4  \cr
\midrule
Trojan & 94.0 & \pc 99.9 & 82.8 & \pc 64.2 & 93.3 & \grc 1.3 & 90.7 & \grc 3.1 & 86.4 & \grc 4.7 & 84.6 & \pc 68.1 & 93.8 & \grc 0.3 & 90.9 & \grc 0.0 & 78.8 & \grc 7.6 & 93.0 & \grc 8.4  \cr
\midrule
CL & 94.1 & \pc 99.9 & 83.1 & \pc 87.4 & 93.5 & \grc 0.5 & 91.3 & \grc 1.1 & 86.0 & \grc 7.5 & 84.7 & \pc 41.5 & 94.0 & \grc 0.2 & 91.0 & \grc 3.6 & 77.8 & \grc 0.1 & 93.1 & \grc 10.6  \cr
\midrule
SIG & 93.8 & \pc 82.7 & 82.1 & \pc 58.3 & 93.5 & \pc 86.5 & 90.8 & \pc 65.8 & 85.6 & \grc 5.8 & 84.5 & \pc 72.8 & 93.8 & \grc 14.2 & 90.7 & \pc 43.5 & 77.1 & \pc 28.5 & 92.9 & \grc 1.6  \cr
\midrule
Dynamic & 93.9 & \pc 99.1 & 82.7 & \pc 86.1 & 93.1 & \grc 6.7 & 91.4 & \pc 56.9 & 85.9 & \pc 22.4 & 84.5 & \pc 20.9 & 89.1 & \pc 74.2 & 90.8 & \grc 0.4 & 79.9 & \grc 1.1 & 93.0 & \grc 16.3  \cr
\midrule
ISSBA & 93.9 & \pc 99.9 & 83.6 & \grc 0.1 & 93.6 & \grc 1.7 & 91.2 & \pc 56.3 & 85.6 & \grc 1.9 & 84.6 & \grc 14.3 & 93.9 & \grc 0.1 & 90.8 & \grc 0.0 & 79.7 & \grc 1.5 & 93.0 & \grc 1.1  \cr
\midrule
WaNet & 93.1 & \pc 93.7 & 81.1 & \grc 2.6 & 93.0 & \pc 81.6 & 90.1 & \grc 14.0  & 85.7 & \grc 1.8 & 83.7 & \pc 86.9 & 93.1 & \pc 23.8 & 90.1 & \pc 86.5 & 76.3 & \pc 53.9 & 92.2 & \grc 2.0  \cr
\midrule
BPP & 89.7 & \pc 99.8 & 78.0 & \grc 8.3 & 89.8 & \pc 33.8 & 88.8 & \grc 1.9 & 89.7 & \grc 0.9 & 80.7 & \pc 91.9 & 89.7 & \grc 2.4 & 86.8 & \grc 0.0 & 74.7 & \grc 14.1 & 88.9 & \grc 0.2  \cr
\midrule
\multicolumn{19}{c}{\centering \textbf{Post-development Attacks}} \cr
\midrule
FT & 93.2 & \pc 99.5 & 82.3 & \pc 95.8 & 92.4 & \pc 46.1 & 91.5 & \pc 93.6 & 86.5 & \grc 9.4 & 83.8 & \grc 16.1 & 93.2 & \grc 18.2 & 90.1 & \grc 11.7 & 79.2 & \pc 22.1 & 92.3 & \grc 4.1  \cr
\midrule
TrojanNN & 93.8 & \pc 100.0 & 83.2 & \pc 91.5 & 92.2 & \grc 0.9 & 91.2 & \pc 41.9 & 86.3 & \grc 10.1 & 84.4 & \grc 0.1 & 93.6 & \grc 0.1 & 90.7 & \grc 0.0 & 80.6 & \grc 0.0 & 92.8 & \grc 7.0  \cr
\midrule
SRA & 90.3 & \pc 99.9 & 79.3 & \pc 100.0 & 91.9 & \grc 1.2 & 91.2 & \grc 1.1 & 82.2 & \grc 2.2 & 81.3 & \pc 88.1 & 90.3 & \grc 0.5 & 88.2 & \grc 0.0 & 73.7 & \pc 68.4 & 89.4 & \grc 0.4  \cr
\midrule
\midrule
\textbf{Average} & 93.2 & 97.2 & 82.1 & 65.2 & \textbf{92.8} & 29.6 & 90.9 & 34.9 & 86.1 & 6.7 & 83.9 & 48.7 & 92.3 & 13.2 & 90.2 & 13.0 & 77.6 & 16.9 & \textbf{92.3} & \textbf{5.1}  \cr
\midrule


\multicolumn{3}{c}{CA Drop (smaller is better)} & $\downarrow$11.1 & & \textbf{$\downarrow$0.4} &  & $\downarrow$2.3 &  & $\downarrow$7.2 &  & $\downarrow$9.3 &  & $\downarrow$0.9 &  & $\downarrow$3.0 &  & $\downarrow$15.7 & & \textbf{$\downarrow$0.9} & \cr
\midrule

\multicolumn{3}{c}{ASR Drop (larger is better)} &  & $\downarrow$32.0 &  & $\downarrow$67.6 & & $\downarrow$62.3 & &  \textbf{$\downarrow$90.5} & & $\downarrow$48.5 & & $\downarrow$84.0 & & $\downarrow$84.1 & & $\downarrow$80.3 & & \textbf{$\downarrow$92.1} \cr
\bottomrule

\end{tabular}
} 
\label{tab:main_cifar10}
\end{table}

\begin{table}[t]
\centering
\caption{Defensive results on CIFAR10 (AUROC).}
\resizebox{0.99\linewidth}{!}{ 
\begin{tabular}{lccccccccccccc}
\toprule
 \textbf{AUROC} ($\%$) & 
\multicolumn{1}{c}{BadNet} &
\multicolumn{1}{c}{Blend} &
\multicolumn{1}{c}{Trojan} &
\multicolumn{1}{c}{CL} &
\multicolumn{1}{c}{SIG} &
\multicolumn{1}{c}{Dynamic} &
\multicolumn{1}{c}{ISSBA} &
\multicolumn{1}{c}{WaNet} &
\multicolumn{1}{c}{Bpp} &
\multicolumn{1}{c}{FT} &
\multicolumn{1}{c}{TrojanNN} &
\multicolumn{1}{c}{SRA} &
\multicolumn{1}{c}{\textbf{Average}}\cr
\midrule

STRIP & 99.1 & 47.1 & 72.1 & 84.7 & 40.3 & 85.2 & 68.1 & 49.8 & 50.1 & 91.8 & 99.3 & 54.8 & 70.2 \cr
\midrule

AC & 100.0 & 54.1 & \textbf{99.7} & \textbf{99.9} & 53.6 & 77.9 & 84.4 & 47.2 & 98.4 & 58.7 & \textbf{99.9} & 99.6 & 81.1 \cr
\midrule

Frequency & 75.1 & 73.5 & 75.1 & 74.5 & 68.0 & 74.9 & 75.1 & 62.8 & 75.0 & 73.5 & 75.0 & 75.1 & 73.1 \cr
\midrule

SCALE-UP & 96.4 & 80.9 & 91.2 & 96.3 & 69.6 & 96.0 & 71.6 & 66.3 & 87.9 & 89.6 & 96.5 & 59.9 & 83.5 \cr
\midrule

\textbf{BaDExpert} & \textbf{100.0} & \textbf{99.2} & 99.2 & 99.0 & \textbf{99.8} & \textbf{99.1} & \textbf{96.1} & \textbf{99.7} & \textbf{100.0} & \textbf{99.6} & 99.3 & \textbf{100.0} & \textbf{99.0} \cr
\bottomrule

\end{tabular}
} 
\label{tab:auroc_cifar10}
\end{table}








\subsubsection{Ablation Studies}
\label{subsubsec:ablation-studies}

\begin{wrapfigure}{r}{0.37\textwidth}
    \centering
         \includegraphics[width=.37\textwidth]{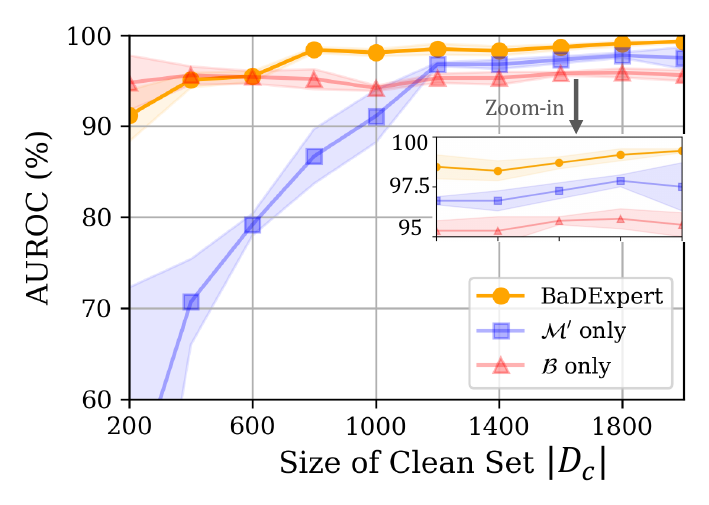}
    \vspace{-6mm}
    \caption{AUROC of BaDExpert (both $\Mprime$ and $\B$), $\Mprime$ only, and $\B$ only, with different reserved clean set sizes.
    }
    \label{fig:ablation_clean_samples_num}
\vspace{-3mm}
\end{wrapfigure}

\paragraph{Size of the Reserved Clean Set $|D_c|$.} In our defense pipeline, we assume a small reserved clean set $D_c$ (default to $5\%$ size of the training dataset in the primary experiment, i.e., $2,000$ samples) to construct both $\mathcal B$ and $\mathcal M'$. To investigate how minimal $|D_c|$ could be, we evaluate BaDExpert with different sizes ($200 \sim 1,800$) of this clean set. The AUROC of BaDExpert (against Blend attack on CIFAR10) is reported in Fig~\ref{fig:ablation_clean_samples_num} (orange line with the circle marker). As shown, as $|D_c|$ becomes smaller, the AUROC of BaDExpert mostly remains higher than $98\%$, and slightly drops (to $95\%$ or $90\%$) when the number of clean samples is extremely limited ($400$ or $200$).
To obtain a clearer view, in Appendix~\ref{appendix:ablation-comparing-with-STRIP-and-ScaleUp-with-fewer-clean-samples}, we compare BaDExpert with ScaleUp and STRIP side-by-side, where they are all assigned to such highly limited amount ($100$, $200$ and $400$) of clean samples. Notably, BaDExpert still outperforms the baselines in most scenarios. Alongside, we also justify how practitioners can acquire a small $D_c$ with possible data-collecting approaches (\cite{287222}).

\paragraph{Necessity of both $\B$ and $\Mprime$.} As mentioned, we ensemble both $\mathcal B$ and $\mathcal M'$ to decide if an input is backdoored or not. However, one may directly attempt to detect backdoor inputs based on the metric of the single confidence of either $\mathcal B$ or $\mathcal M'$. Here, we study whether both of them are necessary for the success of BaDExpert. The blue (square) and red (triangle) lines in Fig~\ref{fig:ablation_clean_samples_num} correspond to the detection AUROC if we only adopt $\mathcal M'$ or $\mathcal B$. There is a three-fold observation: 1) backdoor expert $\mathcal B$ consistently provides a high AUROC ($\sim95\%$), implying that it indeed fulfills the backdoor functionality by assigning backdoor input with higher confidence~(Eq~\eqref{eqn:soft-decision-rule-backdoor-expert}); 2) The finetuned auxiliary model $\mathcal M'$ itself also provides a high AUROC ($>95\%$) when we have more than $1,000$ clean reserved samples, but degrades when the size of the clean set gets smaller (AUROC $<90\%$ when clean samples $<1,000$); 3) Taking advantage from both sides, BaDExpert achieves the highest AUROC in most cases, indicating the necessity of coupling both $\B$ and $\Mprime$.

\paragraph{Ensembling with Existing Model-repairing Defenses.} In our pipeline, we choose finetuning as the default method to obtain an auxiliary model $\Mprime$ where the backdoor is diminished. Importantly, our backdoor expert methodology is also \textbf{orthogonal} to the extensive suite of model-repairing defenses, meaning that the auxiliary model $\Mprime$ can effectively be any model that has undergone baseline repair. For instance, when we combined backdoor experts with models repaired by the NAD technique, we achieve an average AUROC of $98.8\%$ on the CIFAR10 dataset, a result that aligns closely with our primary findings presented in Table~\ref{tab:auroc_cifar10}. For a more detailed discussion, please refer to Appendix~\ref{appendix:ensembling-with-other-defenses}.

\paragraph{Miscellaneous.} In Appendix~\ref{appendix:constructing-backdoor-experts}, we validate that Alg~\eqref{alg:alg-training-BE} consistently isolates the backdoor functionality regardless of different number (1 to 2,000) of clean samples it leverages. In Appendix~\ref{appendix:ablation-unlearning-rates}, we study the choice of (un-)learning rate $\eta$ in Alg~\eqref{alg:alg-training-BE}, showing that BaDExpert's performance remains insensitive to the selection of $\eta$ across a diverse lens. In Appendix~\ref{appendix:ablation-poison-rates}, we demonstrate that BaDEpxert is similarly successful even when the adversary employs different poison rates. In Appendix~\ref{appendix:other-mislabeling-strategies}, we discuss other possible mislabelling strategies used in Line~\eqref{alg:mislabel} of Alg~\eqref{alg:alg-training-BE}.

\begin{wraptable}{r}{0.28\linewidth}
\centering
\vspace{-5mm}
\caption{BaDExpert generalizability on other architetures.}
\resizebox{\linewidth}{!}{ 
\begin{tabular}{lcc}
\toprule
 \textbf{AUROC} ($\%$) & 
\multicolumn{1}{c}{VGG16} &
\multicolumn{1}{c}{MobileNetV2} \cr
\midrule
BadNet & 99.7 & 99.5 \cr
\midrule
Blend & 97.7 & 97.6 \cr
\midrule
Trojan & 98.6 & 97.8 \cr
\midrule
CL & 97.7 & 99.2 \cr
\midrule
SIG & 98.4 & 98.9 \cr
\midrule
Dynamic & 96.7 & 98.1 \cr
\midrule
WaNet & 98.2 & 98.0 \cr
\midrule
FT & 99.3 & 98.7 \cr
\midrule
\textbf{Average} & \textbf{98.3} & \textbf{98.5} \cr
\bottomrule

\end{tabular}
} 
\label{tab:ablation_arch_auroc}
\vspace{-6mm}
\end{wraptable}





\subsection{Generalizability and Scalability}
\label{subsec:generalizability-and-scalability}

\paragraph{Generalizability on Other Model Architectures.}
We first show that BaDExpert works similarly well for two other architectures, VGG-16 and MobileNetV2 in Table~\ref{tab:ablation_arch_auroc} on CIFAR10 (some attacks in Table~\ref{tab:main_cifar10} are ineffective and not shown). As shown, the average AUROC on both architectures achieve $>98\%$, similar to our major results in Table~\ref{tab:main_cifar10}.

\paragraph{Scalability on ImageNet.}
Table~\ref{tab:main_imagenet} reflects the effectiveness of BaDExpert on ImageNet. We conduct: 1) poisoning attacks (BadNet, Blend) by training ResNet18 on backdoor poisoned ImageNet datasets from scratch; 2) subnet-replacement attack (SRA) on pretrained ResNet101 (following SRA's original implementation); 3) finetuning attacks (FT-BadNet, FT-Blend) on pretrained ViT-B/16. We only reserve $\sim 6,000$ clean samples (equivalent to $0.5\%$ size of ImageNet-1000 training set) to BaDExpert. In all scenarios, our BaDExpert can effectively detect backdoor samples ($\sim 100\%$ AUROC and $<5\%$ ASR), costing only negligible CA drop ($\downarrow 0.1\%$). These results confirm the scalability of BaDExpert.

\begin{table}[t]
\centering
\caption{Defensive results of BaDExpert on ImageNet.}
\resizebox{0.8\linewidth}{!}{ 
\begin{tabular}{llcccccccc}
\toprule
 & ($\%$) & 
\multicolumn{3}{c}{ResNet18} &
\multicolumn{2}{c}{ResNet101} &
\multicolumn{3}{c}{ViT-B/16} \cr

\cmidrule(lr){3-5} \cmidrule(lr){6-7} \cmidrule(lr){8-10}
 & &
\multicolumn{1}{c}{No Attack} &
\multicolumn{1}{c}{BadNet} &
\multicolumn{1}{c}{Blend} &
\multicolumn{1}{c}{No Attack} &
\multicolumn{1}{c}{SRA} &
\multicolumn{1}{c}{No Attack} &
\multicolumn{1}{c}{FT-BadNet} &
\multicolumn{1}{c}{FT-Blend} \cr

\multirow{2}*{\shortstack{Without Defense}}
& CA & 69.0 & 68.4 & 69.0 & 77.1 & 74.7 & 81.9 & 82.3 & 81.8 \cr
& ASR & - & 100.0 & 92.2 & - & 100.0 & - & 99.5 & 93.2 \cr 
\midrule

\multirow{3}*{\shortstack{\textbf{BaDExpert}}}
& CA & 68.9 & 68.3 & 68.9 & 77.0 & 74.6 & 81.8 & 82.2 & 81.7 \cr
& ASR & - & \grc 0.0 & \grc 4.9 & - & \grc 0.0 & - & \grc 0.0 & \grc 0.2 \cr
& AUROC & - &  100.0 & 99.9 & - & 100.0 & - & 100.0 & 100.0 \cr
\bottomrule

\end{tabular}
} 
\label{tab:main_imagenet}
\end{table}

\subsection{The Resistance to Adaptive Attacks}
\label{subsubsec:adaptive-analysis}

To thoroughly study the potential risk underlying our defense, we also consider adaptive adversaries that strategically employ backdoor attacks designed specifically to bypass BaDExpert's detection.

One possible adaptive strategy is to \textbf{deliberately establish dependencies between the backdoor and normal functionality}, which may undermine the implicit assumption of our defense — the core concept of \textit{isolating the backdoor functionality from the normal functionality via unlearning}.
Essentially, if the backdoor functionality depends on the normal functionality, the erasure of the normal functionality would subsequently lead to the degradation of the backdoor functionality, potentially reducing BaDExpert's effectiveness.
In fact, there exist several adaptive backdoor attacks tailored to this end. Here, we examine \textbf{TaCT}~\cite{tang2021demon}, \textbf{Adap-Blend}, and \textbf{Adap-Patch}~\cite{qi2023revisiting}, which employ different poisoning strategies to create dependencies between backdoor and normal predictions.
We also consider an \textbf{All-to-All} attack scenario, where each sample originating from any class $i\in [C]$ is targeted to class $(i - 1) \mod C$ --- here, the backdoor predictions rely on both the backdoor trigger and clean semantics, thereby forcing the backdoor functionality to depend on the normal one.
Eventually, we evaluate BaDExpert against \textbf{Natural} backdoor~\cite{zhao2021deep} existing in benign models, where the backdoor triggers are unconsciously learned from normal data.

An alternative perspective that may be exploited by adaptive adversaries to bypass our defense would be to \textbf{utilize specifically constructed asymmetric triggers} at inference time (different from the ones used during model production). We first study a simple scenario where the adversary deliberately use weakened triggers (e.g. blending triggers with lower opacity, dubbed ``\textbf{Low-Opacity}'') to activate the backdoor at inference time. More profoundly, we design a novel adaptive attack tailored against BaDExpert (``\textbf{BaDExpert-Adap-BadNet}''), where the adversary optimizes an asymmetric trigger by minimizing the activation of the backdoor expert model $\B$. Refer to Appendix~\ref{appendix:additional-adaptive-analysis-weakened-triggers}-\ref{appendix:tailored-adaptive-attack} for details.

As shown in Table~\ref{tab:adaptive_attacks}, BaDEpxert's effectiveness indeed experiences certain degradation in (AUROC becomes as low as $87.0\%$), in comparison with Table~\ref{tab:auroc_cifar10} ($99.0\%$ average AUROC).
Nevertheless, we can see that BaDEpxert still demonstrates considerable resilience against all these adaptive efforts.
We recommend interested readers to Appendix~\ref{appendix:additional-adaptive-analysis} for more details in our adaptive analysis.

\begin{table}
\centering
\caption{Defense results of adaptive attacks against BaDExpert.}
\resizebox{0.83\linewidth}{!}{ 
\begin{tabular}{lccccccc}
\toprule
 (\%)& 
\multicolumn{1}{c}{TaCT} &
\multicolumn{1}{c}{Adap-Blend} &
\multicolumn{1}{c}{Adap-Patch} &
\multicolumn{1}{c}{All-to-All} &
\multicolumn{1}{c}{Natural} &
\multicolumn{1}{c}{Low-Opacity} &
\multicolumn{1}{c}{BaDExpert-Adap-BadNet}\cr
\cmidrule{2-8}
ASR (before defense) & 99.3 & 85.4 & 99.4 & 89.1 & 99.2 & 51.6 & 73.5 \cr
\midrule
\textbf{AUROC} & 97.5 & 99.6 & 99.2 & 95.5 & 92.3 & 98.4 & 87.0 \cr
\bottomrule

\end{tabular}
} 
\label{tab:adaptive_attacks}
\end{table}

\section{Related Work}
\label{sec:related}

\paragraph{Backdoor Attacks.} 
Backdoor attacks are typically studied in the data poisoning threat model~\cite{Chen2017TargetedBA,nguyen2021wanet,gao2023not,qi2023revisiting}, where adversaries inject a few poison samples into the victim's training dataset. Victim models trained on the poisoned dataset tend to learn spurious correlations between backdoor triggers and target classes encoded in poison samples and get backdoored. Besides data poisoning, backdoor attacks can be implemented in alternative ways, such as manipulating training process~\cite{bagdasaryan2021blind,li2021backdoor}, supplying backdoored pre-trained models~\cite{yao2019latent,shen2021backdoor}, as well as weights tampering~\cite{liu2017fault,qi2021subnet,qi2022towards,dong2023one}, etc. There are also backdoor attacks that are adaptively designed to evade defenses~\cite{tang2021demon, qi2023revisiting}.

\paragraph{Development-Stage Backdoor Defenses.} The existing literature has extensively explored defensive strategies against backdoor attacks, with a significant focus on {{development-stage defenses}}. These defenses primarily target data-poisoning-based attacks~\cite{goldblum2022dataset} and are presumed to be implemented by model vendors. They either \textit{identify and remove the poison samples} from the dataset before training~\cite{tran2018spectral, tang2021demon,jebreel2023defending,qi2023proactive}, or \textit{suppress the learning of backdoor correlations} during training~\cite{li2021anti,huang2022backdoor,wang2022NONE}. Notably, the security of these approaches heavily relies on the integrity of model vendors, and they cannot prevent backdoor injection after the model development.

\paragraph{Post-Development Backdoor Defenses.} {{Post-development defenses}} \textit{operate independently of model development}.
They typically assume only access to the (potentially backdoored) model intended to be deployed and a small number of reserved clean data for defensive purposes. One category of such approaches attempts to directly \textit{remove the backdoor} from a backdoor model via pruning~\cite{liu2018fine}, distillation~\cite{li2021neural}, forgetting~\cite{zhu2022selective}, finetuning~\cite{sha2022fine}, unlearning reconstructed triggers~\cite{wang2019neural,tao2022model}, etc.
Alternatively, \textit{model diagnosis defenses}~\cite {xu2019Meta,kolouri2020universal} attempt to make a binary diagnosis on whether a model is backdoored.
There are also approaches attempting to detect and \textit{filter backdoor inputs} at inference time~\cite{gao2019strip, zeng2021rethinking, guo2023scaleup} and thus prevent the backdoor from being activated.
The defense we propose in this work falls within this category.
Meanwhile, our idea of backdoor extraction is also relavent to the trigger-reconstruction-based defenses~\cite{wang2019neural,tao2022model} in the sense of backdoor reverse engineering, but different in that we directly extract the backdoor functionality as opposed to backdoor trigger patterns.

\section{Conclusion}

In this study, we introduce a novel post-development defense strategy against backdoor attacks on DNNs. Inspired by the defenses that conduct trigger reverse engineering, we propose a distinctive method that directly extracts the backdoor functionality from a compromised model into a designated backdoor expert model. This extraction process is accomplished by leveraging a simple yet effective insight: finetuning the backdoor model on a set of intentionally mislabeled reserved clean samples allows us to erase its normal functionality while preserving the backdoor functionality. We further illustrate how to apply this backdoor expert model within the framework of backdoor input detection, leading us to devise an accurate and resilient detector for backdoor inputs during inference-time, known as \textbf{BaDExpert}.
Our empirical evaluations show that BaDExpert is effective across different attacks, datasets and model architectures.
Eventually, we provide an adaptive study against BaDExpert, finding that BaDExpert is resilient against diverse adaptive attacks, including a novelly tailored one.

\newpage

\section*{Ethics Statement}

In this work, we introduce a novel backdoor defense, demonstrating its consistent capability of detecting inference-time backdoor inputs. Our defense proposal to thwart potential malicious adversaries should not raise ethical concerns.
Nevertheless, we want to avoid overstating the security provided by our method, since our evaluation on the effectiveness of our defense is empirical, and that our defense comes without any certified guarantee (most prior backdoor defenses share this limitation).
After all, the field of backdoor attacks and defenses is still a long-term cat-and-mouse game, and it is essential for practitioners to exercise  caution when implementing backdoor defenses in real-world scenarios.
Rather, we hope our comprehensive insights into the concept of ``\textit{extracting backdoor functionality}'' can serve as a valuable resource to guide future research in related domains. 

It is also important to note that our work involves a tailored adaptive attack against our proposed defense. However, we emphasize that the sole purpose of this adaptive attack is to rigorously assess the effectiveness of our defense strategy. We strictly adhere to ethical guidelines in conducting this research, ensuring that all our experiments are conducted in a controlled isolated environment.


\section*{Acknowledgements}

This work was supported in part by NSF grants CNS-1553437 and CNS-1704105, the ARL’s Army Artificial Intelligence Innovation Institute (A2I2), the Office of Naval Research Young Investigator Award, the Army Research Office Young Investigator Prize, Schmidt DataX award, Princeton E-affiliates Award, Princeton Francis Robbins Upton Fellowship, and Princeton Gordon Y. S. Wu Fellowship. Any opinions, findings, and conclusions or recommendations expressed in this material are those of the author(s) and do not necessarily reﬂect the views of the funding agencies.

\bibliographystyle{unsrt}
\bibliography{reference}

\begin{thebibliography}{10}

\bibitem{gu2017badnets}
Tianyu Gu, Brendan Dolan-Gavitt, and Siddharth Garg.
\newblock Badnets: Identifying vulnerabilities in the machine learning model supply chain.
\newblock {\em arXiv preprint arXiv:1708.06733}, 2017.

\bibitem{li2022backdoor}
Yiming Li, Yong Jiang, Zhifeng Li, and Shu-Tao Xia.
\newblock Backdoor learning: A survey.
\newblock {\em IEEE Transactions on Neural Networks and Learning Systems}, 2022.

\bibitem{goldblum2022dataset}
Micah Goldblum, Dimitris Tsipras, Chulin Xie, Xinyun Chen, Avi Schwarzschild, Dawn Song, Aleksander M{\k{a}}dry, Bo~Li, and Tom Goldstein.
\newblock Dataset security for machine learning: Data poisoning, backdoor attacks, and defenses.
\newblock {\em IEEE Transactions on Pattern Analysis and Machine Intelligence}, 45(2):1563--1580, 2022.

\bibitem{qi2022towards}
Xiangyu Qi, Tinghao Xie, Ruizhe Pan, Jifeng Zhu, Yong Yang, and Kai Bu.
\newblock Towards practical deployment-stage backdoor attack on deep neural networks.
\newblock In {\em Proceedings of the IEEE/CVF Conference on Computer Vision and Pattern Recognition}, pages 13347--13357, 2022.

\bibitem{tran2018spectral}
Brandon Tran, Jerry Li, and Aleksander Madry.
\newblock Spectral signatures in backdoor attacks.
\newblock {\em Advances in neural information processing systems}, 31, 2018.

\bibitem{li2021anti}
Yige Li, Xixiang Lyu, Nodens Koren, Lingjuan Lyu, Bo~Li, and Xingjun Ma.
\newblock Anti-backdoor learning: Training clean models on poisoned data.
\newblock {\em Advances in Neural Information Processing Systems}, 34, 2021.

\bibitem{huang2022backdoor}
Kunzhe Huang, Yiming Li, Baoyuan Wu, Zhan Qin, and Kui Ren.
\newblock Backdoor defense via decoupling the training process.
\newblock In {\em International Conference on Learning Representations}, 2022.

\bibitem{qi2023proactive}
Xiangyu Qi, Tinghao Xie, Jiachen~T Wang, Tong Wu, Saeed Mahloujifar, and Prateek Mittal.
\newblock Towards a proactive $\{$ML$\}$ approach for detecting backdoor poison samples.
\newblock In {\em 32nd USENIX Security Symposium (USENIX Security 23)}, pages 1685--1702, 2023.

\bibitem{wang2019neural}
Bolun Wang, Yuanshun Yao, Shawn Shan, Huiying Li, Bimal Viswanath, Haitao Zheng, and Ben~Y Zhao.
\newblock Neural cleanse: Identifying and mitigating backdoor attacks in neural networks.
\newblock In {\em 2019 IEEE Symposium on Security and Privacy (SP)}, pages 707--723. IEEE, 2019.

\bibitem{li2021neural}
Yige Li, Xixiang Lyu, Nodens Koren, Lingjuan Lyu, Bo~Li, and Xingjun Ma.
\newblock Neural attention distillation: Erasing backdoor triggers from deep neural networks.
\newblock In {\em International Conference on Learning Representations}, 2021.

\bibitem{gao2019strip}
Yansong Gao, Change Xu, Derui Wang, Shiping Chen, Damith~C Ranasinghe, and Surya Nepal.
\newblock Strip: A defence against trojan attacks on deep neural networks.
\newblock In {\em Proceedings of the 35th Annual Computer Security Applications Conference}, pages 113--125, 2019.

\bibitem{guo2023scaleup}
Junfeng Guo, Yiming Li, Xun Chen, Hanqing Guo, Lichao Sun, and Cong Liu.
\newblock {SCALE}-{UP}: An efficient black-box input-level backdoor detection via analyzing scaled prediction consistency.
\newblock In {\em The Eleventh International Conference on Learning Representations}, 2023.

\bibitem{tao2022model}
Guanhong Tao, Yingqi Liu, Guangyu Shen, Qiuling Xu, Shengwei An, Zhuo Zhang, and Xiangyu Zhang.
\newblock Model orthogonalization: Class distance hardening in neural networks for better security.
\newblock In {\em 2022 IEEE Symposium on Security and Privacy (SP)}, pages 1372--1389. IEEE, 2022.

\bibitem{Wang2022rethinking}
Zhenting Wang, Kai Mei, Hailun Ding, Juan Zhai, and Shiqing Ma.
\newblock Rethinking the reverse-engineering of trojan triggers.
\newblock In {\em Advances in Neural Information Processing Systems}, volume~35, pages 9738--9753, 2022.

\bibitem{Chen2017TargetedBA}
Xinyun Chen, Chang Liu, Bo~Li, Kimberly Lu, and Dawn~Xiaodong Song.
\newblock Targeted backdoor attacks on deep learning systems using data poisoning.
\newblock {\em ArXiv}, abs/1712.05526, 2017.

\bibitem{nguyen2021wanet}
Anh Nguyen and Anh Tran.
\newblock Wanet--imperceptible warping-based backdoor attack.
\newblock In {\em The Eleventh International Conference on Learning Representations}, 2021.

\bibitem{neyman1933}
Jerzy Neyman and Egon~Sharpe Pearson.
\newblock Ix. on the problem of the most efficient tests of statistical hypotheses.
\newblock {\em Philosophical Transactions of the Royal Society of London. Series A, Containing Papers of a Mathematical or Physical Character}, 231(694-706):289--337, 1933.

\bibitem{krizhevsky2012cifar}
Alex Krizhevsky.
\newblock Learning multiple layers of features from tiny images.
\newblock {\em University of Toronto}, 2012.

\bibitem{stallkamp2012man}
Johannes Stallkamp, Marc Schlipsing, Jan Salmen, and Christian Igel.
\newblock Man vs. computer: Benchmarking machine learning algorithms for traffic sign recognition.
\newblock {\em Neural networks}, 32:323--332, 2012.

\bibitem{deng2009imagenet}
Jia Deng, Wei Dong, Richard Socher, Li-Jia Li, Kai Li, and Li~Fei-Fei.
\newblock Imagenet: A large-scale hierarchical image database.
\newblock In {\em 2009 IEEE conference on computer vision and pattern recognition}, pages 248--255. Ieee, 2009.

\bibitem{resnet}
Kaiming He, Xiangyu Zhang, Shaoqing Ren, and Jian Sun.
\newblock Deep residual learning for image recognition.
\newblock In {\em Proceedings of the IEEE conference on computer vision and pattern recognition}, pages 770--778, 2016.

\bibitem{vgg}
Karen Simonyan and Andrew Zisserman.
\newblock Very deep convolutional networks for large-scale image recognition.
\newblock {\em arXiv preprint arXiv:1409.1556}, 2014.

\bibitem{sandler2018mobilenetv2}
Mark Sandler, Andrew Howard, Menglong Zhu, Andrey Zhmoginov, and Liang-Chieh Chen.
\newblock Mobilenetv2: Inverted residuals and linear bottlenecks.
\newblock In {\em Proceedings of the IEEE conference on computer vision and pattern recognition}, pages 4510--4520, 2018.

\bibitem{dosovitskiy2020image}
Alexey Dosovitskiy, Lucas Beyer, Alexander Kolesnikov, Dirk Weissenborn, Xiaohua Zhai, Thomas Unterthiner, Mostafa Dehghani, Matthias Minderer, Georg Heigold, Sylvain Gelly, et~al.
\newblock An image is worth 16x16 words: Transformers for image recognition at scale.
\newblock In {\em International Conference on Learning Representations}, 2020.

\bibitem{liu2017trojaning}
Yingqi Liu, Shiqing Ma, Yousra Aafer, W.~Lee, Juan Zhai, Weihang Wang, and X.~Zhang.
\newblock Trojaning attack on neural networks.
\newblock In {\em NDSS}, 2018.

\bibitem{turner2019label}
Alexander Turner, Dimitris Tsipras, and Aleksander Madry.
\newblock Label-consistent backdoor attacks.
\newblock {\em arXiv preprint arXiv:1912.02771}, 2019.

\bibitem{barni2019new}
Mauro Barni, Kassem Kallas, and Benedetta Tondi.
\newblock A new backdoor attack in cnns by training set corruption without label poisoning.
\newblock In {\em 2019 IEEE International Conference on Image Processing (ICIP)}, pages 101--105. IEEE, 2019.

\bibitem{nguyen2020input}
Tuan~Anh Nguyen and Anh Tran.
\newblock Input-aware dynamic backdoor attack.
\newblock {\em Advances in Neural Information Processing Systems}, 33:3454--3464, 2020.

\bibitem{li2021invisible}
Yuezun Li, Yiming Li, Baoyuan Wu, Longkang Li, Ran He, and Siwei Lyu.
\newblock Invisible backdoor attack with sample-specific triggers.
\newblock In {\em Proceedings of the IEEE/CVF international conference on computer vision}, pages 16463--16472, 2021.

\bibitem{wang2022bppattack}
Zhenting Wang, Juan Zhai, and Shiqing Ma.
\newblock Bppattack: Stealthy and efficient trojan attacks against deep neural networks via image quantization and contrastive adversarial learning.
\newblock In {\em Proceedings of the IEEE/CVF Conference on Computer Vision and Pattern Recognition}, pages 15074--15084, 2022.

\bibitem{zeng2021rethinking}
Yi~Zeng, Won Park, Z~Morley Mao, and Ruoxi Jia.
\newblock Rethinking the backdoor attacks' triggers: A frequency perspective.
\newblock In {\em Proceedings of the IEEE/CVF International Conference on Computer Vision}, pages 16473--16481, 2021.

\bibitem{chen2018activationclustering}
Bryant Chen, Wilka Carvalho, Nathalie Baracaldo, Heiko Ludwig, Benjamin Edwards, Taesung Lee, Ian Molloy, and Biplav Srivastava.
\newblock Detecting backdoor attacks on deep neural networks by activation clustering.
\newblock {\em arXiv preprint arXiv:1811.03728}, 2018.

\bibitem{liu2018fine}
Kang Liu, Brendan Dolan-Gavitt, and Siddharth Garg.
\newblock Fine-pruning: Defending against backdooring attacks on deep neural networks.
\newblock In {\em International Symposium on Research in Attacks, Intrusions, and Defenses}, pages 273--294. Springer, 2018.

\bibitem{wu2021adversarial}
Dongxian Wu and Yisen Wang.
\newblock Adversarial neuron pruning purifies backdoored deep models.
\newblock {\em Advances in Neural Information Processing Systems}, 34:16913--16925, 2021.

\bibitem{zeng2021adversarial}
Yi~Zeng, Si~Chen, Won Park, Z~Morley Mao, Ming Jin, and Ruoxi Jia.
\newblock Adversarial unlearning of backdoors via implicit hypergradient.
\newblock In {\em International conference on learning representations}, 2021.

\bibitem{fawcett2006introduction}
Tom Fawcett.
\newblock An introduction to roc analysis.
\newblock {\em Pattern recognition letters}, 27(8):861--874, 2006.

\bibitem{287222}
Yi~Zeng, Minzhou Pan, Himanshu Jahagirdar, Ming Jin, Lingjuan Lyu, and Ruoxi Jia.
\newblock {Meta-Sift}: How to sift out a clean subset in the presence of data poisoning?
\newblock In {\em 32nd USENIX Security Symposium (USENIX Security 23)}, pages 1667--1684, 2023.

\bibitem{tang2021demon}
Di~Tang, XiaoFeng Wang, Haixu Tang, and Kehuan Zhang.
\newblock Demon in the variant: Statistical analysis of dnns for robust backdoor contamination detection.
\newblock In {\em 30th $\{$USENIX$\}$ Security Symposium ($\{$USENIX$\}$ Security 21)}, 2021.

\bibitem{qi2023revisiting}
Xiangyu Qi, Tinghao Xie, Yiming Li, Saeed Mahloujifar, and Prateek Mittal.
\newblock Revisiting the assumption of latent separability for backdoor defenses.
\newblock In {\em The eleventh international conference on learning representations}, 2023.

\bibitem{zhao2021deep}
Shihao Zhao, Xingjun Ma, Yisen Wang, James Bailey, Bo~Li, and Yu-Gang Jiang.
\newblock What do deep nets learn? class-wise patterns revealed in the input space.
\newblock In {\em International Conference on Learning Representations}, 2022.

\bibitem{gao2023not}
Yinghua Gao, Yiming Li, Linghui Zhu, Dongxian Wu, Yong Jiang, and Shu-Tao Xia.
\newblock Not all samples are born equal: Towards effective clean-label backdoor attacks.
\newblock {\em Pattern Recognition}, 139:109512, 2023.

\bibitem{bagdasaryan2021blind}
Eugene Bagdasaryan and Vitaly Shmatikov.
\newblock Blind backdoors in deep learning models.
\newblock In {\em 30th USENIX Security Symposium (USENIX Security 21)}, pages 1505--1521, 2021.

\bibitem{li2021backdoor}
Yiming Li, Tongqing Zhai, Yong Jiang, Zhifeng Li, and Shu-Tao Xia.
\newblock Backdoor attack in the physical world.
\newblock {\em arXiv preprint arXiv:2104.02361}, 2021.

\bibitem{yao2019latent}
Yuanshun Yao, Huiying Li, Haitao Zheng, and Ben~Y Zhao.
\newblock Latent backdoor attacks on deep neural networks.
\newblock In {\em Proceedings of the 2019 ACM SIGSAC Conference on Computer and Communications Security}, pages 2041--2055, 2019.

\bibitem{shen2021backdoor}
Lujia Shen, Shouling Ji, Xuhong Zhang, Jinfeng Li, Jing Chen, Jie Shi, Chengfang Fang, Jianwei Yin, and Ting Wang.
\newblock Backdoor pre-trained models can transfer to all.
\newblock In {\em Proceedings of the 2021 ACM SIGSAC Conference on Computer and Communications Security}, pages 3141--3158, 2021.

\bibitem{liu2017fault}
Yannan Liu, Lingxiao Wei, Bo~Luo, and Qiang Xu.
\newblock Fault injection attack on deep neural network.
\newblock In {\em 2017 IEEE/ACM International Conference on Computer-Aided Design (ICCAD)}, pages 131--138. IEEE, 2017.

\bibitem{qi2021subnet}
Xiangyu Qi, Jifeng Zhu, Chulin Xie, and Yong Yang.
\newblock Subnet replacement: Deployment-stage backdoor attack against deep neural networks in gray-box setting.
\newblock {\em arXiv preprint arXiv:2107.07240}, 2021.

\bibitem{dong2023one}
Jianshuo Dong, Qiu Han, Yiming Li, Tianwei Zhang, Yuanjie Li, Zeqi Lai, Chao Zhang, and Shu-Tao Xia.
\newblock One-bit flip is all you need: When bit-flip attack meets model training.
\newblock In {\em ICCV}, 2023.

\bibitem{jebreel2023defending}
Najeeb~Moharram Jebreel, Josep Domingo-Ferrer, and Yiming Li.
\newblock Defending against backdoor attacks by layer-wise feature analysis.
\newblock In {\em Pacific-Asia Conference on Knowledge Discovery and Data Mining}, pages 428--440. Springer, 2023.

\bibitem{wang2022NONE}
Zhenting Wang, Hailun Ding, Juan Zhai, and Shiqing Ma.
\newblock Training with more confidence: Mitigating injected and natural backdoors during training.
\newblock {\em Advances in Neural Information Processing Systems}, 35:36396--36410, 2022.

\bibitem{zhu2022selective}
Rui Zhu, Di~Tang, Siyuan Tang, XiaoFeng Wang, and Haixu Tang.
\newblock Selective amnesia: On efficient, high-fidelity and blind suppression of backdoor effects in trojaned machine learning models.
\newblock In {\em 2023 IEEE Symposium on Security and Privacy (SP)}, pages 1220--1238. IEEE Computer Society, 2022.

\bibitem{sha2022fine}
Zeyang Sha, Xinlei He, Pascal Berrang, Mathias Humbert, and Yang Zhang.
\newblock Fine-tuning is all you need to mitigate backdoor attacks.
\newblock {\em arXiv preprint arXiv:2212.09067}, 2022.

\bibitem{xu2019Meta}
Xiaojun Xu, Qi~Wang, Huichen Li, Nikita Borisov, Carl~A Gunter, and Bo~Li.
\newblock Detecting ai trojans using meta neural analysis.
\newblock In {\em Proceedings of the IEEE Symposium on Security and Privacy (May 2021)}, 2021.

\bibitem{kolouri2020universal}
Soheil Kolouri, Aniruddha Saha, Hamed Pirsiavash, and Heiko Hoffmann.
\newblock Universal litmus patterns: Revealing backdoor attacks in cnns.
\newblock In {\em Proceedings of the IEEE/CVF Conference on Computer Vision and Pattern Recognition}, pages 301--310, 2020.

\end{thebibliography}

\newpage

\appendix

\section{Implementation and Configuration}
\label{appendix:implementation-and-configuration}

\subsection{BaDExpert Implementation Details}
\label{appendix:badexpert-implementation-details}

\subsubsection{Unlearning and Finetuning Configuration}
\label{appendix:unlearning-and-finetuning-configuration}

During unlearning (Alg~\ref{alg:alg-training-BE}), we select a small while effective (un)learning rate $\eta$. For our major experiments on ResNet18, $\eta=10^{-4}$ for CIFAR10 and $\eta=2.5 \cdot 10^{-5}$ for GTSRB. As for other architectures (CIFAR10), $\eta=8 \cdot 10^{-5}$ for VGG16, $\eta=8 \cdot 10^{-5}$ for MobileNetV2, and $\eta=10^{-2}$ for ResNet110 (SRA attack). On ImageNet, $\eta=10^{-4}$ for ResNet18 and ResNet101, and $\eta=10^{-6}$ for pretrained vit\_b\_16 (\texttt{IMAGENET1K\_SWAG\_LINEAR\_V1} version). We conduct unlearning using Adam optimizer for only 1 epoch, with a batch size of 128.

\textbf{Selection of $\eta$.} While we recommend using a conservatively small (u)nlearning rate $\eta$ ($10^{-4}$ in our major experiments) in Alg~\eqref{alg:alg-training-BE}, we also show that BaDExpert's defense performance is not sensitive to the choice of $\eta$ across a wide range (from $5\cdot10^{-5}$ to $5\cdot10^{-3}$) in Appendix~\ref{appendix:ablation-unlearning-rates}.

During the clean finetuning, we select a series of relatively larger learning rates ($\eta'$) in order to diminish the model's backdoor. The initial learning rate is $0.1$ for primary experiments on CIFAR10 with ResNet18, and $0.05$ on GTSRB. As for other architectures (CIFAR10), the initial learning is $0.2$ for MobileNetV2 and VGG16, and $0.05$ for ResNet110 (SRA attack). On ImageNet, the initial learning rate is $0.05$ for ResNet18, $10^{-5}$ for ResNet101, and $5 \cdot 10^{-4}$ for pretrained vit\_b\_16 (\texttt{IMAGENET1K\_SWAG\_LINEAR\_V1} version). We conduct the clean finetuning with SGD optimizer for 10 epochs, and reduce the learning rate to its $10\%$ after every two epochs, with a batch size of 64.

\textbf{Selection of $\eta'$.} In our major experiments, we directly follow the standard finetuning hyperparameters adopted in prior work~\cite{li2021neural}. Selecting the initial clean finetuning learning rate $\eta'$ can also be done via a manual search by analyzing the CA trend. Specifically, we observe that clean finetuning can best diminish the backdoor functionality with a large learning rate, where the finetuned model's CA drops to $\sim 0\%$ in the first one or two epochs, and then recovers to a significant value in the following epochs (the recovered CA depends on the size $|D_c|$; e.g. $\sim80\%$ when $|D_c| = 2,000$, but $\sim60\%$ when $|D_c| = 1,000$).

\subsection{Decision Rule}
\label{appendix:decision-rule}

\begin{figure}[t]
\centering
    \centering
         \includegraphics[width=.5\textwidth]{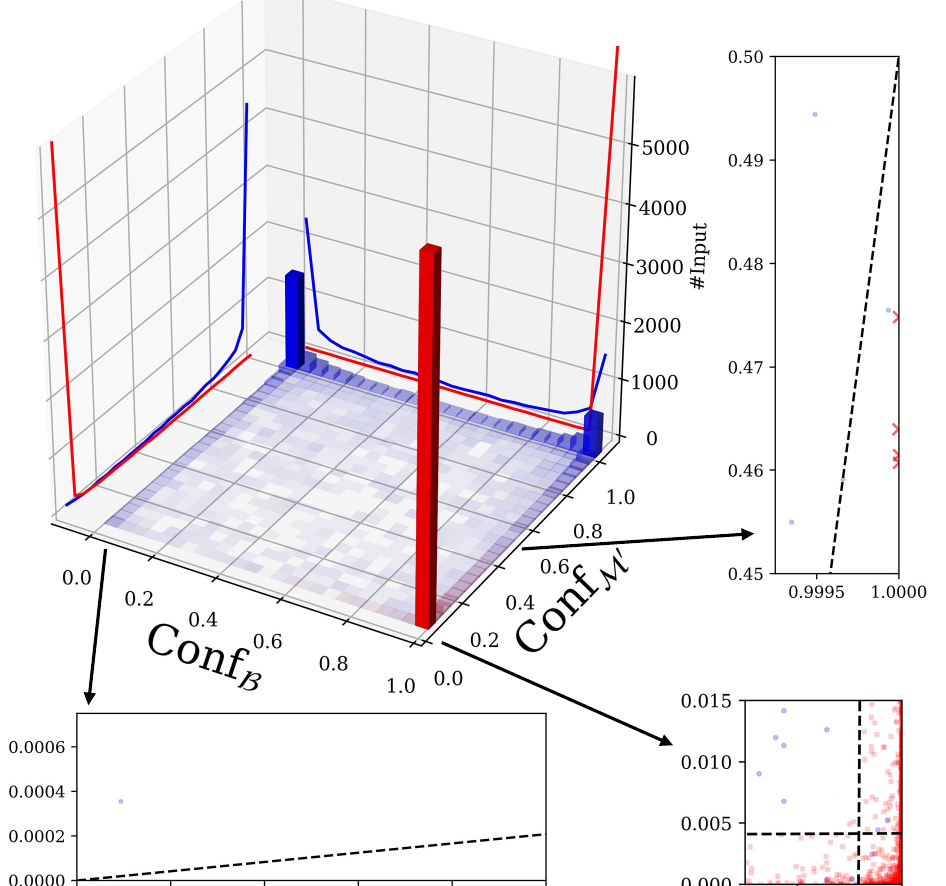}
        \caption{Clean (\blue{blue}) and backdoor (\red{red}) inputs on the $(\text{Conf}_\B, \text{Conf}_\mathcal{M'})$ plane. Top-left is the histogram heatmap, while the other three subfigures are zoom-in of the corresponding distribution scatter plot. Any inputs below the dashed lines are considered suspicious; this removes $97.54\%$ backdoor inputs while only results in $<1\%$ FPR.} 
        \label{fig:distribution-demo-with-decision-line}
\end{figure}

\paragraph{Distribution of clean and backdoor inputs on the 2D joint-confidence plane.}
Fig~\ref{fig:distribution-demo-with-decision-line} (top-left) demonstrates the $(\text{Conf}_\B, \text{Conf}_\mathcal{M'})$ 2D distribution histogram heatmap of clean inputs and backdoor inputs (WaNet~\cite{nguyen2021wanet} attack on CIFAR10). A deeper \blue{blue} grid represents more clean inputs and a deeper \red{red} grid represents more backdoor inputs.

As shown, \textit{clean and backdoor inputs could be easily distinguished} --- since most backdoor inputs locate nearby the bottom-right corner $(1.0, 0.0)$ (>90\% backdoor inputs are in the deepest red grid at the bottom-right corner), while the clean inputs do not (only <0.5\% clean inputs are in this grid).
Intuitively, 1) if $x$ is a \blue{clean} input, $\text{Conf}_\B$ should be small (the backdoor expert $\B$ should not vote for the predicted class of clean inputs) and $\text{Conf}_\mathcal{M'}$ should be relatively high (the auxiliary model $\Mprime$ should also recognize a clean input as $\M$ does); 2) if $x$ is a \red{backdoor} input that successfully triggers the backdoor, $\text{Conf}_\B$ should be high ($\B$ should recognize backdoor inputs as $\M$ does) and $\text{Conf}_\mathcal{M'}$ should be relatively low (finetuning has diminished the backdoor). 

\paragraph{Modified Decision Rule.}
As mentioned, the simple likelihood ratio score in Eq~\eqref{eq:decision-rule-simplified} is only optimal when both $\B$ and $\Mprime$ are well-calibrated, which cannot be guaranteed since the actual construction procedures for them involve deep learning techniques.
Therefore, in our practical implementation, we slightly modify Eq~\eqref{eq:decision-rule-simplified} according to the observed confidence distribution in the 2D plane as mentioned in the previous paragraph. Specifically, we score each input-prediction pair $(x, \predM)$ alternatively as follows:

\newcommand{\score}{\text{Score}(x, \Tilde{y})}

\vspace{-5mm}
\begin{small}
\begin{align}
    \text{Reject input } x \text{ if Score}(x, \Tilde{y}) := \min \left(
    \cfrac{\text{Conf}_\Mprime(\Tilde{y} | x)}
    {\gamma\cdot \text{Conf}_\B(\Tilde{y}|x)},
    \cfrac{1 - \text{Conf}_\B(\Tilde{y}|x)}
    {\Big(\gamma - \text{Conf}_\Mprime(\Tilde{y}|x)\Big)^+} \right) \le \alpha \text{ (threshold)}
    \label{eq:decision-rule}
\end{align}
\end{small}
where $(\cdot)^+ := \max(\cdot, 10^{-8})$ represents the operation of numerical positive clamping, and $\gamma$ is a hyperparameter (fixed to 0.5 through our major experiments).
This rule functions similarly to Eq~\eqref{eq:decision-rule-simplified}: a backdoor input $x$ tends to have a high $\confB( \predM | x )$ (i.e., $\B$ agrees with $\M$) and a low $\confMprime( \predM | x)$ (i.e., $\Mprime$ disagrees with $\M$), and therefore a low $\frac{\confMprime( \predM | x)}{\gamma\cdot \confB( \predM | x )}$ and $\frac{1 - \text{Conf}_\B(\Tilde{y}|x)}{(\gamma - \text{Conf}_\Mprime(\Tilde{y}|x))^+}$.
This modified score formulation is designed accordingly to best capture the actual $(\text{Conf}_\B, \text{Conf}_\mathcal{M'})$ distribution nature, as shown in Fig~\ref{fig:distribution-demo-with-decision-line}.
In the following paragraph, we will desribe a detailed empirical geometric interpretation of Eq~\eqref{eq:decision-rule}.

\paragraph{Empirically understanding the modified decision rule.}
In Fig~\ref{fig:distribution-demo-with-decision-line}, an obviously straightforward decision rule for backdoor detection is to remove any inputs dropped into the right-corner grid region.
However, since both $\B$ and $\Mprime$ may possibly make mistakes (i.e., not well-calibrated or suboptimal), some backdoor inputs would lie beyond this grid.
Therefore, we further smooth out this grid region into two \textit{triangle-shaped} regions (connected by the dashed lines and the borders in Fig~\ref{fig:distribution-demo-with-decision-line} top-right and bottom-left), and claim any inputs dropped into these two triangles to be suspicious for backdoor. Fig~\ref{fig:distribution-demo-with-decision-line} (bottom-right) reveals that our decision regions indeed capture the majority of backdoor inputs that locate around the bottom-right corner $(1.0, 0.0)$. Furthermore, our decision regions also capture a majority of backdoor outliers that distribute alongside the $\text{Conf}_\B = 1$ and $\text{Conf}_\mathcal{M'} = 0$ axes. Formally, this geometric decision rule is equivalent to calculating a score for any input $x$ and rejecting inputs with a score lower than a selected threshold $\alpha$, which has already been described in Eq~\eqref{eq:decision-rule}.

\paragraph{Selection of $\alpha$.}
In Table~\ref{tab:main_cifar10}, the threshold $\alpha$ is selected dynamically such that the false positive rate is always $1\%$. This is very much following previous work (e.g. STRIP~\cite{gao2019strip}) where defense results are reported when the false positive rates are fixed. Meanwhile, a fairer and widely-adopted way to report the results of such threshold-based input detectors would be to report the AUROC, which is threshold-free. Intuitively, a detector / classifier with a higher AUROC is usually considered better in pratical. To fairly present the effectiveness of our proposed defense (BaDExpert), we report both 1) ASR and CA when fixing FPR to $1\%$ and, as shown in Table~\ref{tab:main_cifar10}; 2) AUROC, which does not involve threshold selection, as shown in Table~\ref{tab:auroc_cifar10}.

For practitioners, an empirical and simple way for threshold selection would be to calculate a set of BaDExpert scores on $D_c$, and then determine the threshold $\alpha$ to be the highest $1^\text{st}$ percentile (or any other FPR) score among this set. Alternatively, the defender could also select an appropriate threshold based on the desired FPR by observing the score distribution of a small number of manually inspected benign inputs at inference time. According to our experimental results, deployers can reasonably anticipate BaDExpert to provide robust defense against potential backdoor attacks with a low permissible FPR (e.g. $1\%$ in our major experiment); and as the permissible FPR increases, the effectiveness of our defense mechanism is anticipated to further improve.

As for the sensitivity w.r.t. threshold selection, it appears that for differently trained and attacked models, $\alpha$ may need to be selected accordingly. However, an interesting quantative results on CIFAR10 would be: even if we set $\alpha$ to a fixed number (e.g. $-0.003$), the defense performance would not vary too much across different attacks (all ASR $<20\%$ while CA drop no more than $5\%$).

\paragraph{Selection of $\gamma$.}
In Fig~\ref{fig:distribution-demo-with-decision-line}, the hyperparameter $\gamma \in (0,1]$ corresponds to the intersection y-coordination of the top-right dashed line with the vertical border (e.g., the intersection point $(1.0, 0.5)$ corresponds to $\gamma=0.5$). $\gamma$ could be selected based on the confidence distribution of $\Mprime$ --- if $\Mprime$ assigns high confidences to most (clean) inputs, then a larger $\gamma$ would not induce too much FPR, while possibly incorporating more backdoor outliers (vice versa). Nevertheless, we find fixing $\gamma$ to $0.5$ already provides a consistently good performance through all our major experiments.

\subsection{Baseline Attacks Configurations}
\label{appendix:baseline-attacks-configuration}

Our detailed configurations for baseline attacks (CIFAR10) are listed as follow:
\begin{itemize}
    \item \textbf{BadNet}: $0.3\%$ poison ratio, using the 3x3 BadNet patch trigger~\cite{gu2017badnets} at the right-bottom corner.
    \item \textbf{Blend}: $0.3\%$ poison ratio, $20\%$ blending alpha, using the 32x32 Hellokitty trigger~\cite{Chen2017TargetedBA} at the right-bottom corner.
    \item \textbf{Trojan}: $0.3\%$ poison ratio, using the 8x8 TrojanNN patch trigger~\cite{liu2017trojaning} at the right-bottom part of the image.
    \item \textbf{CL}: $0.3\%$ poison ratio, adversarial perturbation on poisoned images bounded with $\ell_2$-norm of $600$, using four duplicates of the 3x3 BadNet patch trigger~\cite{gu2017badnets} at the four corners (for a considerable ASR).
    \item \textbf{SIG}: $2\%$ poison ratio (to achieve a considerable ASR).
    \item \textbf{Dynamic}: $0.3\%$ poison ratio.
    \item \textbf{ISSBA}: $2\%$ poison ratio (to achieve a stably considerable ASR).
    \item \textbf{WaNet}: $5\%$ poison ratio and $10\%$ cover ratio (recommended configurations).
    \item \textbf{BPP}: training-time poisoning, $20\%$ injection ratio, $20\%$ negative ratio (recommended configurations).
    \item \textbf{FT}: finetuning with the full training set (for both high CA and high ASR), $20\%$ injection ratio, $20\%$ blending alpha, using the 32x32 Hellokitty trigger~\cite{Chen2017TargetedBA} at the right-bottom corner.
    \item \textbf{TrojanNN}: finetuning with the full training set (for both high CA and high ASR), $10\%$ injection ratio, trojan trigger generated following the procedure in \cite{liu2017trojaning}.
    \item \textbf{SRA}: directly using the authors pretrained clean models and backdoor subnets (ResNet110) to conduct post-development backdoor injection.
\end{itemize}

We adopt the standard training pipeline to obtain backdoor models: SGD optimizer with a momentum of 0.9, a weight decay of $10^{-4}$, a batch size of 128, 100 epochs in total, initial learning rate of $0.1$ (decayed to its $10\%$ at the 50th and 75th epoch), with \texttt{RandomHorizontalFlip} and \texttt{RandomCrop} as the data augmentation.

\subsection{Baseline Defenses Configurations}
\label{appendix:baseline-defenses-configuration}

Our detailed configurations for baseline attacks (CIFAR10) are listed as follow:
\begin{itemize}
    \item \textbf{FP}: We forward a 2,000 reserved clean samples to the model, and keep pruning as much inactive neurons in the last convolutional layer as possible, until the CA drop reaches $10\%$.
    \item \textbf{NC}: Reverse engineer a trigger for each class with a 2,000-sample reserved clean set. Then an anomaly index is estimated for every class. The class with the highest anomaly index $> 2$ (whose mask norm is also smaller than the median mask norm) is determined as the target class for unlearning. Its reversed trigger is then attached to the same 2,000 clean samples (correctly labeled), on which the model is retrained to unlearn the backdoor (learning rate is $10^{-2}$ for one epoch).
    \item \textbf{MOTH}: Similarly, the trigger reverse engineering and model reparing are performed on a 2,000-sample reserved clean set. The learning rate for the model repairing process is default to $10^{-3}$ (for 2 epochs). 
    \item \textbf{NAD}: First train a teacher model in 10 epochs via finetuning (initial learning rate 0.1, decrease to its $10\%$ every two epochs), and use it to distill a student model in 20 epochs (learning rate is 0.1 for the first two epochs and 0.05 for the rest). NAD uses a 2,000 clean set to perform both finetuning and distillation.
    \item \textbf{STRIP}: Calculate an entropy for each sample is calculated by superimposing it with $N = 100$ randomly sampled clean samples, and consider inputs with the higher entropy to be backdoored; in Table~\ref{tab:main_cifar10}, the FPR is fixed to $10\%$ to show its effectiveness.
    \item \textbf{AC}: Gather all inputs for each class, perform a 2-clustering based on their latent representation, then assign each class a silhouette score. The class with the highest silhouette score is suspected, and the inputs within its larger cluster is considered as backdoored. The silhouette scores are used to report AUROC.
    \item \textbf{Frequency}: We directly use their official pretrained model to perform detection. The difference between output 1 and output 0 is used to report AUROC.
    \item \textbf{SCALE-UP}: Each input is scaled up 5 times (\texttt{scale\_set} $= \{3, 5, 7, 9, 11\}$), and the score corresponds to the fraction of the model's scaled predictions that equal to the prediction on the original input. The threshold in Table~\ref{tab:main_cifar10} is set to 0.5.
\end{itemize}

\paragraph{Fairness Considerations in Comparison.}
We mostly follow the baselines' original implementations if available. Moreover, to ensure their hyperparameters and implementations work in our settings (model architecture, optimizers, etc.), we also try to tune their hyperparameters if necessary, in order to report their best overall results. Most of these baseline defenses (other than those require no clean samples or those not sensitive to the number of clean samples) are given access to the exactly same clean reserved data (2,000 samples) as BaDExpert, which further ensures fairness in our comparison.

\subsection{Computational Environments}
\label{appendix:environments}

We run all experiments on a 4-rack cluster equipped with 2.8 GHz Intel Ice Lake CPUs and Nvidia A100 GPUs. Our major experiment requires training 63 models ($\sim$50 GPU hours in total), with an additional $>100$ GPU hours for ablation studies (e.g. training ImageNet models).

\section{Discussions}
\label{appendix:discussions}

\subsection{Formulation of Agreement Measurement (Hard-Label Decision Rules)}
\label{appendix:formulation-of-agreement-measurement}

Let us fist consider an ideal backdoor expert $\B$ that completely unlearns the normal functionality of the backdoored model $\M$ while fully preserving its backdoor functionality, i.e.,
{\small\begin{align}
    & \mathbb P_{(x,y)\sim \mathcal{P}} \Big[\B(\mathcal{T}(x)) = t |\mathcal M(\mathcal{T}(x)) = t\Big] \approx 1,\label{eqn:full_backdoor}\\ 
    & \mathbb P_{(x,y)\sim \mathcal{P}} \Big[\B(x) \ne y |\mathcal M(x) = y\Big] \approx 1 \label{eqn:full_unlearn}
\end{align}}
Under this condition: \textbf{1)} we can fully inhibit the embedded backdoor in $\M$ from activation (i.e., reduce the ASR of $\M$ to $0\%$) by simply {\textit{rejecting all inputs wherein predictions of $\M$ and $\B$ fall within an agreement}}. This is because $\B$ and $\M$ will always \textit{agree} with each other on a backdoor input $\mathcal{T}(x)$ that can exploit $\M$~(Eqn~\ref{eqn:full_backdoor}); \textbf{2)} Meanwhile, this rejection rule will not impede the CA of $\M$, because $\B$ will always \textit{disagree} with $\M$ on clean inputs $x$ that $\M$ correctly predict~(Eqn~\ref{eqn:full_unlearn}). This example thus suggests the feasibility of \textit{performing accurate backdoor input detection via measuring whether the predictions of the backdoored model and backdoored expert concur}.

\subsection{Formulation of and Insights into the Soft Decision Rules}
\label{appendix:soft-decision-rules}

As discussed in Sec~\ref{subsec:backdoor-detection}, for practical implementation, we can generalize the hard-label conditions to a soft version that is based on the soft-label (confidence-level) predictions. We can derive soft decision rules for both the backdoor expert $\B$ and the auxiliary model $\Mprime$.

Trivially, for any backdoor expert $\B$, the following soft conditions must establish:
\begin{align}
    & \exists \tau_1,\tau_2 \in [0,1], \text{s.t.} \\
    & \mathbb P_{(x,y)\sim \mathcal{P}} \Big[ \text{Conf}_\B ( t | \mathcal{T}(x)) \ge \tau_1  \Big|\mathcal M(\mathcal{T}(x)) = t\Big] \approx 1,\\ 
    & \mathbb P_{(x,y)\sim \mathcal{P}} \Big[ \text{Conf}_\B ( y | x) < \tau_2 \Big|\mathcal M(x) = y\Big] \approx 1,
\end{align}

\begin{figure}[t]
\centering
    \begin{subfigure}{0.49\textwidth}
    \centering
         \includegraphics[width=\textwidth]{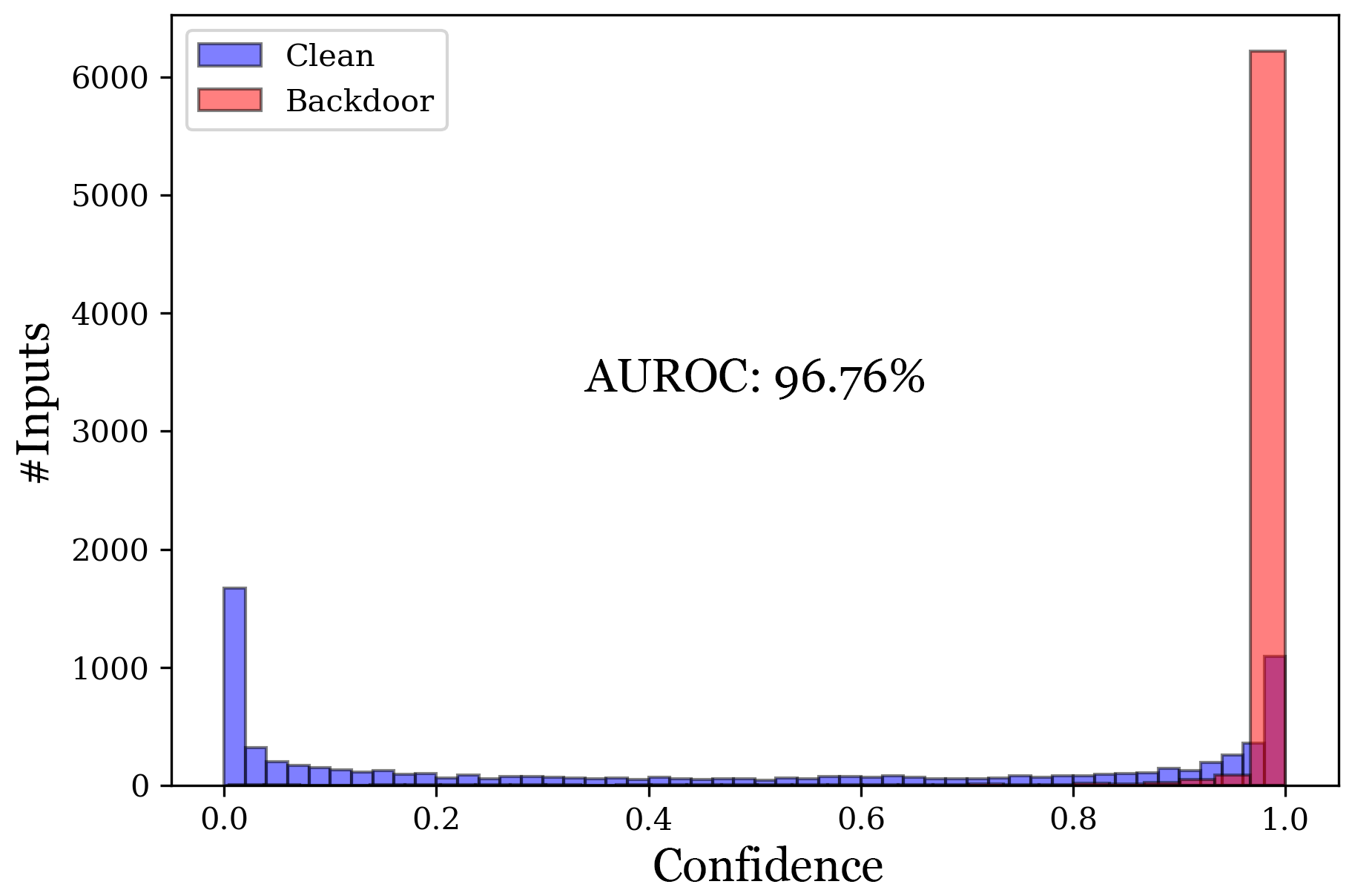}
        \caption{Backdoor expert $\B$.}
        \label{fig:B-conf-demo}
    \end{subfigure}
    \begin{subfigure}{0.49\textwidth}
    \centering
         \includegraphics[width=\textwidth]{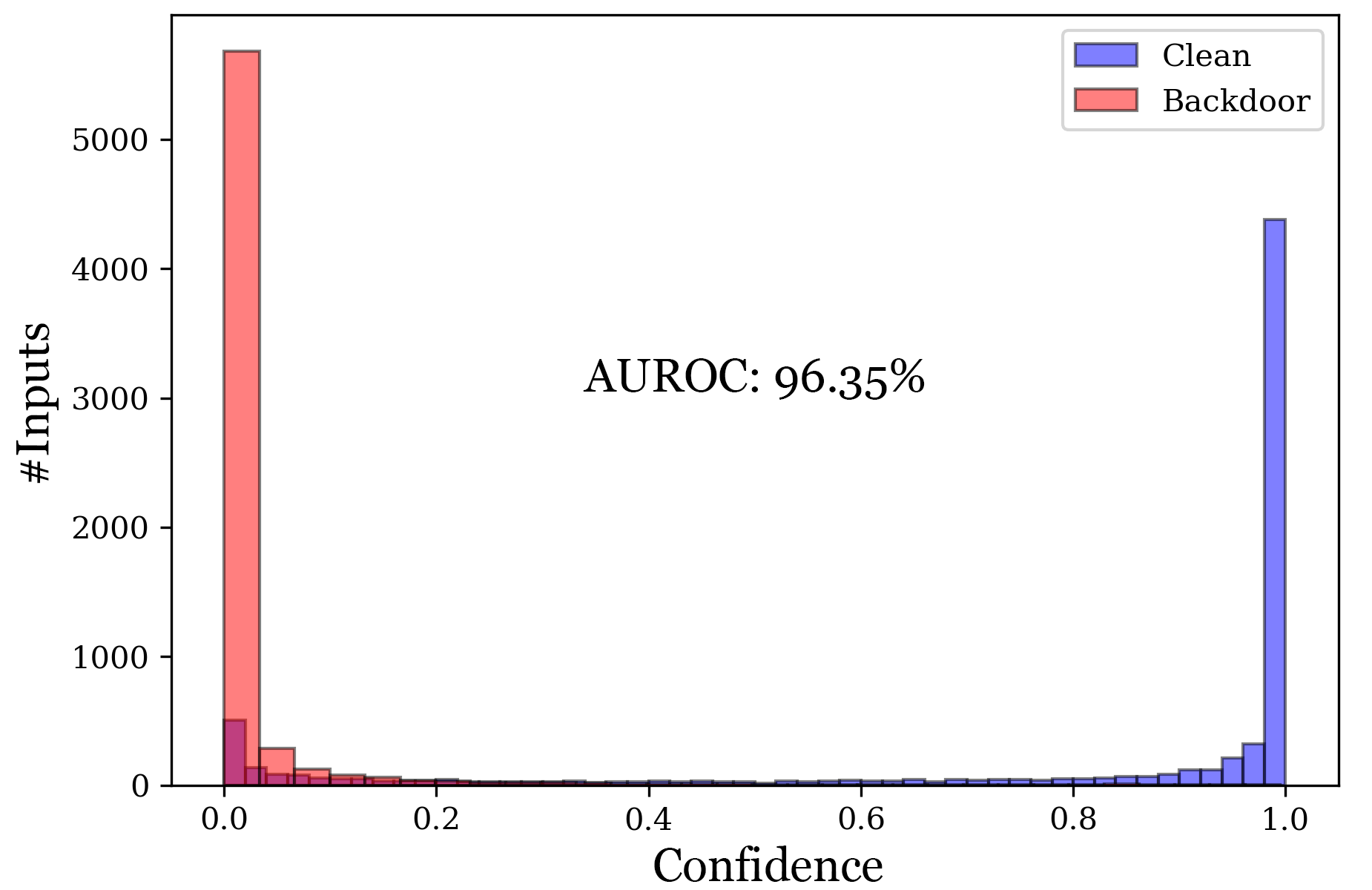}
        \caption{Auxiliary model $\Mprime$}
        \label{fig:Mprime-conf-demo}
    \end{subfigure}
    \caption{Confidence distribution for clean and backdoor (Blend attack on CIFAR10) inputs with regards to the intial predictions of the originally backdoored model $\M$. As shown, $\B$ tends to assign higher confidence to backdoor inputs and lower confidence to clean inputs, while $\Mprime$ does the contrary. When we directly apply the described soft rule, using either $\B$ or $\Mprime$, to distinguish backdoor inputs from clean inputs, we can achieve a high AUROC ($>96\%$).}
    \label{fig:conf-demo}
\end{figure}

Given an input $\Tilde{x}$, we can define a soft decision rule that rejects $\Tilde{x}$ conditional on $\text{Conf}_\B ( \M(\Tilde{x}) | \Tilde{x}) \ge \tau$. If $\tau_1 \ge \tau_2$, applying a $\tau \in [\tau_2,\tau_1]$ will still result in a perfect backdoor input detector. In suboptimal cases where $\tau_1 < \tau_2$, we will have a trade-off between the TPR and FPR during detection. Generally, $\B$ could serve as a good backdoor input detector if $\B$ tends to assign higher confidences for backdoor inputs and lower confidences for clean inputs, which is practically true (Fig~\ref{fig:B-conf-demo}). If we directly apply this confidence-level rule to detect backdoor inputs with a backdoor expert $\B$ (Blend attack on CIFAR10), we can achieve a $96.76\%$ AUROC.

On the contrary, an auxiliary model $\Mprime$ must satisfy the following conditions:
\begin{align}
    & \exists \tau_3,\tau_4 \in [0,1], \text{s.t.} \\
    & \mathbb P_{(x,y)\sim \mathcal{P}} \Big[ \text{Conf}_\Mprime ( t | \mathcal{T}(x)) \le \tau_3  \Big|\mathcal M(\mathcal{T}(x)) = t\Big] \approx 1,\\ 
    & \mathbb P_{(x,y)\sim \mathcal{P}} \Big[ \text{Conf}_\Mprime ( y | x) > \tau_4 \Big|\mathcal M(x) = y\Big] \approx 1,
\end{align}
where $\tau_3,\tau_4 \in [0,1]$. Given an input $\Tilde{x}$, we can define a soft decision rule that rejects $\Tilde{x}$ conditional on $\text{Conf}_\Mprime ( \M(\Tilde{x}) | \Tilde{x}) \le \tau'$. If $\tau_3 < \tau_4$, applying a $\tau' \in [\tau_3,\tau_4]$ will still result in a perfect backdoor input detector. In suboptimal cases where $\tau_3 \ge \tau_4$, we will have a trade-off between the TPR and FPR during detection. Similarly, $\Mprime$ could serve as a good backdoor input detector if $\Mprime$ tends to assign lower confidences for backdoor inputs and higher confidences for clean inputs, which is also practically true (Fig~\ref{fig:Mprime-conf-demo}). Analogously, if we directly apply this confidence-level rule to detect the same backdoor inputs, but with a finetuned auxiliary model $\Mprime$, we can achieve a $96.35\%$ AUROC.

To sum up, we see that these soft decision rules, based on the confidence-level information of $\B$ and $\Mprime$, can already detect backdoor inputs effectively ($>96\%$ AUROC). Our BaDExper framework, built on top of both the two models $\B$ and $\Mprime$ via ensembling, achieves an \textbf{even better detection performance (AUROC $> 99\%$)}.

\subsection{Other Possible Mislabeling Strategies}
\label{appendix:other-mislabeling-strategies}

In Alg~\ref{alg:alg-training-BE}, we specifically mislabel clean samples to their neighboring classes, i.e., $Y' \gets (Y + 1) \% C$. In our preliminary experiment, we have actually explored three different mislabeling strategies:

\begin{enumerate}
    \item Shifting $Y' \gets (Y + 1) \% C$ (adopted in Alg~\ref{alg:alg-training-BE});
    \item Randomly mapping $Y$ to any $Y'$ as long as $Y \ne Y'$;
    \item Change the one-hot label $Y = [0,0,\dotsm, 0, 1, 0, \dots, 0]$ to $Y = [\epsilon,\epsilon,\dotsm, \epsilon, 0, \epsilon, \dots, \epsilon]$ ($0 < \epsilon \le 1$) in a soft-label fashion.
\end{enumerate}

Surprisingly, we find the phenomenon -- ``\textit{finetuning a backdoored model on a few mislabeled clean samples can cause the model to forget its regular functionality, resulting in low clean accuracy, but remarkably, its backdoor functionality remains intact, leading to a high attack success rate}'' -- exists regardless of the adopted mislabeling strategy choice. And as a matter of fact, BaDExpert with each of the three strategies would be similarly effective against diverse set of attacks. We finally settled at the first mislabeling choice mostly due to a stable set of hyperparameters are easier to determined than the other two strategies.

\subsection{Comparing BaDExpert with Confusion Training~\cite{qi2023proactive}}
\label{appendix:comparing-badexpert-with-confusion-training}

\cite{qi2023proactive} introduces a novel backdoor poison training set cleanser based on the technique of ``confusion training'', where they train an inference model jointly on the poisoned dataset and a small number of \textit{mislabeled} clean samples (similar to our Alg~\eqref{alg:alg-training-BE}). Nevertheless, we highlight several critical differences between our work and theirs.

\paragraph{Problems.} \cite{qi2023proactive} focuses on \textbf{poisoned training set inspection} and aims at identifying poison samples within the training set. Instead, our work focuses on \textbf{identifying backdoor inputs during inference time}. The two problems have completely different setups.

\paragraph{Methods.} Our method and \cite{qi2023proactive}'s are different at two critical levels:
\begin{enumerate}
\item \textbf{Access of Information}:
    \begin{itemize}
        \item \cite{qi2023proactive} relies on the necessary \textit{access of poisoned training samples} (i.e., requiring information about the backdoor), so that their detection model can capture the backdoor correlation.
        \item Our method, on the other hand, operates independently of how the model is generated --- \textit{does not rely on any information about the backdoor samples or the poisoned dataset}, which is a significantly \textit{more challenging} scenario.
    \end{itemize}
\item \textbf{Principle}:
    \begin{itemize}
        \item During the training on a poisoned dataset, \cite{qi2023proactive} \textit{disrupts the fitting of clean training samples using a ``confusion batch'' of mislabeled clean data (i.e., counteracts the gradient updates learned from the normal training samples), so that the resulting model can only capture the backdoor correlation}. They then utilize this resulting model to identify poisoned training samples by seeing which data points are correctly fitted (by comparing the model's predictions with the data points' ground-truth labels).
        \item Distinguishly, our method is more related to catastrophic forgetting --- we \textit{only finetune the original backdoor model $\M$ on the mislabeled clean data} (without any access to poisoned samples), resulting in a backdoor expert model $\B$ that loses the normal functionality but retains the backdoor functionality. We then measure the agreements between the resulting backdoor expert model $\B$ and the original backdoor model $\M$, in order to identify the backdoor inputs at inference time. Notice that our approach operates without access to the ground-truth labels of inference-time inputs.
    \end{itemize}
\end{enumerate}

\subsection{Comparing BaDExpert with SEAM~\cite{zhu2022selective}}
\label{appendix:comparing-badexpert-with-seam}

\cite{zhu2022selective} introduces a novel model-repairing backdoor defense (SEAM). In the first phase, they finetune the backdoored model on a small number of \textit{mislabeled} clean samples (similar to our Alg~\eqref{alg:alg-training-BE}), observing that \textit{both the CA and ASR would diminish}. In the second phase, they finetune the resultant model (after phase one) on a portion of correctly labeled samples from the training set, by which the CA will gradually recover, but the ASR will not.

Interestingly, when finetune the backdoored model on mislabeled clean samples, \cite{zhu2022selective}'s observation (\textit{both CA and ASR decrease}) seems to be different from ours (\textit{CA drops but ASR retains}). Nevertheless, we argue that our observations are actually not contradictory to theirs.

In our method and experiments, we suggest using a conservatively small (un-)learning rate $\eta$, with which only the normal functionality degrades but the backdoor functionality retains. However, as shown in Fig~\ref{fig:backdoor-expert-badnet} (and Fig~\ref{fig:unlearning-curves} in Appendix~\ref{appendix:constructing-backdoor-experts}), when the (un-)learning rate $\eta$ is large enough (e.g., $10^{-3}$), both the normal and backdoor functionality would be lost (both CA and ASR $\to 0$) --- which corresponds to \cite{zhu2022selective}'s observation. In summary, the different observations between our work and \cite{zhu2022selective} are possibly due to different selections of the (un-)learning rate.

\section{Additional Results}
\label{appendix:additional-results}

\subsection{Effectiveness of BaDExpert on GTSRB}
\label{appendix:effectiveness-gtsrb}

Our primary results on GTSRB are shown in Table~\ref{tab:main_gtsrb} and Table~\ref{tab:auroc_gtsrb}. As a general post-development defense, BaDExpert effectively defends against all attacks (average ASR $= 2.0\%$), with a CA drop as negligible as $0.1\%$; Meanwhile, other baseline defenses fail against at least one backdoor attack. As a backdoor input detector, BaDExpert achieves an average $100\%$ detection AUROC, and outperforms other baseline detectors in every scenario.

\begin{table}[t]
\centering
\caption{Defensive results on GTSRB (CA and ASR).}
\resizebox{0.99\linewidth}{!}{
\begin{tabular}{ccccccccccccccccccccc}
\toprule
\textbf{Defenses$\rightarrow$} &
\multicolumn{2}{c}{No Defense} &
\multicolumn{2}{c}{FP} &
\multicolumn{2}{c}{NC} & 
\multicolumn{2}{c}{MOTH} & 
\multicolumn{2}{c}{NAD} & 
\multicolumn{2}{c}{STRIP} & 
\multicolumn{2}{c}{AC} & 
\multicolumn{2}{c}{Frequency} & 
\multicolumn{2}{c}{SCALE-UP} & 
\multicolumn{2}{c}{\textbf{BaDExpert}} \cr
\cmidrule(lr){2-3} \cmidrule(lr){4-5} \cmidrule(lr){6-7} \cmidrule(lr){8-9} \cmidrule(lr){10-11} \cmidrule(lr){12-13} \cmidrule(lr){14-15} \cmidrule(lr){16-17} \cmidrule(lr){18-19} \cmidrule(lr){20-21}
\textbf{Attacks} $\downarrow$ & CA & ASR & CA & ASR & CA & ASR & CA & ASR & CA & ASR & CA & ASR & CA & ASR & CA & ASR & CA & ASR & CA & ASR \cr
\midrule
No Attack & 97.1 & - & 85.4 & - & 97.1 & - & 82.2 & - & 96.3 & - & 87.4 & - & 96.8 & - & 65.0 & - & 60.8 & - & 97.0 & - \cr
\midrule
\multicolumn{19}{c}{\centering \textbf{Development-Stage Attacks}} \cr
\midrule
BadNet & 97.3 & \pc 100.0 & 85.7 & \grc 16.1 & 97.8 & \grc 0.0 & 92.0 & \pc 100.0 & 96.1 & \grc 0.5 & 86.9 & \pc 30.5 & 96.3 & \pc 100.0 & 64.6 & \grc 0.0 & 59.7 & \grc 0.0 & 97.2 & \grc 0.0 \cr
\midrule
Blend & 96.9 & \pc 97.5 & 86.7 & \pc 97.1 & 97.0 & \grc 6.9 & 88.2 & \pc 97.5 & 96.5 & \pc 49.2 & 87.3 & \pc 57.2 & 96.7 & \pc 97.5 & 64.9 & \grc 7.8 & 59.5 & \pc 59.6 & 96.8 & \grc 4.6 \cr
\midrule
Trojan & 97.1 & \pc 100.0 & 87.1 & \pc 98.3 & 98.1 & \grc 0.1 & 82.0 & \pc 100.0 & 96.4 & \pc 34.4 & 87.4 & \pc 64.5 & 96.9 & \pc 100.0 & 65.1 & \grc 0.0 & 60.4 & \pc 95.0 & 96.9 & \grc 3.7 \cr
\midrule
Dynamic & 97.1 & \pc 100.0 & 87.0 & \pc 100.0 & 97.4 & \pc 34.5 & 79.0 & \pc 100.0 & 96.9 & \pc 64.4 & 87.4 & \pc 37.1 & 96.8 & \pc 100.0 & 65.0 & \grc 12.0 & 59.3 & \pc 24.6 & 97.0 & \grc 0.0 \cr
\midrule
WaNet & 95.7 & \pc 91.4 & 84.1 & \grc 5.4 & 98.2 & \grc 17.4 & 94.5 & \pc 89.5 & 97.0 & \grc 0.4 & 86.2 & \pc 84.0 & 95.5 & \pc 91.4 & 64.1 & \pc 61.8 & 56.7 & \pc 79.2 & 95.6 & \grc 0.2 \cr
\midrule

\multicolumn{19}{c}{\centering \textbf{Post-development Attacks}} \cr
\midrule
FT & 95.5 & \pc 99.8 & 85.4 & \grc 0.0 & 94.0 & \pc 39.9 & 83.6 & \pc 98.0 & 96.4 & \grc 13.5 & 86.0 & \grc 19.3 & 95.2 & \pc 99.9 & 63.9 & \grc 7.9 & 62.3 & \pc 66.8 & 95.4 & \grc 3.7 \cr
\midrule
TrojanNN & 96.2 & \pc 98.4 & 84.2 & \grc 14.8 & 96.2 & \grc 0.2 & 78.9 & \pc 97.7 & 96.1 & \grc 0.3 & 86.6 & \pc 40.5 & 96.0 & \pc 98.4 & 64.5 & \grc 0.0 & 62.8 & \grc 2.1 & 96.4 & \grc 1.9 \cr
\midrule
\midrule
\textbf{Average} & 96.6 & 98.2 & 85.7 & 47.4 & 97.0 & 14.1 & 85.1 & 97.5 &  96.5 & 23.2 & 86.9 & 47.6 & 96.3 & 98.2 & 64.6 & 12.8 & 60.2 & 46.7 & \textbf{96.5} & \textbf{2.0} \cr
\midrule

\multicolumn{3}{c}{CA Drop (smaller is better)} & $\downarrow$10.9 & & \textbf{$\uparrow$0.3} &  & $\downarrow$11.6 &  & $\downarrow$0.1 &  & $\downarrow$9.7 &  & $\downarrow$0.3 &  & $\downarrow$32.0 &  & $\downarrow$36.4 & & \textbf{$\downarrow$0.1} & \cr
\midrule

\multicolumn{3}{c}{ASR Drop (larger is better)} &  & $\downarrow$50.8 &  & $\downarrow$84.0 & & $\downarrow$0.6 &  & $\downarrow$74.9 & & $\downarrow$50.6 & & $\downarrow$0.0 & & \textbf{$\downarrow$85.4} & & $\downarrow$51.4 & & \textbf{$\downarrow$96.1} \cr
\bottomrule

\end{tabular}
} 
\label{tab:main_gtsrb}
\end{table}

\begin{table}
\centering
\caption{Defensive results on GTSRB (AUROC).}
\resizebox{0.7\linewidth}{!}{ 
\begin{tabular}{lccccccccccccc}
\toprule
 \textbf{AUROC} ($\%$) & 
\multicolumn{1}{c}{BadNet} &
\multicolumn{1}{c}{Blend} &
\multicolumn{1}{c}{Trojan} &
\multicolumn{1}{c}{Dynamic} &
\multicolumn{1}{c}{WaNet} &
\multicolumn{1}{c}{FT} &
\multicolumn{1}{c}{TrojanNN} &
\multicolumn{1}{c}{\textbf{Average}}\cr
\midrule

STRIP & 93.4 & 74.8 & 76.5 & 86.2 & 41.6 & 92.2 & 88.7 & 79.0 \cr
\midrule

AC & 52.7 & 50.2 & 32.4 & 62.7 & 34.2 & 48.3 & 30.9 & 44.5 \cr
\midrule

Frequency & 75.3 & 73.6 & 75.3 & 72.8 & 61.3 & 73.6 & 75.3 & 72.4 \cr
\midrule

SCALE-UP & 88.6 & 54.2 & 34.0 & 81.1 & 34.1 & 56.4 & 90.4 & 62.7 \cr
\midrule

\textbf{BaDExpert} & \textbf{100.0} & \textbf{99.9} & \textbf{100.0} & \textbf{100.0} & \textbf{100.0} & \textbf{100.0} & \textbf{99.9} & \textbf{100.0} \cr
\bottomrule

\end{tabular}
} 
\label{tab:auroc_gtsrb}
\end{table}

\subsection{Ensembling with Other Defenses}
\label{appendix:ensembling-with-other-defenses}

\begin{table}[t]
\centering
\caption{AUROC of BaDExpert with baseline (FP, NC, MOTH and NAD) repaired models as $\Mprime$ (``w/ Backdoor Expert'') on CIFAR10. We use the exact ensembling decision rule in our primary experiment. 
Results of directly deploying the repaired models, following the soft rule described in Sec~\ref{appendix:soft-decision-rules}, are shown in ``w/o Backdoor Expert'' rows. Obviously, our backdoor experts in the BaDExpert framework serve as effective augmentations (add-ons) for these baseline methods during backdoor input detection.}
\label{tab:auroc-ensembling-with-other-defenses}
\resizebox{0.99\linewidth}{!}{ 
\begin{tabular}{llccccccccccccc}
\toprule
Baseline as $\Mprime$ $\downarrow$ & 
Attacks $\rightarrow$ & 
\multicolumn{1}{c}{BadNet} &
\multicolumn{1}{c}{Blend} &
\multicolumn{1}{c}{Trojan} &
\multicolumn{1}{c}{CL} &
\multicolumn{1}{c}{SIG} &
\multicolumn{1}{c}{Dynamic} &
\multicolumn{1}{c}{ISSBA} &
\multicolumn{1}{c}{WaNet} &
\multicolumn{1}{c}{Bpp} &
\multicolumn{1}{c}{FT} &
\multicolumn{1}{c}{TrojanNN} &
\multicolumn{1}{c}{SRA} &
\multicolumn{1}{c}{\textbf{Average}}\cr
\midrule

\multirow{2}*{FP}
& w/o Backdoor Expert & 10.1 & 64.3 & 64.4 & 40.5 & 90.8 & 8.3 & 97.2 & 97.3 & 99.2 & 64.1 & 66.9 & 0.2 & 58.6 \cr
& w/ Backdoor Expert & 100.0 & 96.9 & 99.5 & 99.6 & 99.8 & 66.6 & 94.8 & 99.3 & 99.9 & 99.5 & 99.4 & 99.9 & \textbf{96.3} \cr

\midrule

\multirow{2}*{NC}
& w/o Backdoor Expert & 99.2 & 53.4 & 99.9 & 99.9 & 66.7 & 99.5 & 98.8 & 52.7 & 99.9 & 93.2 & 99.8 & 99.9 & 88.6 \cr
& w/ Backdoor Expert & 100.0 & 94.8 & 100.0 & 100.0 & 97.6 & 99.9 & 95.3 & 98.7 & 100.0 & 98.3 & 100.0 & 100.0 & \textbf{98.7} \cr

\midrule

\multirow{2}*{MOTH}
& w/o Backdoor Expert & 99.7 & 32.1 & 99.0 & 99.6 & 37.0 & 94.2 & 52.6 & 95.8 & 99.4 & 63.3 & 93.2 & 99.8 & 80.5 \cr
& w/ Backdoor Expert & 100.0 & 88.6 & 99.8 & 99.9 & 87.4 & 98.8 & 92.3 & 99.7 & 100.0 & 94.7 & 98.2 & 100.0 & \textbf{96.6} \cr

\midrule

\multirow{2}*{NAD}
& w/o Backdoor Expert & 98.9 & 95.9 & 96.8 & 94.9 & 98.6 & 92.9 & 98.0 & 99.1 & 99.6 & 97.1 & 92.6 & 99.2 & 97.0 \cr
& w/ Backdoor Expert & 100.0 & 98.6 & 99.2 & 98.6 & 99.8 & 98.0 & 95.1 & 99.8 & 100.0 & 99.5 & 97.6 & 100.0 & \textbf{98.8} \cr

\bottomrule

\end{tabular}
} 
\end{table}

As discussed in Sec~\ref{subsubsec:ablation-studies}, we can apply any baseline-repaired models as $\Mprime$ in our BaDExpert framework, to ensemble with our backdoor experts $\B$.
We demonstrate the ensembling results in Table~\ref{tab:auroc-ensembling-with-other-defenses} (``w/ Backdoor Expert'' rows). For an insightful comparison, we also show the results when only the baseline-repaired model is used for backdoor input detection (``w/o Backdoor Expert'' rows), following the soft decision rule for $\Mprime$ described in Sec~\ref{appendix:soft-decision-rules}.

As shown, BaDExpert can achieve overall $\sim 99\%$ AUROCs when ensembling with NC and NAD, which align well with our major results in Table~\ref{tab:auroc_cifar10}. 
When combined with FP (failed against 9 of 12 attacks in Table~\ref{tab:main_cifar10}) and MOTH (failed against 6 of 12 attacks in Table~\ref{tab:main_cifar10}), BaDExpert slightly degrades to $\sim 96.5\%$, due to the significant failures of the baselines themselves (which can also be told from that deploying $\Mprime$ without backdoor expert can sometimes barely achieve AUROC $<50\%$ --- worse than random guessing).
Moreover, in almost all cases, BaDExpert (``w/ Backdoor Expert'') achieves higher AUROCs compared to deploying $\Mprime$ alone (``w/o Backdoor Expert''). In other words, our backdoor expert models and the BaDExpert framework could serve as effective augmentations (or add-ons) to existing model-repairing backdoor defense baselines, during backdoor input detection.

\subsection{Backdoor Experts Construction}
\label{appendix:constructing-backdoor-experts}

\begin{figure}
\centering

    \begin{subfigure}{0.32\textwidth}
        \includegraphics[width=\textwidth]{fig/unlearning_curves/unlearning_curve_BadNet.png}
        \caption{BadNet} \label{fig:unlearning_curve_BadNet}
    \end{subfigure}
    \begin{subfigure}{0.32\textwidth}
        \includegraphics[width=\textwidth]{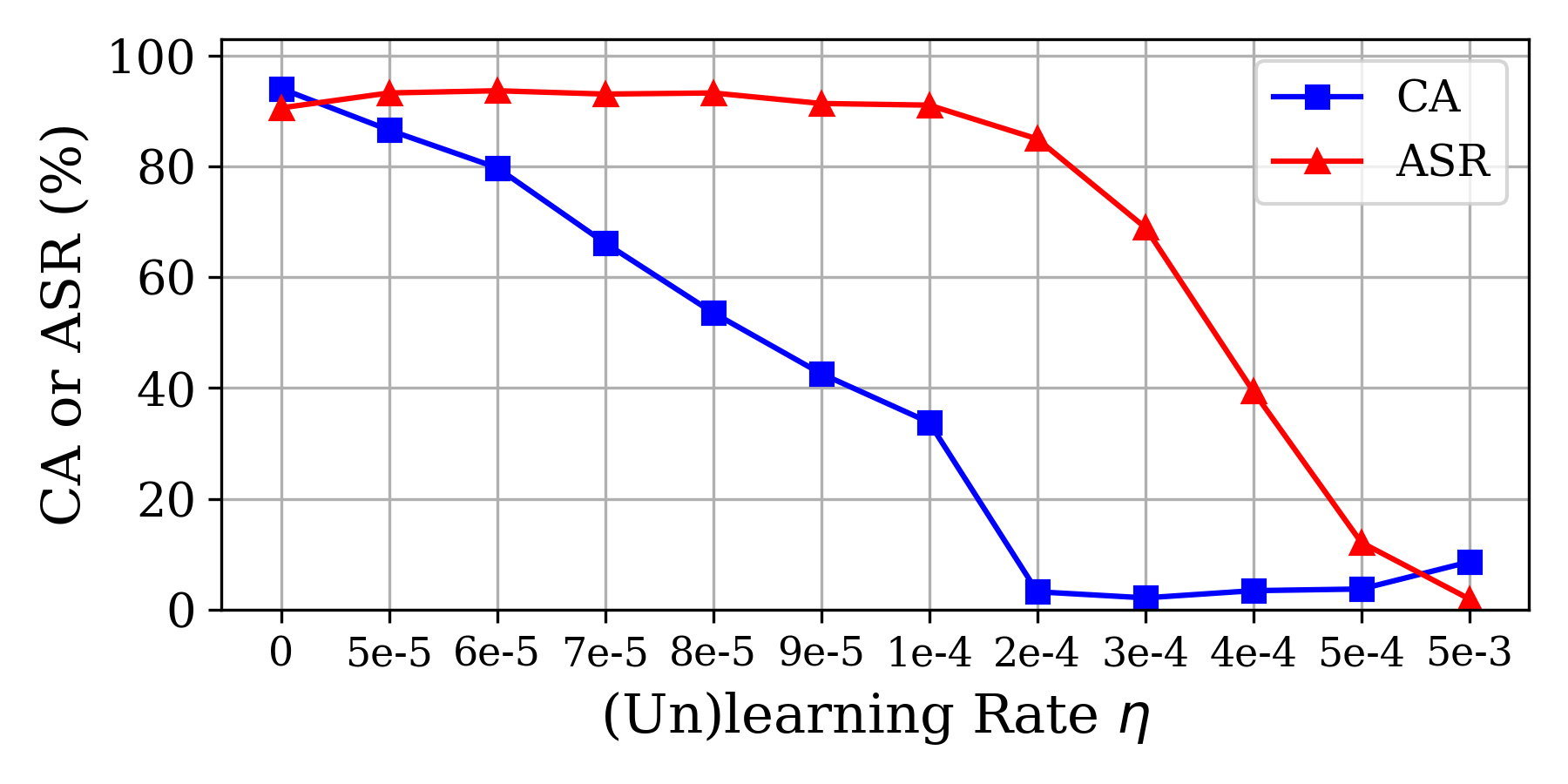}
        \caption{Blend}
    \end{subfigure}
    \begin{subfigure}{0.32\textwidth}    
        \includegraphics[width=\textwidth]{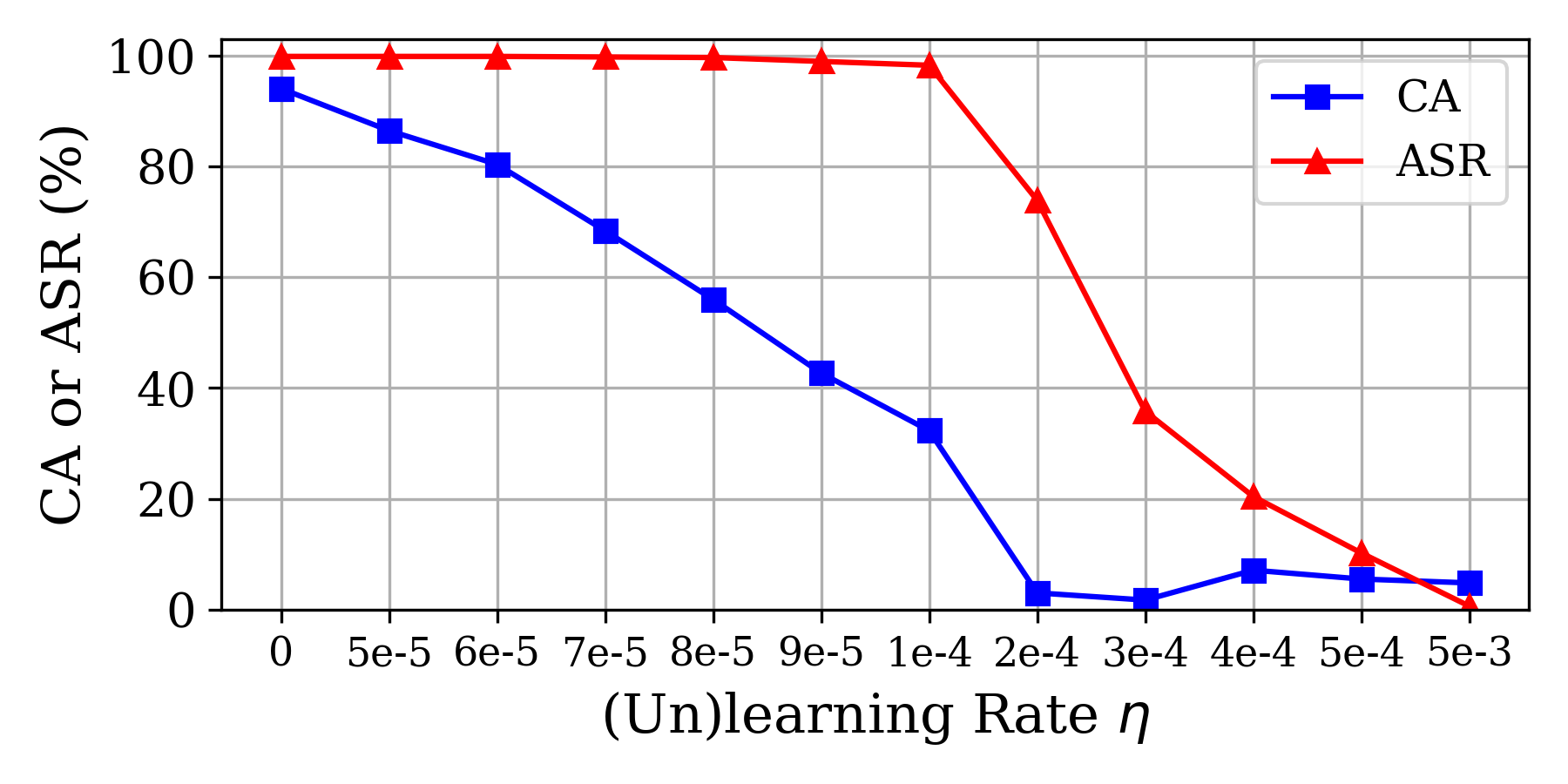}
        \caption{Trojan}
    \end{subfigure}
    \\
    \begin{subfigure}{0.32\textwidth}    
        \includegraphics[width=\textwidth]{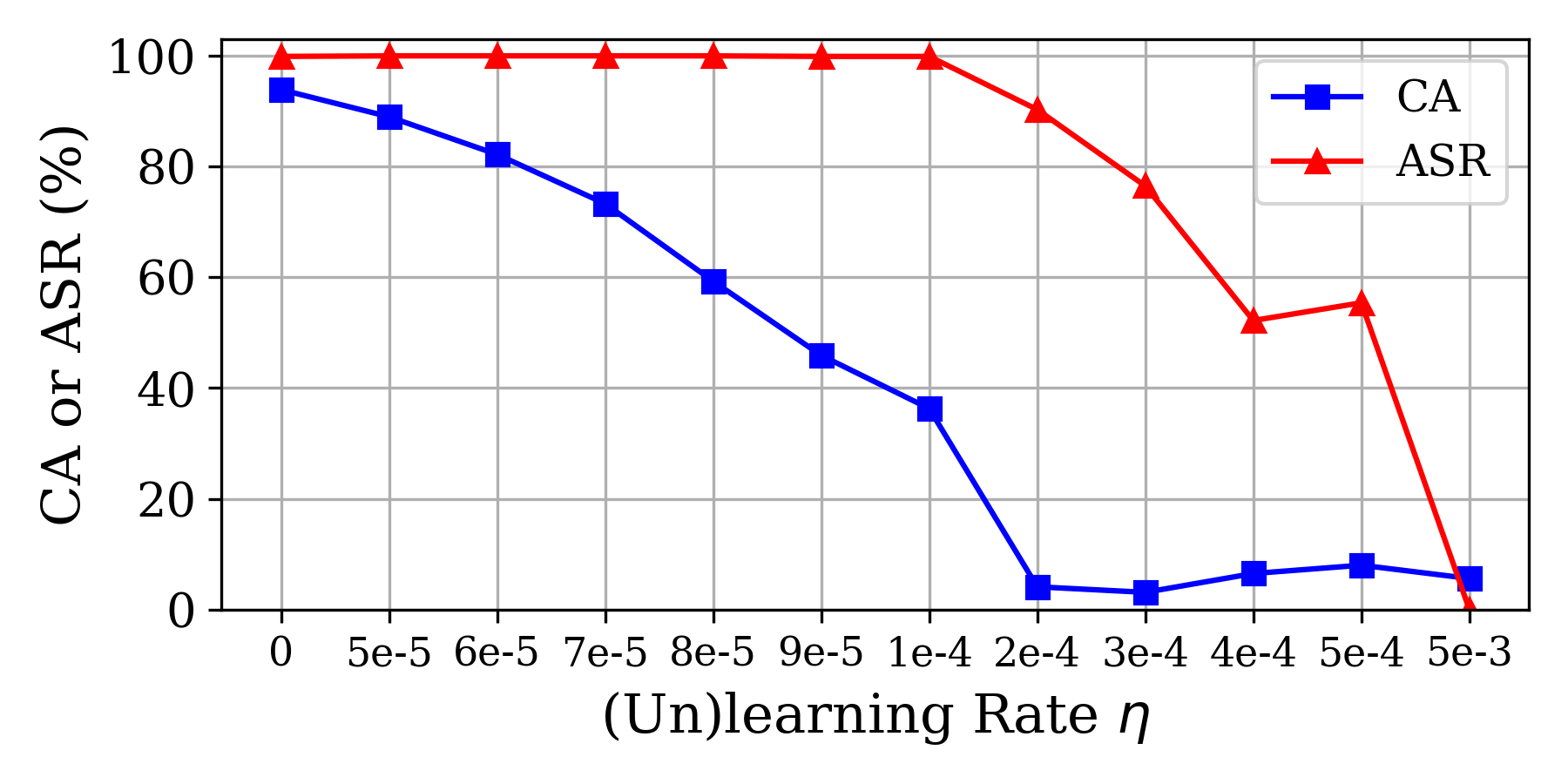}
        \caption{CL}
    \end{subfigure}
    \begin{subfigure}{0.32\textwidth}    
        \includegraphics[width=\textwidth]{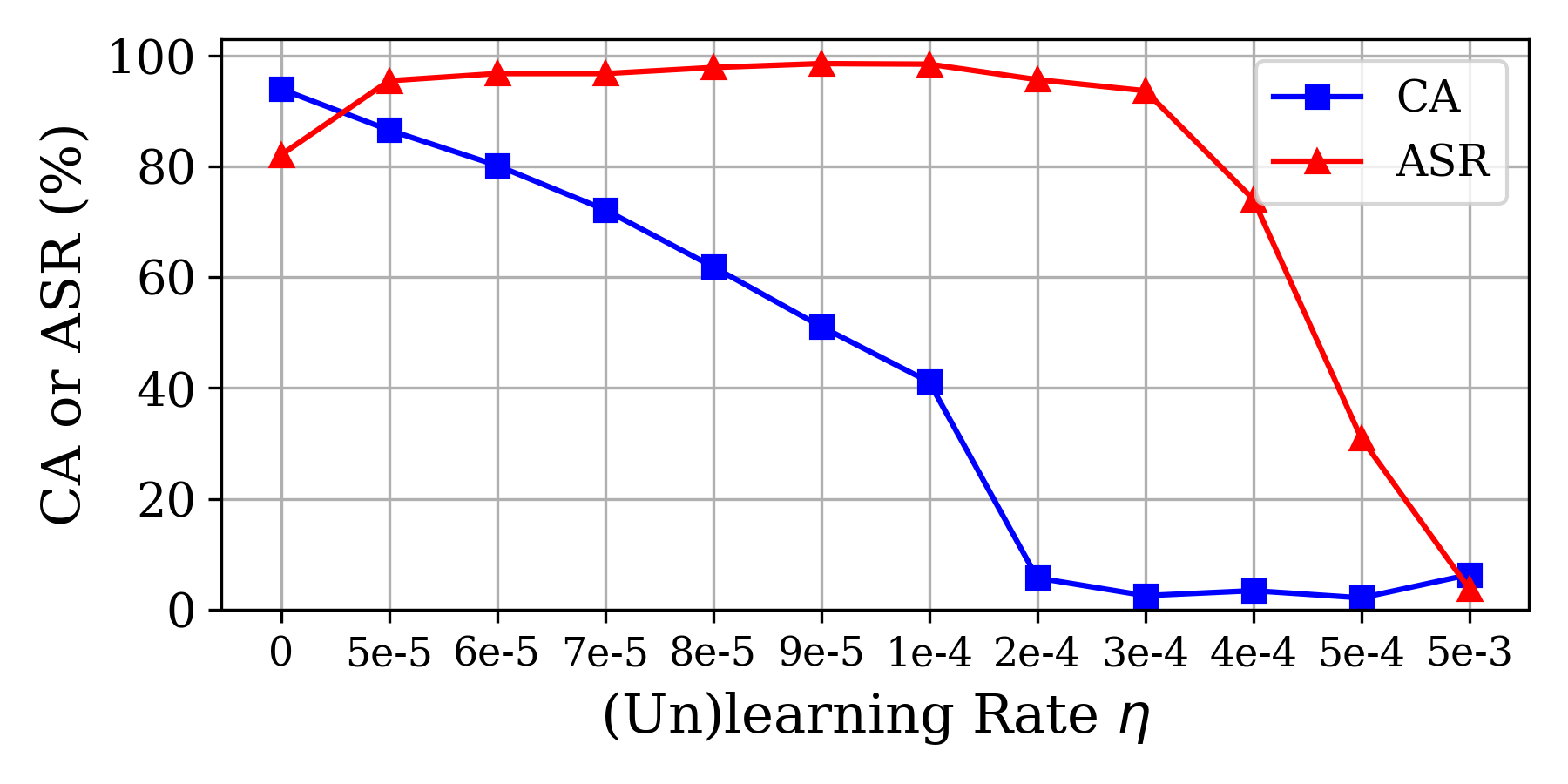}
        \caption{SIG}
    \end{subfigure}
    \begin{subfigure}{0.32\textwidth}    
        \includegraphics[width=\textwidth]{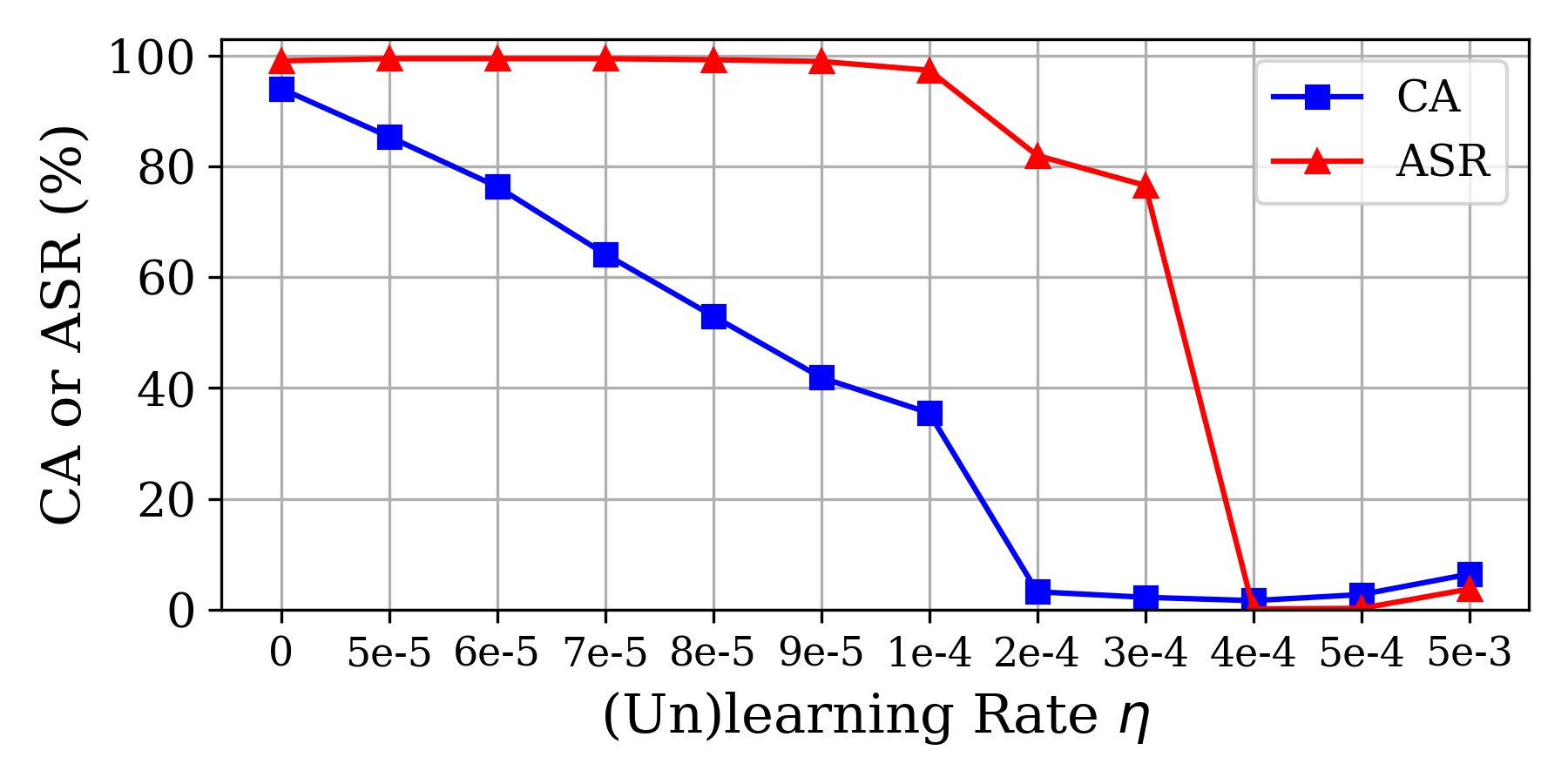}
        \caption{Dynamic}
    \end{subfigure}
    \\
    \begin{subfigure}{0.32\textwidth}    
        \includegraphics[width=\textwidth]{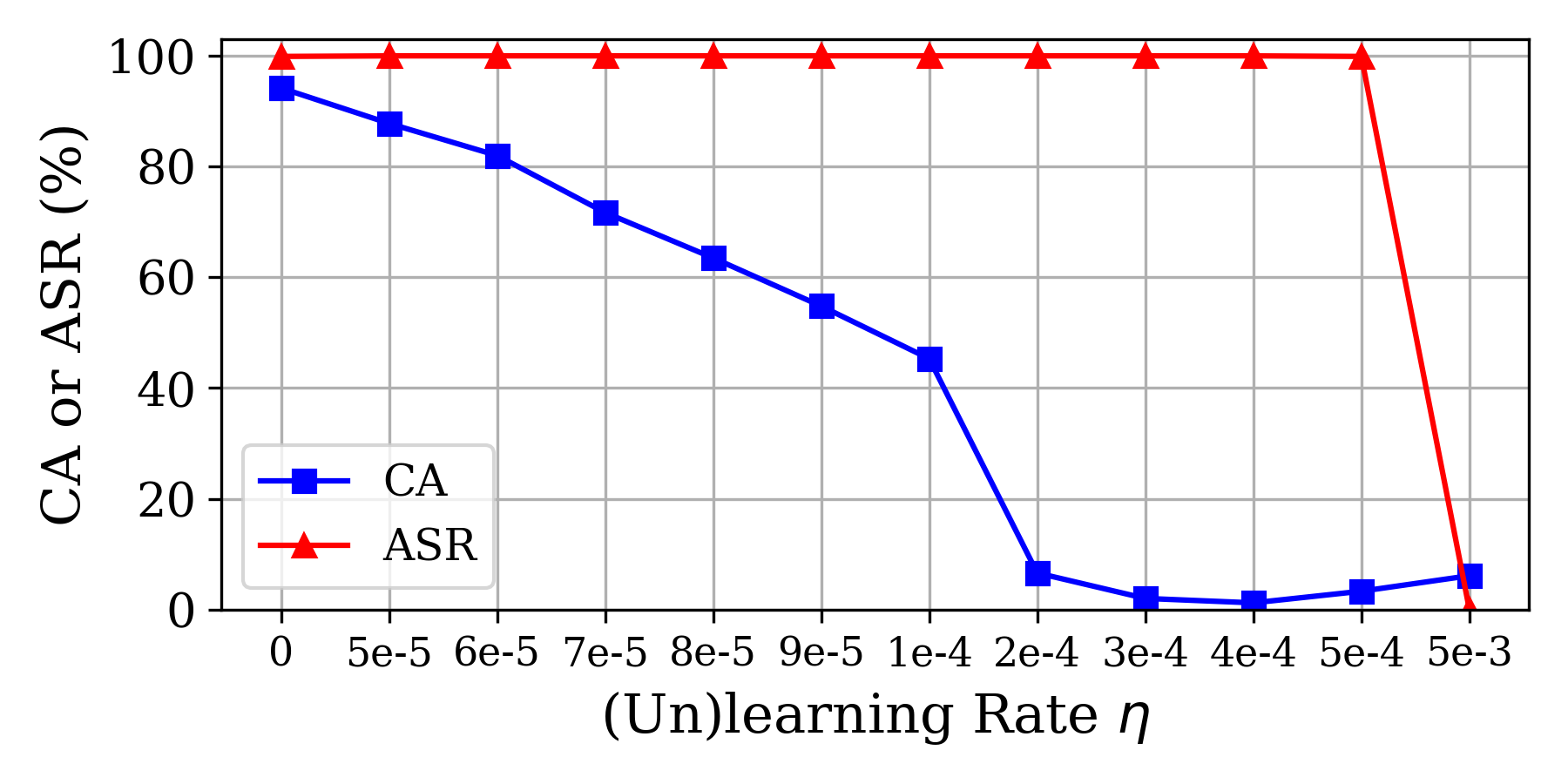} 
        \caption{ISSBA}
    \end{subfigure}
    \begin{subfigure}{0.32\textwidth}
        \includegraphics[width=\textwidth]{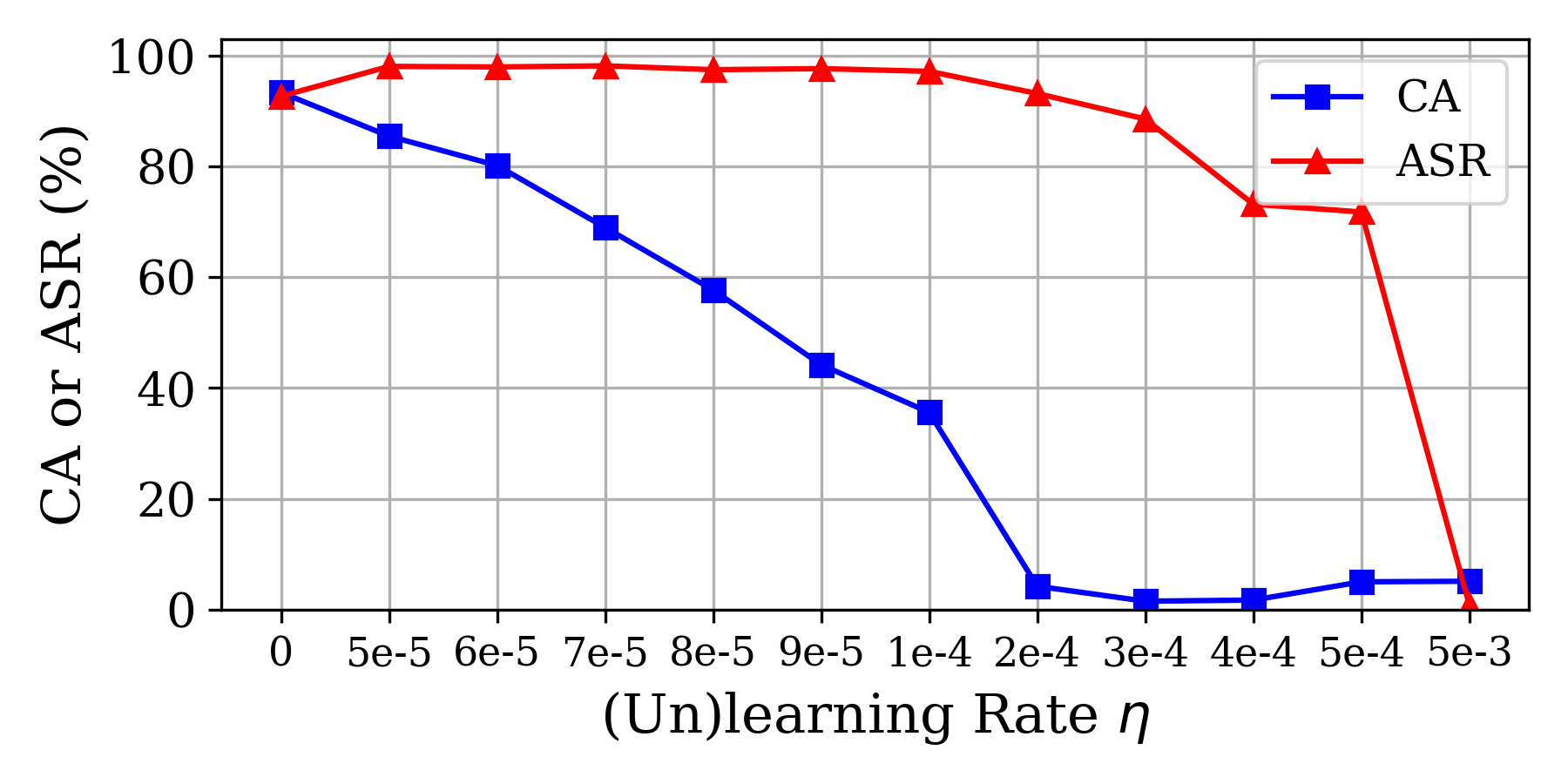}
        \caption{WaNet}
    \end{subfigure}
    \begin{subfigure}{0.32\textwidth}
        \includegraphics[width=\textwidth]{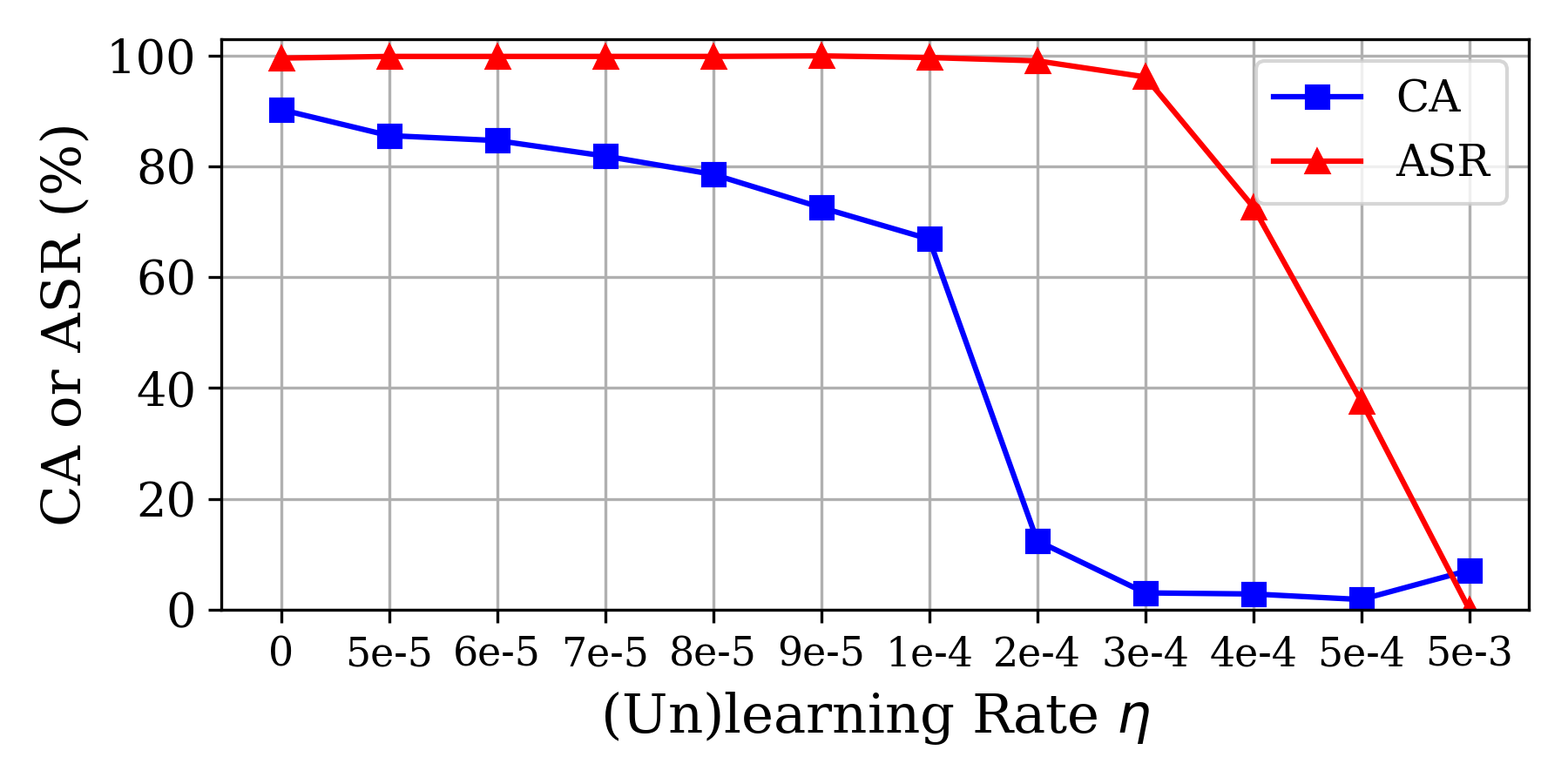}
        \caption{BPP}
    \end{subfigure}
    \\
    \begin{subfigure}{0.32\textwidth}
        \includegraphics[width=\textwidth]{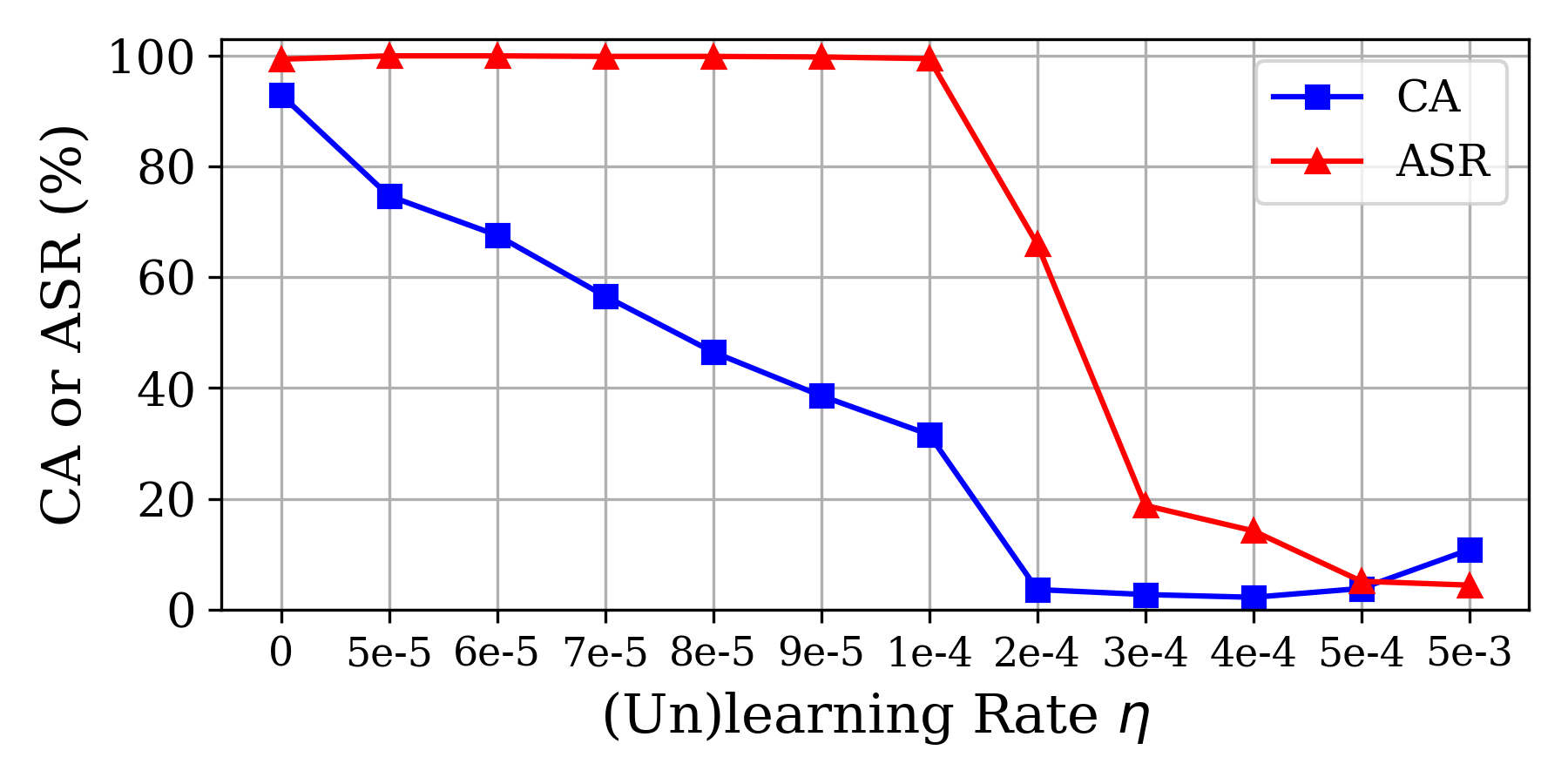}
        \caption{FT}
    \end{subfigure}
    \begin{subfigure}{0.32\textwidth}
        \includegraphics[width=\textwidth]{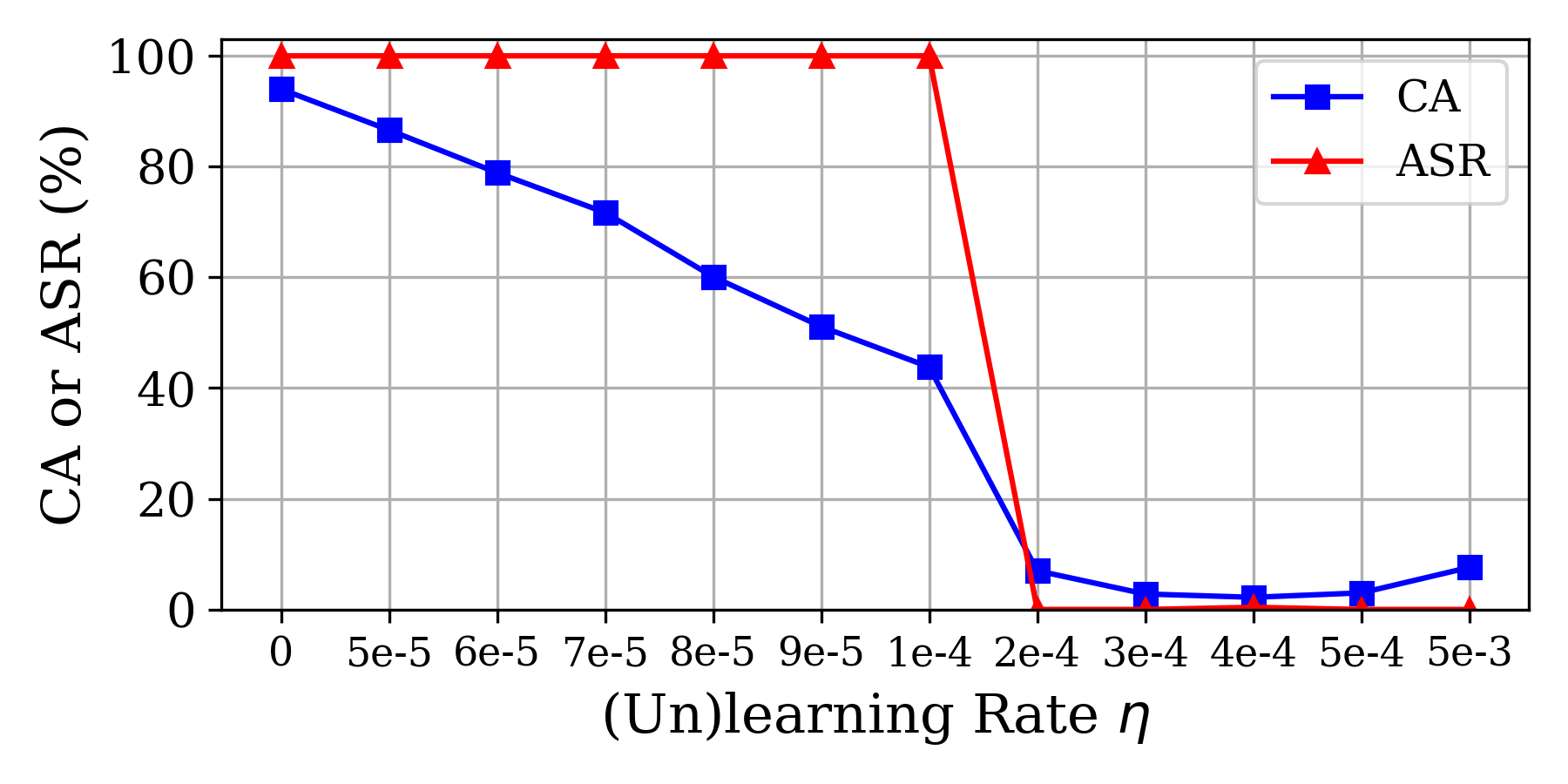} 
        \caption{TrojanNN}
    \end{subfigure}
    \begin{subfigure}{0.32\textwidth}
        \includegraphics[width=\textwidth]{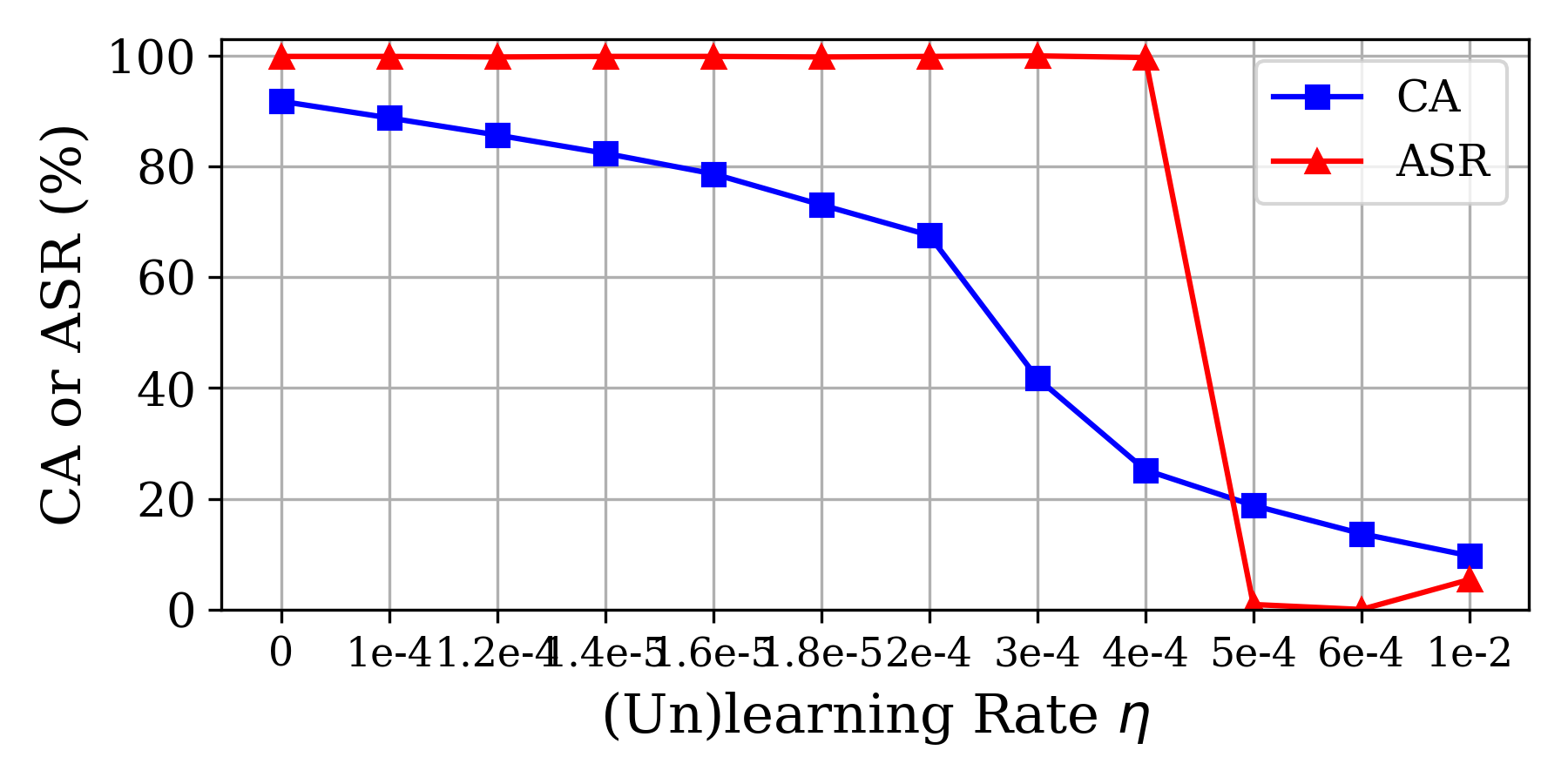}
        \caption{SRA} \label{fig:unlearning_curve_SRA}
    \end{subfigure}
    \\
    \begin{subfigure}{0.32\textwidth}
        \includegraphics[width=\textwidth]{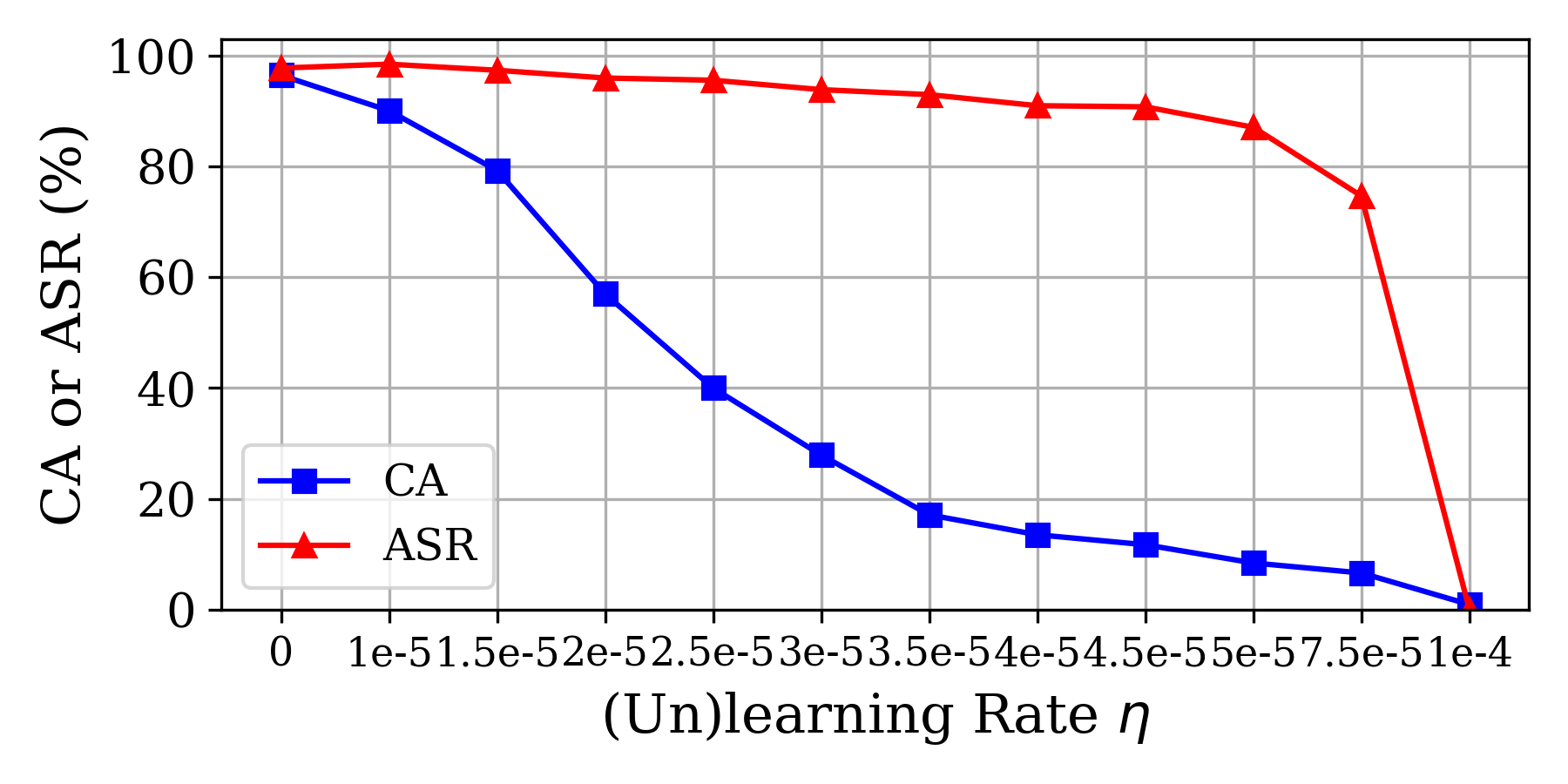}
        \caption{GTSRB} \label{fig:unlearning_curve_GTSRB}
    \end{subfigure}
    \begin{subfigure}{0.32\textwidth}
        \includegraphics[width=\textwidth]{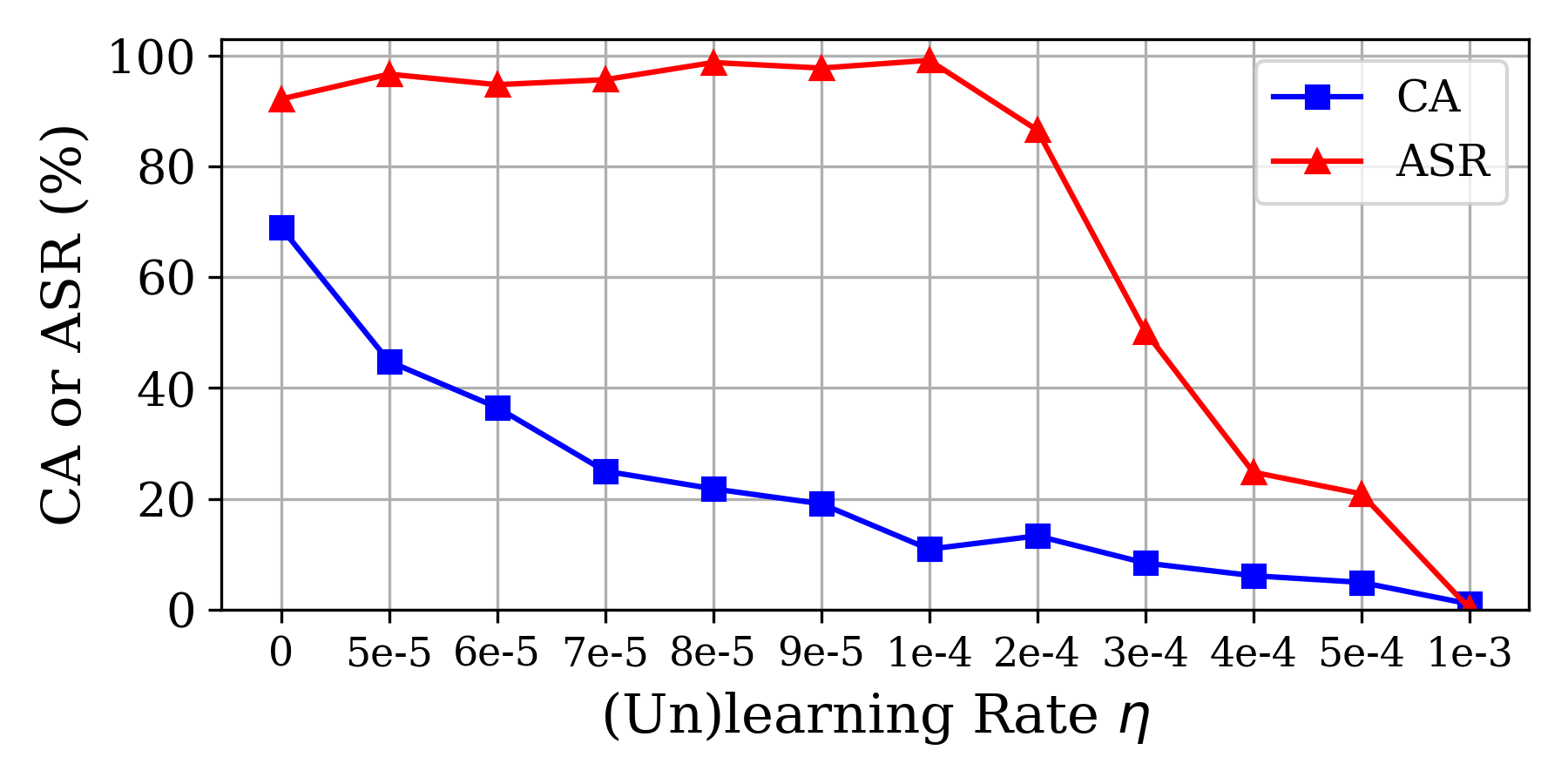}
        \caption{ImageNet} \label{fig:unlearning_curve_ImageNet}
    \end{subfigure}
    \begin{subfigure}{0.32\textwidth}
        \includegraphics[width=\textwidth]{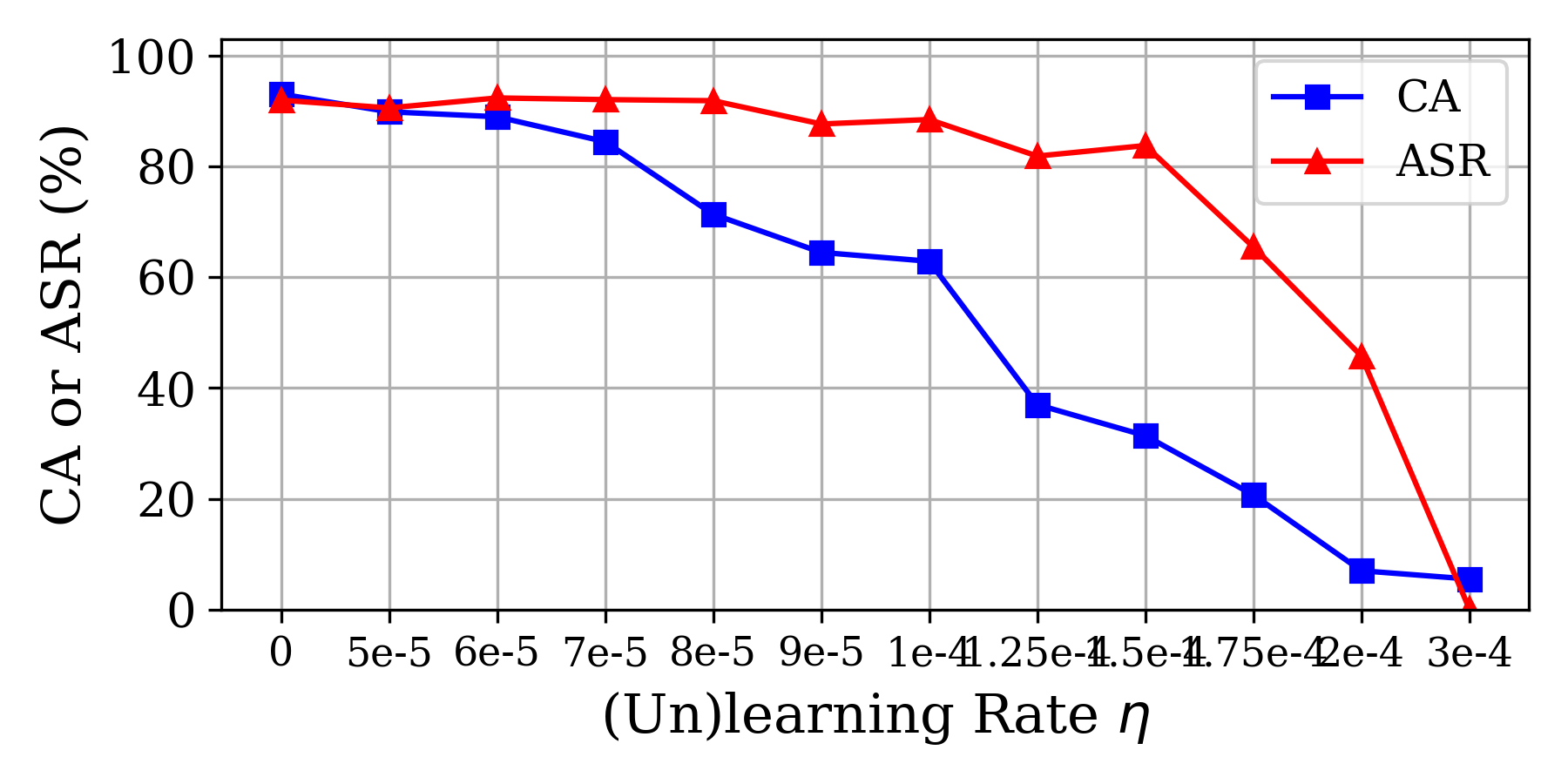}
        \caption{VGG16} \label{fig:unlearning_curve_VGG}
    \end{subfigure}

    
    
    
    
    
    
    

    
    
    

\caption{Unlearning curves with different $\eta$ of diverse scenarios. Fig~\ref{fig:unlearning_curve_BadNet}$\sim$\ref{fig:unlearning_curve_SRA} correspond to attacks conducted on CIFAR10 (ResNet18); Fig~\ref{fig:unlearning_curve_GTSRB} and \ref{fig:unlearning_curve_ImageNet} correspond to Blending attacks on GTSRB and ImageNet (ResNet18), respectively; Fig~\ref{fig:unlearning_curve_VGG} corresponds to the Blending attack on CIFAR10 (VGG16 instead of ResNet18).}
\label{fig:unlearning-curves}
\end{figure}

For all 12 attacks evaluated in our primary experiments on CIFAR10, we visualize in Fig~\ref{fig:unlearning_curve_BadNet}$\sim$\ref{fig:unlearning_curve_SRA} the constructed backdoor experts' (Alg~\ref{alg:alg-training-BE}) CA and ASR, when we unlearn the originally backdoored model $\M$ with different $\eta$'s. As depicted, with a conservatively small $\eta$ (e.g. $10^{-4}$), we can always enforce the resulting backdoor expert to lose a significant amount of normal functionality (CA drop $\sim 50\%$), while retaining a similar backdoor functionality (ASR drop $\sim 0\%$). However, if we choose a large $\eta$ (e.g. 1e-3), both functionalities would be erased (both CA and ASR $\approx 0\%$). Actually, we sometimes may have to tradeoff between the maintenance of the backdoor functionality and the unlearning of the normal functionality. But overall, we can see that unlearning the backdoor functionality is \textbf{\textit{slower}} than unlearning the normal functionality. More crucially, we find this phenomenon to \textbf{consistently exist across datasets and architectures}~(Fig~\ref{fig:unlearning_curve_GTSRB}$\sim$~\ref{fig:unlearning_curve_VGG}).

An intuitive justification for such a phenomenon can be referred to and derived from \cite{qi2023proactive}, where the authors show that during poison training, clean samples could be forgotten faster than poison samples in the context of catastrophic forgetting (for simplified settings of training overparameterized linear model).


\begin{figure}
\centering
    \includegraphics[width=0.37\textwidth]{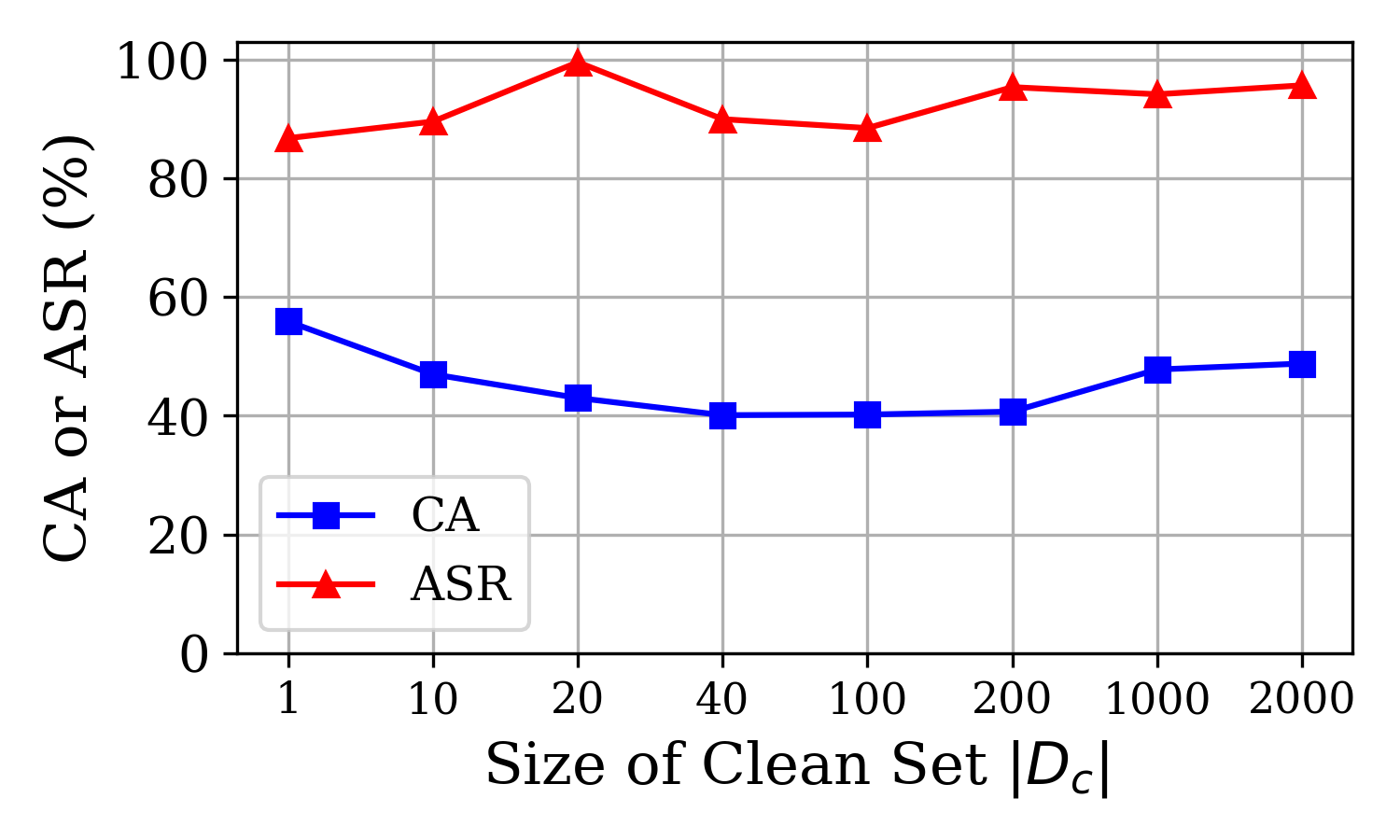}
    \caption{Backdoor experts constructed with reserved clean sets of different sizes.}
    \label{fig:asr_ca_clean_set_size}
\end{figure}

We also notice that, this property is \textit{not sensitive w.r.t. the number of clean samples ($|D_c|$)} we have. No matter how many clean samples (from 1 or 10 to 1,000 or 2,000) we use to conduct Alg~\ref{alg:alg-training-BE}, with an appropriately small learning rate $\eta$ (e.g. by selecting a small $\eta$ such that the resulting CA only drops to $\sim 40\%$), we can still separate the CA and ASR by a certain extent (Fig~\ref{fig:asr_ca_clean_set_size}).

\subsection{Ablation Studies on Different (Un-)Learning Rates $\eta$}
\label{appendix:ablation-unlearning-rates}

\begin{table}
\centering
\caption{Ablation results on different (un-)learning rates $\eta$. AUROCs of BaDExpert are reported on CIFAR10.}
\resizebox{0.99\linewidth}{!}{
\begin{tabular}{lcccccccccccc}
\toprule
 \textbf{(Un-)learning Rate $\eta\rightarrow$} & $5 * 10^{-5}$ & $6 * 10^{-5}$ & $7 * 10^{-5}$ & $8 * 10^{-5}$ & $9 * 10^{-5}$ & $1 * 10^{-4}$ & $2 * 10^{-4}$ & $3 * 10^{-4}$ & $4 * 10^{-4}$ & $5 * 10^{-4}$ & $1 * 10^{-3}$ & $5 * 10^{-3}$ \cr
\midrule

BadNet & 100.0 & 100.0 & 100.0 & 100.0 & 100.0 & 100.0 & 100.0 & 100.0 & 100.0 & 100.0 & 99.9 & 90.8 \cr

\midrule

Blend & 97.6 & 98.1 & 98.6 & 98.7 & 99.0 & 99.2 & 99.2 & 99.0 & 99.1 & 99.2 & 95.2 & 92.5 \cr
\bottomrule

\end{tabular}
} 
\label{tab:ablation_unlearning_rates}
\end{table}

Our method is not sensitive to the choice of (un-)learning rate $\eta$. In this ablation study, we evaluate BadExpert against BadNet and Blend attacks on CIFAR10, across a wide range of $\eta$ (from $5\cdot 10^{-5}$ to $5 \cdot 10^{-3}$). As shown in Table~\ref{tab:ablation_unlearning_rates}, the AUROC of our detection remains stable across different $\eta$.

\subsection{Ablation Studies on Different Poison Rates}
\label{appendix:ablation-poison-rates}

\begin{table}
\centering
\caption{Ablation results on different poison rates. AUROCs of BaDExpert are reported on CIFAR10 against Blend attack.}
\resizebox{0.6\linewidth}{!}{ 
\begin{tabular}{lcccccccc}
\toprule
 \textbf{Poison Rate $\rightarrow$} & 
0.1\% & 0.2\% & 0.3\% & 0.4\% & 0.5\% & 1.0\% & 5.0\% & 10.0 \cr
\midrule

CA (before defense) & 93.6 & 94.1 & 94.1 & 93.5 & 94.2 & 93.1 & 93.9 & 94.0 \cr
\midrule

ASR (before defense) & 80.8 & 87.3 & 90.6 & 95.1 & 94.5 & 98.5 & 99.8 & 99.9 \cr
\midrule

AUROC & 99.2 & 99.2 & 99.2 & 99.1 & 99.4 & 98.9 & 100.0 & 100.0 \cr
\bottomrule

\end{tabular}
} 
\label{tab:ablation_poison_rates}
\end{table}

As shown in Table~\ref{tab:ablation_poison_rates}, BaDExpert is insensitive to the variation of poison rate. Even when the poison rate is extremely low (poison rate = $0.1\%$, equivalent to only 50 poison samples) and ASR drops to $\sim80\%$, our BaDExpert is still manifesting near-perfect effectiveness.

\subsection{Comparing BaDExpert with STRIP and ScaleUp in Scenarios with Fewer Clean Samples}
\label{appendix:ablation-comparing-with-STRIP-and-ScaleUp-with-fewer-clean-samples}

\begin{table}
\centering
\caption{Comparison of BaDExpert with STRIP and ScaleUp alongside, when they are all given equal access to different number of clean samples. AUROCs of all defenses are reported on CIFAR10 against Blend attack.}
\resizebox{0.45\linewidth}{!}{
\begin{tabular}{lcccccccc}
\toprule
Number of Clean Samples & Defense & BadNet & Blend \cr
\midrule

\multirow{3}{*}{100}
& ScaleUp & 95.7 & 79.8 \cr
& STRIP & 99.2 & 42.5\cr
& \textbf{BaDExpert} & \textbf{99.9} & 72.9 \cr
\midrule

\multirow{3}{*}{200}
& ScaleUp & 95.7 & 79.9 \cr
& STRIP & 99.3 & 44.4 \cr
& \textbf{BaDExpert} & \textbf{100.0} & \textbf{91.2} \cr
\midrule

\multirow{3}{*}{400}
& ScaleUp & 95.7 & 79.9 \cr
& STRIP & 99.3 & 44.3 \cr
& \textbf{BaDExpert} & \textbf{100.0} & \textbf{95.1} \cr
\midrule

\multirow{3}{*}{2000}
& ScaleUp & 95.8 & 80.0 \cr
& STRIP & 99.4 & 44.6 \cr
& \textbf{BaDExpert} & \textbf{100.0} & \textbf{99.2} \cr
\bottomrule

\end{tabular}
} 
\label{tab:comparing_with_STRIP_and_ScaleUp}
\end{table}

We further analyze in details how our BaDExpert would perform when there are fewer accessible clean reserved samples. We compare it alongside with two strong baseline detectors: 1) STRIP, which requires only $N=100$ clean samples by default, and 2) ScaleUp, which by default can work without clean samples (meanwhile, ScaleUp can also utilize available clean samples for its SPC value estimation). To make a fair comparison in our additional study, we assign an equal number of clean samples (100, 200, 400 and 2000) to all three defenses in each scenario.

In Table~\ref{tab:comparing_with_STRIP_and_ScaleUp}, we can quickly notice that \textbf{under most circumstances (number of clean samples = $200$, $400$ and $2,000$), BaDExpert performs the best}. And when the number of clean samples is extremely restrited to $100$, BaDExpert becomes less effective on Blend ($72.9\%$ AUROC), but still performs similar to ScaleUp ($79.8\%$ AUROC).

\paragraph{Approaches to Obtain Reserved Clean Samples.} There exist various possiblitites to acquire such a small clean set $|D_c|$:
\begin{itemize}
    \item A developed model usually comes with an independent set of test samples in order to validate the model' utility, the defender could simply use (a partition) of this test samples as $|D_c|$;
    \item Collect data (e.g. taking photos) in a secure environment;
    \item Manually inspect and isolate clean samples from online inputs.
\end{itemize}
Furthermore, there exists a trending line of work on isolating clean samples from a poisoned set of samples, e.g. META-SIFT~\cite{287222}. Such work can be directly applied here to aid the defender to obtain reserved clean samples. Overall, we argue that obtaining a small amount of clean samples (e.g. $200$, $400$, $2000$) would not be a major bottleneck for defenders in practice.

\subsection{Additional Adaptive Analysis Details}
\label{appendix:additional-adaptive-analysis}

\subsubsection{Natural Backdoor}
\label{appendix:additional-adaptive-analysis-natural-backdoor}

Natural backdoors (e.g. \cite{zhao2021deep}) may exist in models trained on only clean data, which can also be considered as an adaptive backdoor attack against BaDExpert. This is because the natural backdoor functionality learned by a clean model is strongly correlated with the normal functionality -- therefore, such natural backdoors also directly challenge our insight that ``\textit{backdoor functionality can be isolated from normal functionality}''.

We construct such natural backdoor attacks~\cite{zhao2021deep} on normally trained models (on clean data) and achieve ASR = 99.2\% on CIFAR10 and ASR = 99.3\% on GTSRB. We find that \textbf{BaDExpert can defend the natural backdoor attacks on both datasets with AUROC = 92.3\% and AUROC = 92.8\%}, respectively. While the performance shows a degradation compared to our major results in Table 2 in our paper (average AUROC = 99.0\%), the >92\% AUROC still reflects the nontrivial effectiveness of BaDExpert as an inference-time defense. Besides, we note that our key finding ``when fintuning on mislabled clean data, the backdoor functionality would remain, while normal functionality does not'' \textbf{still stands} against this attack:

\begin{itemize}
    \item Following Alg~\eqref{alg:alg-training-BE} on CIFAR10, the constructed backdoor expert $\B$ retains a high ASR 95.3\% (originally 99.2\%), while the normal functionality degrades significantly (CA drops from 94.2\% to 44.8\%);
    \item On GTSRB, the observation is similar (ASR drops from 99.3\% to 83.0\%, CA drops from 96.8\% to 41.0\%).
\end{itemize}

\subsubsection{Adaptive Attack by Using Weakened Triggers at Inference Time}
\label{appendix:additional-adaptive-analysis-weakened-triggers}

\begin{table}
\centering
\caption{Adaptive attack by using weakened triggers at inference time. (Blend attack on CIFAR10)}
\resizebox{0.85\linewidth}{!}{
\begin{tabular}{lcccccccccccc}
\toprule
 \textbf{Blending Alpha} & 10\% & 11\% & 12\% & 13\% & 14\% & 15\% & 16\% & 17\% & 18\% & 19\% & 20\% (Standard) \cr
\midrule

ASR (before defense) & 5.0 & 10.1 & 17.8 & 28.0 & 39.7 & 51.6 & 62.8 & 72.3 & 80.2 & 86.3 & 90.6 \cr
\midrule

AUROC & 94.4 & 96.2 & 97.3 & 97.8 & 98.1 & 98.4 & 98.6 & 98.7 & 98.9 & 99.0 & 99.2 \cr
\bottomrule

\end{tabular}
} 
\label{tab:adaptive_weakened_triggers}
\end{table}

Empirically, adversary may sometimes bypass existing backdoor defense methods by using a weakened version of the backdoor trigger at inference time. Therefore, we also study how our method reacts to such an adaptive attack by \textbf{decreasing the inference-time trigger blending alpha} of the Blend attack on CIFAR10. As shown in Table~\ref{tab:adaptive_weakened_triggers}, when the adversary uses a lower blending alpha at inference time (instead of the default 20\% during poisoning), the AUROC of BaDExpert indeed degrades, to as low as ~94\%. Nevertheless, the attack ASR drops more rapidly. Overall, we can observe a tradeoff between backdoor inputs' stealthiness (AUROC) and the attack's effectiveness (ASR). Generally speaking, \textbf{the adversary can hardly evade BaDExpert's detection by using a weaker trigger at inference time}, since the ASR will drop rapidly way before BaDExpert becomes unusable (\textbf{AUROC > 97.5\% whenever ASR > 20\%}).

\subsubsection{A Tailored Adaptive Attack against BaDExpert}
\label{appendix:tailored-adaptive-attack}

To provide more valuable insights for future researchers / developers into our method, we also tailored a novel adaptive attack against our proposed BaDExpert.

Concretely, we assume that the adversary (who produced the backdoor model) is aware of our BaDExpert defense. Accordingly, he/she will design an evading strategy via \textbf{using an alternative trigger during inference time} (i.e., \textit{asymmetric} backdoor trigger, which was adopted in prior work like \cite{qi2023revisiting}), such that the alternative trigger satisfies:
1. Can still activate the backdoor functionality of the original backdoor model $\mathcal M$ (i.e., achieving a high ASR)
2. Cannot activate the backdoor functionality of the original backdoor model $\mathcal B$ (i.e., enforcing $\mathcal B$ to provide a low confidence on the backdoor inputs).

Notice that \textbf{the adversary is only assumed to have access to the backdoor model $\mathcal M$}. The adversary, however, could also follow our BaDExpert procedure to construct a surrogate backdoor expert model $\mathcal B'$ (Alg 1) to the actual backdoor expert model $\mathcal B$ used by the victim / defender. (Empirically, we find the attack results using this surrogate backdoor expert model are similar to using the actual $\mathcal B$ of the defender, if they both are trained following the same configuration. We therefore do not explicitly distinguish the notation of $\mathcal B$ and $\mathcal B'$ in the rest of this section.)

The two goals upon can be implemented via optimization on the models $\mathcal M$ and $\mathcal B$. Formally, as the adaptive adversary, we compute an alternative trigger mark $\mathbf \Delta$ and the corresponding trigger mask $\mathbf m$ as follow:
\begin{align}
\min_{\mathbf \Delta, \mathbf m}\quad &
\text{CrossEntropyLoss}\Big(\mathcal M(x^*)_\text{raw}, t\Big) + \lambda_1\cdot \text{Conf}_{\mathcal B}(t|x^*) + \lambda_2 \cdot |\mathbf m|
&(x,y)\sim \mathcal{P}
\\
\text{where}\quad &x^* = (1 - \mathbf m) \odot x + \mathbf m \odot \mathbf \Delta
\end{align}
Specifically, the trigger mark $\mathbf \Delta\in \mathcal X = [0, 1]^{c\times w\times h}$ is a 3-dim real-value matrix (the same shape as the input $x$), while the trigger mask $\mathbf m \in [0, 1]^{w\times h}$ is a 2-dim real-value matrix that decides the opacity of each trigger mark pixel to mix with the corresponding input $x$s' pixel. The $\odot$ operator upon computes the Hadamard product of two matrices. (We adopt the similar trigger optimization setting in \cite{wang2019neural}.)

Now we explain this optimization formula:
\begin{itemize}
    \item Minimizing the first \textbf{CrossEntropyLoss} term will help the adversary approach the first target: the alternative trigger ($\mathbf \Delta, \mathbf m$) can still activate the victim backdoor model $\mathcal M$ with a high ASR.
    \item The second term represents the \textbf{confidence of the backdoor expert model $\mathcal B$}. Minimizing it will \textbf{directly violate our core design insight}, since the resulting backdoor inputs now will not activate the backdoor functionality of the backdoor expert model $\mathcal B$ anymore.
    \item The last term, $\ell_1$ norm of the trigger mask, corresponds to the \textbf{visual stealthiness of the trigger}. The adversary usually prefers to implant a less noticeable trigger, which can be realized via minimizing the magnitude of the trigger mask $\mathbf m$.
\end{itemize}

The two hyperparameter $\lambda_1$ and $\lambda_2$ controls the tradeoff among these three goals.

In Table~\ref{tab:adaptive_tailored}, we report the results of this adaptive attack (``\textbf{BaDExpert-Adap-[ORIGINAL ATTACK]}'') against our BaDExpert defense when $\lambda_1$ and $\lambda_2$ are selected differently. To fully study the potential tradeoff between the adaptive attack effectiveness and its defense evasiveness, we report alongside 1) the adaptive trigger \textbf{Norm} $|\mathbf m|$; 2) the attack success rate (\textbf{ASR}); 3) \textbf{AUROC} of BaDExpert against the tailored adaptive attack.

\begin{table}
\centering
\caption{\small Tailored adaptive attack against BaDExpert. (CIFAR10)}
\resizebox{0.8\linewidth}{!}{
\begin{tabular}{lcccccccccc}
\toprule
 \multirow{3}{*}{\textbf{Attack}}
& $\lambda_2$ &
\multicolumn{2}{c}{0.5} &
\multicolumn{2}{c}{0.1} &
\multicolumn{2}{c}{0.05} &
\multicolumn{2}{c}{0.01} & 
\multirow{3}{*}{\textbf{Average}} \cr
\cmidrule(lr){3-4} \cmidrule(lr){5-6} \cmidrule(lr){7-8} \cmidrule(lr){9-10}
& $\lambda_1$ & 10 & 1 & 10 & 1 & 10 & 1 & 10 & 1 & \cr
\midrule

\multirow{3}{*}{BaDExpert-Adap-BadNet}
& Norm & 4.8 & 1.8 & 16.6 & 5.0 & 36.7 & 5.1 & 59.8 & 5.3 & 16.9 \cr
& ASR & 20.8 & 57.6 & 61.8 & 83.2 & 82.5 & 89.2 & 94.5 & 98.1 & 73.5 \cr
& AUROC & 93.3 & 99.8 & 80.7 & 99.8 & 65.4 & 99.8 & 57.2 & 100.0 & 87.0 \cr
\midrule

\multirow{3}{*}{BaDExpert-Adap-Blend}
& Norm & 11.4 & 10.2 & 32.6 & 29.5 & 44.2 & 38.6 & 59.1 & 53.5 & 34.9 \cr
& ASR & 11.4 & 27.2 & 72.7 & 75.7 & 83.7 & 86.5 & 96.7 & 97.2 & 68.9 \cr
& AUROC & 63.0 & 72.7 & 61.1 & 68.7 & 47.1 & 58.1 & 36.4 & 54.2 & 57.6 \cr
\bottomrule

\end{tabular}
}
\label{tab:adaptive_tailored}
\end{table}

As shown in Table~\ref{tab:adaptive_tailored}, we can observe that the tailored adaptive attack could effectively diminish the effectiveness of BaDexpert -- AUROC becomes as low as $57.2\%$ against BaDExpert-Adap-BadNet and $36.4\%$ against BaDExpert-Adap-Blend. Meanwhile, we can also notice two trends from the table:
\begin{enumerate}
    \item As $\lambda_2$ becomes larger (with $\lambda_1$ fixed), the \textbf{ASR} becomes higher, while the defense \textbf{AUROC} decreases. However, the increasing evasiveness of the adaptive attack comes with a price: the magnitude of the trigger mask (\textbf{Norm}) also increases -- i.e., the backdoor attack becomes less stealthy.
    \item As $\lambda_1$ becomes higher (with $\lambda_2$ fixed), the defense \textbf{AUROC} effectively degrades. However: 1) the \textbf{ASR} drops, and 2) the backdoor attack becomes stealthier since \textbf{Norm} becomes larger.
\end{enumerate}

In brief, we can see that \textbf{the adaptive attack can indeed restrict BaDExpert's performance, at the cost of either attack effectiveness (lower ASR) or stealthiness (higher Norm)}. Overall, BaDExpert still performs nontrivially in most scenarios -- $87.0\%$ average AUROC against BaDExpert-Adap-BadNet and $57.6\%$ against BaDExpert-Adap-Blend.

\subsection{Comparing BaDExpert with Two Additional Recent Baselines: ANP~\cite{wu2021adversarial} and I-BAU~\cite{zeng2021adversarial}}
\label{appendix:comparing_with_ANP_and_IBAU}

\begin{table}
\centering
\caption{Comparing BaDExpert alongside two model-repairing baseline defenses, ANP~\cite{wu2021adversarial} and I-BAU~\cite{zeng2021adversarial}. CA and ASR are reported on CIFAR10.}
\resizebox{0.55\linewidth}{!}{
\begin{tabular}{lccccc}
\toprule
 Attack & & No Defense & ANP~\cite{wu2021adversarial} & I-BAU~\cite{zeng2021adversarial} & \textbf{BaDExpert (Ours)} \cr
\midrule

\multirow{2}{*}{BadNet}
& CA & 94.1 & 80.8 & 88.0 & \textbf{93.1}\cr
& ASR & 100.0 & 0.4 & 0.8 & \textbf{0.0}\cr
\midrule

\multirow{2}{*}{Blend}
& CA & 94.1 & 83.2 & 90.4 & \textbf{93.1}\cr
& ASR & 90.6 & 16.8 & 16.6 & \textbf{11.4}\cr
\bottomrule

\end{tabular}
} 
\label{tab:comparing_with_ANP_and_IBAU}
\end{table}

\begin{table}
\centering
\caption{Ensembling ANP~\cite{wu2021adversarial} and I-BAU~\cite{zeng2021adversarial} with backdoor experts. AUROCs are reported on CIFAR10. Configurations are the same as Table~\ref{tab:auroc-ensembling-with-other-defenses}.}
\resizebox{0.8\linewidth}{!}{
\begin{tabular}{lcccc}
\toprule
 Attack & ANP w/o Backdoor Expert & ANP w/ Backdoor Expert & I-BAU w/o Backdoor Expert & I-BAU w Backdoor Expert \cr
\midrule

BadNet & 94.3 & \textbf{100.0} & 99.6 & \textbf{100.0} \cr
\midrule

Blend & 87.7 & \textbf{98.8} & 97.0 & \textbf{99.1}\cr
\bottomrule

\end{tabular}
} 
\label{tab:ensembling_with_ANP_and_IBAU}
\end{table}

In Table~\ref{tab:comparing_with_ANP_and_IBAU}, we further showcase that BaDExpert outperforms two other recent model-repairing baselines, ANP~\cite{wu2021adversarial} and I-BAU~\cite{zeng2021adversarial}, w.r.t. attack success rate and clean accuracy. Furthermore, as shown in Table~\ref{tab:ensembling_with_ANP_and_IBAU}, when ensembling our backdoor expert models $\B$ with models repaired by ANP and I-BAU (as $\Mprime$), our BaDExpert framework consistently provides extra boostup w.r.t. AUROC.

\subsection{Standard Deviations of Major Experiments}
\label{appendix:stdev-major-experiments}

Standard deviations of our major experiments Table~\ref{tab:main_cifar10}, \ref{tab:auroc_cifar10}, \ref{tab:main_gtsrb} and \ref{tab:auroc_gtsrb} are shown in Table~\ref{tab:stdev_main_cifar10}, \ref{tab:stdev_auroc_cifar10}, \ref{tab:stdev_main_gtsrb} and \ref{tab:stdev_auroc_gtsrb}.

\begin{table}[H]
\centering
\caption{Standard deviation of Table~\ref{tab:main_cifar10}.}
\resizebox{0.99\linewidth}{!}{
\begin{tabular}{ccccccccccccccccccccc}
\toprule
\textbf{Defenses$\rightarrow$} &
\multicolumn{2}{c}{No Defense} &
\multicolumn{2}{c}{FP} &
\multicolumn{2}{c}{NC} & 
\multicolumn{2}{c}{MOTH} & 
\multicolumn{2}{c}{NAD} & 
\multicolumn{2}{c}{STRIP} & 
\multicolumn{2}{c}{AC} & 
\multicolumn{2}{c}{Frequency} & 
\multicolumn{2}{c}{SCALE-UP} & 
\multicolumn{2}{c}{\textbf{BaDExpert}} \cr
\cmidrule(lr){2-3} \cmidrule(lr){4-5} \cmidrule(lr){6-7} \cmidrule(lr){8-9} \cmidrule(lr){10-11} \cmidrule(lr){12-13} \cmidrule(lr){14-15} \cmidrule(lr){16-17} \cmidrule(lr){18-19} \cmidrule(lr){20-21} 
\textbf{Attacks} $\downarrow$ & CA & ASR & CA & ASR & CA & ASR & CA & ASR & CA & ASR & CA & ASR & CA & ASR & CA & ASR & CA & ASR & CA & ASR \cr
\midrule
No Attack & 0.3 & - & 1.1 & - & 1.4 & - & 0.3 & - & 0.6 & - & 0.3 & - & 0.4 & - & 0.2 & - & 0.0 & - & 0.3 & - \cr
\midrule
\multicolumn{19}{c}{\centering \textbf{Development-Stage Attacks}} \cr
\midrule
BadNet & 0.1 & 0.0 & 0.4 & 0.0 & 0.2 & 2.4 & 0.4 & 0.2 & 0.5 & 0.4 & 0.1 & 0.1 & 0.1 & 0.1 & 0.1 & 0.0 & 0.1 & 0.0 & 0.1 & 0.0  \cr
\midrule
Blend & 0.1 & 1.3 & 0.7 & 1.4 & 0.6 & 1.7 & 0.1 & 11.5 & 0.4 & 4.7 & 0.1 & 1.1 & 0.6 & 14.4 & 0.1 & 0.2 & 0.3 & 2.8 & 0.1 & 6.5  \cr
\midrule
Trojan & 0.2 & 0.1 & 0.7 & 36.0 & 0.2 & 0.5 & 0.5 & 2.1 & 0.2 & 2.8 & 0.2 & 15.3 & 0.2 & 0.4 & 0.2 & 0.0 & 2.5 & 5.8 & 0.2 & 3.3  \cr
\midrule
CL & 0.2 & 0.0 & 0.9 & 7.3 & 0.2 & 0.5 & 0.2 & 0.5 & 0.4 & 3.5 & 0.2 & 3.7 & 0.2 & 0.3 & 0.3 & 0.0 & 1.1 & 0.0 & 0.2 & 7.1  \cr
\midrule
SIG & 0.4 & 0.7 & 2.6 & 25.8 & 0.4 & 7.2 & 0.2 & 50.9 & 0.2 & 6.0 & 0.4 & 3.2 & 0.4 & 12.1 & 0.4 & 0.2 & 0.5 & 24.6 & 0.4 & 1.7  \cr
\midrule
Dynamic & 0.2 & 0.2 & 1.3 & 11.8 & 0.1 & 5.0 & 0.5 & 41.2 & 0.5 & 1.3 & 0.3 & 4.5 & 4.3 & 43.2 & 0.2 & 0.0 & 0.9 & 0.2 & 0.2 & 5.4  \cr
\midrule
ISSBA & 0.3 & 0.0 & 0.2 & 0.2 & 0.3 & 2.0 & 0.3 & 34.3 & 0.6 & 0.3 & 0.2 & 2.1 & 0.3 & 0.1 & 0.3 & 0.0 & 0.1 & 0.1 & 0.3 & 0.5  \cr
\midrule
WaNet & 0.6 & 0.9 & 1.8 & 2.6 & 0.8 & 20.3 & 0.4 & 10.1 & 1.3 & 0.4 & 0.5 & 0.7 & 0.6 & 3.8 & 0.5 & 0.9 & 0.1 & 1.6 & 0.5 & 0.3  \cr
\midrule
BPP & 0.5 & 0.2 & 0.4 & 8.8 & 0.3 & 57.3 & 0.2 & 1.5 & 0.7 & 0.2 & 0.4 & 3.7 & 0.5 & 0.6 & 0.6 & 0.0 & 4.7 & 10.2 & 0.5 & 0.1  \cr
\midrule
\multicolumn{19}{c}{\centering \textbf{Post-development Attacks}} \cr
\midrule
FT & 0.2 & 0.2 & 0.8 & 4.6 & 0.4 & 16.2 & 0.5 & 5.1 & 0.8 & 1.0 & 0.3 & 3.3 & 0.2 & 9.6 & 0.2 & 0.2 & 0.2 & 7.4 & 0.2 & 1.8  \cr
\midrule
TrojanNN & 0.2 & 0.0 & 0.4 & 9.6 & 0.1 & 0.3 & 0.4 & 23.4 & 0.3 & 7.7 & 0.2 & 0.1 & 0.2 & 0.0 & 0.2 & 0.0 & 1.4 & 0.0 & 0.3 & 6.6  \cr
\midrule
SRA & 1.4 & 0.0 & 1.2 & 0.0 & 0.4 & 0.6 & 0.4 & 0.5 & 2.6 & 0.4 & 1.2 & 0.9 & 1.4 & 0.0 & 1.0 & 0.0 & 0.8 & 5.3 & 1.4 & 0.4  \cr
\bottomrule

\end{tabular}
} 
\label{tab:stdev_main_cifar10}
\end{table}

\begin{table}[H]
\centering
\caption{Standard deviation of Table~\ref{tab:auroc_cifar10}}
\resizebox{0.85\linewidth}{!}{
\begin{tabular}{lcccccccccccc}
\toprule
 \textbf{AUROC} ($\%$) & 
\multicolumn{1}{c}{BadNet} &
\multicolumn{1}{c}{Blend} &
\multicolumn{1}{c}{Trojan} &
\multicolumn{1}{c}{CL} &
\multicolumn{1}{c}{SIG} &
\multicolumn{1}{c}{Dynamic} &
\multicolumn{1}{c}{ISSBA} &
\multicolumn{1}{c}{WaNet} &
\multicolumn{1}{c}{Bpp} &
\multicolumn{1}{c}{FT} &
\multicolumn{1}{c}{TrojanNN} &
\multicolumn{1}{c}{SRA} \cr
\midrule

STRIP & 0.2 & 2.2 & 10.6 & 3.4 & 6.8 & 2.0 & 1.4 & 0.8 & 13.1 & 1.4 & 0.3 & 0.5 \cr
\midrule

AC & 0.0 & 10.1 & 0.1 & 0.0 & 2.0 & 7.2 & 2.8 & 0.4 & 0.2 & 15.3 & 0.0 & 0.0 \cr
\midrule

Frequency & 0.1 & 0.1 & 0.0 & 0.1 & 0.0 & 0.1 & 0.1 & 0.2 & 0.1 & 0.1 & 0.1 & 0.1 \cr
\midrule

SCALE-UP & 0.6 & 1.5 & 4.4 & 0.7 & 1.8 & 0.4 & 0.4 & 0.8 & 8.9 & 2.6 & 0.8 & 2.4 \cr
\midrule

\textbf{BaDExpert} & 0.0 & 0.1 & 0.2 & 0.4 & 0.1 & 0.1 & 1.0 & 0.0 & 0.0 & 0.1 & 0.6 & 0.0 \cr
\bottomrule

\end{tabular}
} 
\label{tab:stdev_auroc_cifar10}
\end{table}

\begin{table}[H]
\centering
\caption{Standard Deviation of Table~\ref{tab:main_gtsrb}.}
\resizebox{0.99\linewidth}{!}{
\begin{tabular}{ccccccccccccccccccccc}
\toprule
\textbf{Defenses$\rightarrow$} &
\multicolumn{2}{c}{No Defense} &
\multicolumn{2}{c}{FP} &
\multicolumn{2}{c}{NC} & 
\multicolumn{2}{c}{MOTH} & 
\multicolumn{2}{c}{NAD} & 
\multicolumn{2}{c}{STRIP} & 
\multicolumn{2}{c}{AC} & 
\multicolumn{2}{c}{Frequency} & 
\multicolumn{2}{c}{SCALE-UP} & 
\multicolumn{2}{c}{\textbf{BaDExpert}} \cr
\cmidrule(lr){2-3} \cmidrule(lr){4-5} \cmidrule(lr){6-7} \cmidrule(lr){8-9} \cmidrule(lr){10-11} \cmidrule(lr){12-13} \cmidrule(lr){14-15} \cmidrule(lr){16-17} \cmidrule(lr){18-19} \cmidrule(lr){20-21}
\textbf{Attacks} $\downarrow$ & CA & ASR & CA & ASR & CA & ASR & CA & ASR & CA & ASR & CA & ASR & CA & ASR & CA & ASR & CA & ASR & CA & ASR \cr
\midrule
No Attack & 0.4 & - & 1.8 & - & 0.5 & - & 3.7 & - & 0.5 & - & 0.4 & - & 0.4 & - & 0.3 & - & 3.5 & - & 0.4 & - \cr
\midrule
\multicolumn{19}{c}{\centering \textbf{Development-Stage Attacks}} \cr
\midrule
BadNet & 0.6 & 0.0 & 1.7 & 11.5 & 1.0 & 0.0 & 0.6 & 0.0 & 0.3 & 0.1 & 0.8 & 1.7 & 0.8 & 0.1 & 0.6 & 0.0 & 2.7 & 0.1 & 0.6 & 0.1 \cr
\midrule
Blend & 0.5 & 0.3 & 0.7 & 0.8 & 0.1 & 1.1 & 1.2 & 1.1 & 0.3 & 13.0 & 0.5 & 5.6 & 0.5 & 0.2 & 0.4 & 0.1 & 0.9 & 4.3 & 0.5 & 2.9 \cr
\midrule
Trojan & 0.2 & 0.0 & 0.2 & 1.1 & 0.3 & 0.1 & 3.5 & 0.0 & 0.6 & 56.8 & 0.2 & 3.9 & 0.2 & 0.0 & 0.2 & 0.0 & 1.1 & 2.2 & 0.0 & 2.2 \cr
\midrule
Dynamic & 0.1 & 0.0 & 0.1 & 0.0 & 0.6 & 56.7 & 3.5 & 0.0 & 0.3 & 49.2 & 0.1 & 2.9 & 0.1 & 0.0 & 0.1 & 0.0 & 1.6 & 6.6 & 0.1 & 0.1 \cr
\midrule
WaNet & 0.2 & 0.5 & 2.1 & 5.9 & 0.2 & 20.3 & 0.5 & 0.4 & 0.1 & 0.1 & 0.1 & 0.6 & 0.2 & 0.5 & 0.2 & 0.4 & 0.8 & 0.3 & 0.2 & 0.0 \cr
\midrule

\multicolumn{19}{c}{\centering \textbf{Post-development Attacks}} \cr
\midrule
FT & 0.6 & 0.1 & 0.5 & 0.0 & 0.6 & 19.9 & 4.7 & 1.3 & 0.2 & 18.2 & 0.6 & 4.0 & 0.6 & 0.1 & 0.4 & 0.0 & 1.1 & 1.7 & 0.6 & 4.7 \cr
\midrule
TrojanNN & 0.7 & 2.4 & 1.0 & 14.7 & 0.7 & 0.3 & 7.0 & 3.2 & 0.3 & 0.2 & 0.6 & 10.5 & 0.7 & 2.4 & 0.5 & 0.0 & 1.0 & 3.3 & 0.8 & 1.8 \cr
\bottomrule

\end{tabular}
} 
\label{tab:stdev_main_gtsrb}
\end{table}

\begin{table}[H]
\centering
\caption{Standard deviation of Table~\ref{tab:auroc_gtsrb}.}
\resizebox{0.6\linewidth}{!}{
\begin{tabular}{lcccccccccccc}
\toprule
 \textbf{AUROC} ($\%$) & 
\multicolumn{1}{c}{BadNet} &
\multicolumn{1}{c}{Blend} &
\multicolumn{1}{c}{Trojan} &
\multicolumn{1}{c}{Dynamic} &
\multicolumn{1}{c}{WaNet} &
\multicolumn{1}{c}{FT} &
\multicolumn{1}{c}{TrojanNN} \cr
\midrule

STRIP & 1.8 & 2.6 & 2.7 & 1.6 & 1.1 & 1.9 & 3.7 \cr
\midrule

AC & 29.9 & 7.6 & 0.6 & 9.8 & 2.2 & 6.2 & 0.2 \cr
\midrule

Frequency & 0.1 & 0.0 & 0.0 & 0.1 & 0.1 & 0.0 & 0.0 \cr
\midrule

SCALE-UP & 0.6 & 1.9 & 4.9 & 2.8 & 0.3 & 1.2 & 1.5 \cr
\midrule

\textbf{BaDExpert} & 0.0 & 0.2 & 0.0 & 0.0 & 0.0 & 0.0 & 0.0 \cr
\bottomrule

\end{tabular}
} 
\label{tab:stdev_auroc_gtsrb}
\end{table}

\end{document}